\documentclass[aps,pra,twoside,twocolumn,10pt,longbibliography,nofootinbib,floatfix]{revtex4-1}
\usepackage[colorlinks=true, citecolor=red, urlcolor=blue ]{hyperref}
\usepackage{amssymb,amsfonts,amsmath,bm,subfigure,mathtools,amsthm,braket,soul,enumitem,color,geometry,comment}
\usepackage[normalem]{ulem}

\begin{document}

\title{Localizing genuine multipartite entanglement in noisy stabilizer states}
\author{Harikrishnan K. J. and Amit Kumar Pal}
\affiliation{Department of Physics, Indian Institute of Technology Palakkad, Palakkad 678 623, India}
\date{\today}

\begin{abstract}
Characterizing large noisy multiparty quantum states using genuine multipartite entanglement is a challenging task. In this paper, we calculate lower bounds of genuine multipartite entanglement localized over a chosen multiparty subsystem of multi-qubit stabilizer states in the noiseless and noisy scenario. In the absence of noise, adopting a graph-based technique, we perform the calculation for arbitrary graph states as representatives of the stabilizer states, and show that the graph operations required for the calculation has a polynomial scaling with the system size. As demonstrations, we compute the localized genuine multipartite entanglement over subsystems of large graphs having linear, ladder, and square structures. We also extend the calculation for graph states subjected to single-qubit Markovian or non-Markovian Pauli noise on all qubits, and demonstrate, for a specific lower bound of the localizable genuine multipartite entanglement corresponding to a specific Pauli measurement setup, the existence of a critical noise strength beyond which all of the post measured states are biseparable. The calculation is also useful for arbitrary large stabilizer states under noise due to the local unitary connection between stabilizer states and graph states. We demonstrate this by considering a toric code defined on a square lattice, and computing a lower bound of localizable genuine multipartite entanglement over a non-trivial loop of the code. Similar to the graph states, we show the existence of the critical noise strength in this case also, and discuss its interesting features.        
\end{abstract}

\maketitle

\section{Introduction}
\label{sec:intro}

In the last three decades, characterizing multiparty quantum states using multipartite entanglement~\cite{horodecki2009,guhne2009}  as figure of merit has been established as one of the most important and yet challenging problems. These quantum states are used as resource in quantum information processing tasks, such as measurement-based quantum computation~\cite{raussendorf2001,walther2005,briegel2009}, quantum cryptography~\cite{ekert1991,Karlsson1999,jennewein2000,Naik2000,gisin2002}, and quantum dense coding~\cite{bennett1992,mattle1996,sende2010} involving multiple senders and receivers~\cite{bruss2004,bruss2006,das2014,das2015}. Additionally, multipartite entanglement has also been proved a crucial ingredient in problems related to quantum many-body systems~\cite{amico2008,DeChiara_2018}, photosynthetic complexes~\cite{sarovar2010,zhu2012,lambert2013}, and even AdS/CFT correspondence in quantum gravity~\cite{Balasubramanian_2014,hubeny2015,pastawski2015,almheiri2015,jahn2017,Harper2021}. These observations have driven the world-wide effort to create multiparty-entangled quantum states in laboratory using  trapped ions~\cite{Leibfried2005,monz2011}, cold atoms and optical lattices~\cite{mandel2003,bloch2005,Treutlein2006,Cramer2013},  photons~\cite{prevedel2009,gao2010,yao2012,wang2018}, nuclear magnetic resonance~\cite{negrevergne2006}, and superconducting qubits~\cite{barends2014}.     

Among the multiparty quantum states of interest, stabilizer states~\cite{nielsen2010,Hayashi2008} form an extremely important class of states containing genuinely multipartite entanglement (GME) ~\cite{horodecki2009}, i.e., these states are not separable in any bipartition.  Stabilizer states are used widely in several areas of quantum information theory. Arguably, the most prominent examples of these states are the graph states~\cite{hein2006}, encompassing (a) the multi-qubit GHZ states~\cite{greenberger1989} having immense applications in quantum communication~\cite{bose1998,hillery1999} and quantum metrology~\cite{giovannetti2004}, and (b) the cluster states~\cite{hein2004} used as resource in one-way quantum computation, and in building quantum networks~\cite{Wallnofer2019}. Stabilizer states are also used in fault-tolerant quantum error corrections~\cite{Gottesman1997,gottesman2010}, and in establishing secure quantum communication~\cite{Dur2005,chen2007}. Experimental realization and manipulation of stabilizer states have also been possible~\cite{Wunderlich2010,nigg2014,bell2014,kelly2015,Greganti2015,linke2017,wright2019}, thereby highlighting the potential for testing theoretical results in the laboratory.  

Stabilizer states with a large number of qubits are required to perform quantum error correction where error occurs on multiple physical qubits. Therefore, it is natural to look for methodologies to characterize large stabilizer states as well as its subsystems under different types of noise using GME. In this respect, graph states have received a great deal of the attention due to the fact that all stabilizer states can be connected to graph states via local unitary transformations~\cite{van-den-nest2004}, leading to translation of the results on GME obtained for the graph states~\cite{hein2004,hein2006} to the case of arbitrary stabilizer states. However, while a few studies exist on characterizing noisy graph states using bipartite~\cite{Cavalcanti2009,Aolita2010} and genuine multipartite entanglement~\cite{hein2005,ali2014,ali2014a}, in general, calculation for mixed states in the presence of noise becomes difficult due to the scarcity of computable GME measures for mixed states~\cite{horodecki2009}. Therefore, a full understanding of the behaviour of GME in noisy stabilizer states and it its different multiparty subsystems, is far from complete. This brings us to the investigation of the GME over chosen subsystems of a multi-qubit system described by a stabilizer state with and without noise, which is the focus of this paper.    

Towards this, we adopt a measurement-based protocol~\cite{divincenzo1998,verstraete2004,popp2005} for quantifying entanglement in subsystems of stabilizer states, which has been established as an appropriate approach~\cite{hein2006,amaro2018,amaro2020a} over the partial-trace based methods~\cite{horodecki2009}. This leads to \emph{localizable entanglement} (LE) -- the maximum average entanglement over the chosen subsystem, maximized over all possible single-qubit projection measurements over all qubits in the rest of the system~\cite{verstraete2004,popp2005} -- defined in the same vein as the \emph{entanglement of assistance}~\cite{divincenzo1998}. Apart from being crucial in quantifying two-qubit entanglement in pure~\cite{hein2006} as well as noisy stabilizer states~\cite{amaro2018,amaro2020a}, LE and related ideas have also been proven a key element in studying the entanglement length~\cite{verstraete2004,popp2005,verstraete2004a,jin2004} and quantum phase transitions~\cite{skrovseth2009,smacchia2011,montes2012,Hk1} in quantum spin models, and in defining entanglement percolation~\cite{acin2007} in quantum networks. Extensive investigations into the features of LE in generic multi-qubit quantum states in the presence of noise have also been performed~\cite{Wang2015,Banerjee2020,Banerjee2022,krishnan2022}.  

The studies on the role of LE in characterizing noiseless and noisy quantum states have mostly considered a bipartite split of the chosen subsystem -- often a two-qubit subsystem~\cite{verstraete2004,popp2005,hein2006,amaro2018,amaro2020a,verstraete2004a,jin2004,skrovseth2009,smacchia2011,montes2012,Hk1,acin2007}. The exponential growth of the Hilbert space with the number of qubits  has also  confined these studies to systems of small number of qubits. While exploration of the behaviour of the localizable GME (LGME) on multiparty subsystems of paradigmatic pure states have been attempted~\cite{sadhukhan2017}, LGME on subsystems of noisy multipartite quantum states of arbitrary size, to the best of our knowledge, remain unexplored.

In this paper, we focus specifically on arbitrary stabilizer states~\cite{nielsen2010,Hayashi2008} in noiseless scenarios as well as under noise corresponding to specific noise models, and compute LGME along with its lower bounds to investigate its features. Selecting \emph{graph states}~\cite{hein2006} as a candidate stabilizer state, we calculate the LGME and its lower bounds over a selected subsystem using a graphical approach for single-qubit measurements~\cite{elliott2008,elliott2009}, and demonstrate the results in typical graphs having linear, ladder, and square graph structures.  We also extend the calculation in situations where the graph states (or, the stabilizer states) are subjected to single-qubit local Pauli noises~\cite{nielsen2010,holevo2012} of Markovian~\cite{Yu2006} and non-Markovian~\cite{Daffer2004,Shrikant2018,Gupta2020} variants, and investigate at which noise strength the localized \emph{genuine} multipartite entanglement cease to exist. The calculations can also be extended for arbitrary stabilizer states due to their connection to graph states by local unitary operations~\cite{van-den-nest2004,lang2012}. We demonstrate this by considering a toric code~\cite{kitaev2001,kitaev2006} defined on a square lattice. A more detailed yet non-technical overview of the main results, discussed in Secs.~\ref{sec:le_noiseless_graph}-\ref{sec:topological}, can be found in Sec.~\ref{sec:overview}. Sec.~\ref{sec:conclusion} contains the concluding remarks and an overview of possible future directions.

\section{Overview of the Main Results}
\label{sec:overview}

In this Section, we present an overview of the subsequent sections containing the technical details and different results. In Sec.~\ref{sec:le_noiseless_graph}, we consider the noiseless scenario, and provide a description of the graph states and their important features in Sec.~\ref{subsec:graph}. The formal definition of LE is given in Sec.~\ref{subsec:le}, and the LGME on a chosen subsystem corresponding to single-qubit Pauli measurements on arbitrary graph states are discussed. Since the measurements are restricted to single-qubit Pauli measurements on all qubits outside the chosen subsystem, our calculation provides a lower bound of the LGME. Specifically, we show that 
\begin{enumerate}
\item[(a)] for a fixed Pauli measurement setup, the resulting graphs corresponding to the post-measured states are independent of the measurement outcomes, and 
\item[(b)] the post-measured graph states differ from each other only by local unitary operations, thereby having identical entanglement properties. 
\end{enumerate} 
These results lead to the lower bound of the LGME on an arbitrary subsystem of a graph to be \emph{the maximum of the GME over the set of connected subgraphs on the chosen subsystem, obtained from the post-measured graphs once the Pauli measurements are incorporated}. In these calculations, we employ graph operations to implement the Pauli measurements, and show that \emph{the graph operations involved in this process has a polynomial scaling with the system size}. By virtue of the local unitary connection between an arbitrary stabilizer state and the graph states, the above results hold for arbitrary stabilizer states also.

We further show that depending on the structure of the graph, the set of all possible single-qubit Pauli measurement can be divided into two classes -- (1) one, where all measurement outcomes are equally probable, and (2) another, where only a subset of outcomes  having equal probability of occurrence are \emph{allowed}, and the rest are \emph{forbidden}. These classifications do not affect the calculation in the case of pure states (see Sec.~\ref{subsubsec:forbidden_sets}). However, in situations where the qubits in the system are subjected to Pauli noise of Markovian and non-Markovian types, which we discuss in Sec.~\ref{sec:le_noisy_graph}, 
\begin{enumerate}
\item[(a)] \emph{the post-measured states corresponding to only the first category of Pauli measurement setups are equally probable and are local unitarily connected to each other}. 
\item[(b)] \emph{On the other hand, if the Pauli measurement setup belongs to the second category, transitions may happen between the allowed and forbidden sets of measurement outcomes, and the post-measured states are not, in general, connected by local unitary operators, thereby loosing the advantages in calculation.} 
\end{enumerate}
A brief description of the specific noise channels used in this paper can be found in Sec.~\ref{subsec:noise}, while the interesting features of the Pauli measurements on noisy graph states and the subsequent calculation of the lower bound of LGME are discussed in Sec.~\ref{subsec:pauli_noise_pauli_measurement_le}. 

In Sec.~\ref{subsec:gme_examples_mixed_graphs}, we construct specific Pauli measurement setups belonging to the first category for typical graph states, such as the linear graph, ladder graph, and square graph, under Markovian and non-Markovian Pauli noise, and determine a lower bound for LGME. The main results summarized in this section are as follows. 
\begin{enumerate}
\item[(a)] \emph{There exists a critical noise strength for all four types of Pauli noise considered in this paper, beyond which all the post-measured states corresponding to the chosen Pauli measurement setup become biseparable}. 
\item[(b)] \emph{From the numerical data, we estimate the dependence of this critical noise strength on the degree of non-Markovianity, quantified by the non-Markovianity parameter} 
\item[(c)] \emph{We further infer that the critical noise strength is independent of the system size}.
\end{enumerate} 

The methodology to compute lower bounds of LGME, as adopted in this paper, holds for arbitrary stabilizer states due to their local unitary connection to the graph states. To demonstrate this, in Sec.~\ref{sec:topological}, we consider the toric code defined on a square lattice, and compute a lower bound of the LGME on a non-trivial loop representing a logical operator in both noisy and noiseless scenario by constructing an appropriate Pauli measurement setup. We show that 
\begin{enumerate}
   \item[(a)] \emph{it is always possible to localize GME over a non-trivial loop of the toric code in the absence of noise}, and  
   \item[(b)] \emph{similar to the noisy graphs, a critical noise strength exists for each type of Pauli noise such that all of the post-measured states corresponding to the chosen Pauli measurement setup become biseparable beyond it}. 
\end{enumerate}
In the case of the BF noise, we determine the form of the critical noise strength analytically as a function of the non-Markovianity parameter, and prove its system-size independence. In the case of other Pauli noise, our data suggests a decrease in the value of the critical noise strength with increasing system size, and increasing non-Markovianity parameter. We also construct an appropriate Pauli measurement setup that can provide a lower bound of localizable bipartite entanglement between two non-trivial loops of the toric code.

\section{LGME in graph states}
\label{sec:le_noiseless_graph}

In this Section, we discuss salient features of graph states, and present the definition of LE. We also calculate LE and one of its lower bounds corresponding to localization maximized for Pauli measurements.

\subsection{Graph states}
\label{subsec:graph}

A graph~\cite{Diestel2000,West2001} $G(V,L)$ is a collection of $N$ nodes $V\equiv\{1,2,\cdots,N\}$,  and a set of links $L\equiv\{(i,j)\}$, $i\neq j$, $i,j\in V$. We are interested in graphs that are \emph{simple}, i.e., have only one edge between two nodes, \emph{connected}, i.e., each node is connected to at least another node in the graph, and \emph{undirected}, i.e., both nodes corresponding to a link are equivalent. Assuming that a qubit is situated on every node of such as graph $G$, a graph state $\ket{G}$ is defined with respect to $G$ as~\cite{hein2006}
\begin{eqnarray}
\ket{G}&=&\left[\otimes_{(i,j)\in G}C_{(i,j)}^Z\right]\ket{+}^{\otimes N}. 
\label{eq:graph_state}
\end{eqnarray}
Here, $\ket{+}=(\ket{0}+\ket{1})/\sqrt{2}$ is the eigenstate of $\sigma^1$ corresponding to the eigenvalue $+1$, and    
\begin{eqnarray}
C_{(i,j)}^Z &=& \frac{1}{2}\left[(\sigma^0_{i}+\sigma^3_{i})\sigma^0_{j}+(\sigma^0_{i}-\sigma^3_{i})\sigma^3_{j}\right], 
\end{eqnarray}
with $\sigma^\alpha$, $\alpha=0,1,2,3$, being the $2\times 2$ identity matrix and the $x$, $y$, and $z$ components of the Pauli matrices respectively.  
Graph states can also be considered as multiqubit stabilizer states~\cite{Gottesman1997,hein2006,nielsen2010} defined by a set of stabilizer generators $\{g_{a_i}\}\subset \mathcal{P}^N$ with 
\begin{eqnarray}
g_{i}=\sigma^1_{i}\otimes_{j\in\mathcal{N}_{i}}\sigma^3_j,
\end{eqnarray}
where $\mathcal{P}^N$ is the $N$-qubit Pauli group, and $\mathcal{N}_{i}$ is the neighborhood of the node $i$ constituted of nodes having a direct link with the node $i$. Here, $\text{supp}(g_{i})=i\cup\mathcal{N}_{i}$, called the \emph{support} of $g_{i}$, are the qubits on which $g_{i}$ acts non-trivially, i.e., $\sigma_{j}\neq I_j$ $\forall$ $j \in i\cup \mathcal{N}_{i}$, with $\sigma_j$ being the contribution in $g_{i}$ corresponding to the node $j$.  The generators are mutually commuting, and therefore share a common eigenspace, given by
\begin{eqnarray}
\ket{\text{G}_{\alpha}}=Z_\alpha\ket{G}=\otimes_{i\in V}\left(\sigma_i^3\right)^{\alpha_{i}}\ket{G},
\label{eq:graph_state_basis}
\end{eqnarray}
where $\alpha_i=0,1$ $\forall i\in V$, $Z_\alpha=\otimes_{i\in V}\left(\sigma_i^3\right)^{\alpha_{i}}$, and $\alpha\equiv \alpha_1\alpha_2\cdots\alpha_N$ is a multi-index. Note that $\ket{G}$ is given  $\ket{\text{G}_0}$, corresponding to $\alpha_i=0$ $\forall i\in V$, which is the eigenstate having eigenvalue $+1$ for all $g_i$, i.e., 
\begin{eqnarray}
g_{{i}}\ket{G}=(+1)\ket{G}\quad \forall i\in V.  
\end{eqnarray}
The states $\{\ket{\text{G}_\alpha};\alpha=0,1,2,\cdots,2^N-1\}$ form a complete basis of the Hilbert space of the $N$-qubit system, and any state, $\rho$, of the system that is diagonal in this basis,  having the form 
\begin{eqnarray}
\rho=\sum_{\alpha=0}^{2^N-1}p_{\alpha}\ket{\text{G}_\alpha}\bra{\text{G}_\alpha},
\label{eq:GDstate}
\end{eqnarray}
is called the graph-diagonal (GD) state for an arbitrary probability distribution $\{p_\alpha\}$, with $\sum_{\alpha}p_\alpha=1$.

\begin{figure*}
    \centering
    \includegraphics[width=\textwidth]{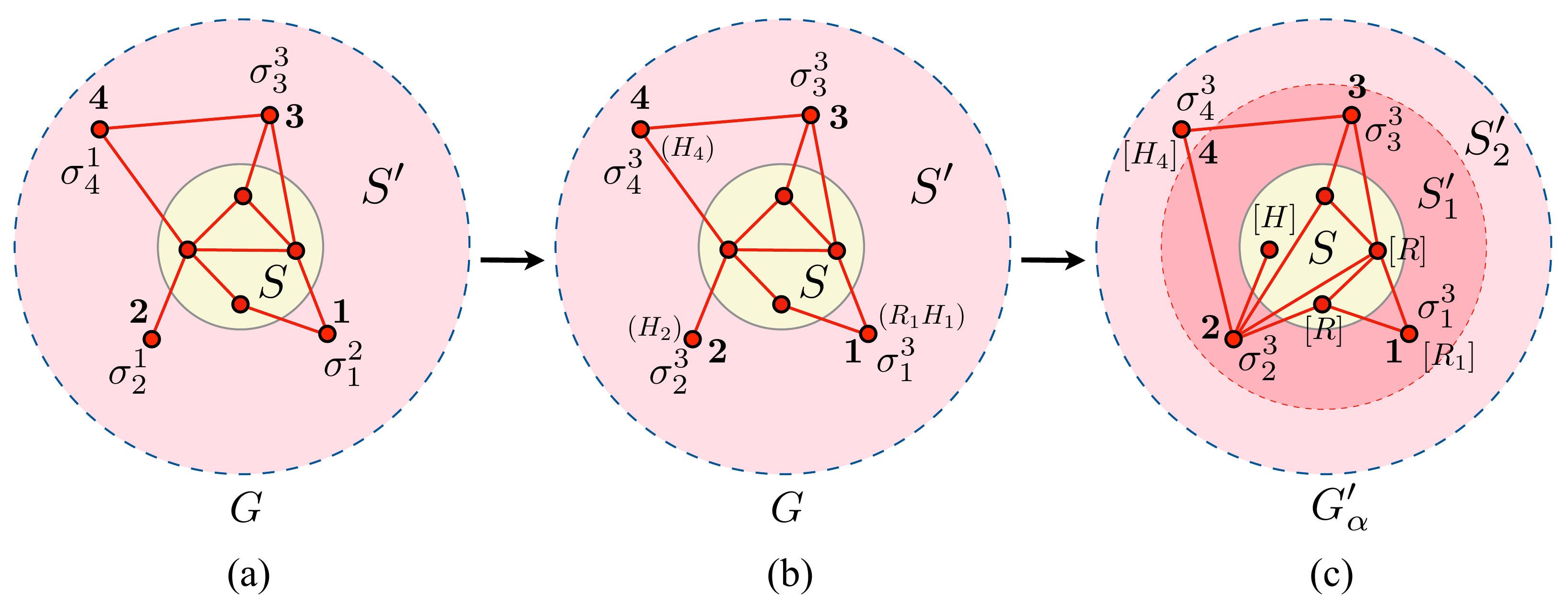}
    \caption{(a) Subsystems $S$ and $S^\prime$ of an arbitrary graph $G$. The nodes in $S^\prime$ are labelled as $j=1,2,3,4$ (see Sec.~\ref{subsubsec:pauli_measurement_le}), and a Pauli measurement setup $\sigma^2_1\sigma^1_2\sigma^3_3\sigma^1_4$, denoted by $\alpha\equiv\alpha_1\alpha_2\alpha_3\alpha_4=2131$ is assumed for demonstration. (b) The Pauli measurement setup $\sigma^2_1\sigma^1_2\sigma^4_3\sigma^1_4$ ($\alpha=2131$) can be transformed to $\sigma^3_1\sigma^3_2\sigma^3_3\sigma^3_4$ corresponding to $\alpha=\beta=3333$ via the application of the unitary operators $U^2_1$, $U^1_2$, and $U^1_4$ (definitions given in Eq.~(\ref{eq:definition_of_U})), which are denoted within parenthesis in the figure. The connectivity of the graph remains unchanged. (c) The graph state $\rho_\alpha$ (depicted in (b)) can be transformed to $\rho^\prime_\alpha$ (see Eq.~(\ref{eq:reduced_form})), where $G^\prime_\alpha$ is a graph on the same nodes with different connectivity, as shown in (c). The unitary operators $V_j$, defined through the Eqs.~(\ref{eq:product_unitary_V})-(\ref{eq:V_rules}), are shown within square brackets. The subsystem $S^\prime_1$ is constituted of the nodes that belong to the (combined) neighborhood of all the nodes in $S$, while $S^\prime_2$ is the set of the rest of the nodes in $S^\prime$ ($S^\prime_1\cup S^\prime_2=S^\prime$, and $S^\prime_1\cap S^\prime_2=\emptyset$). Note that the unitary operators $V_j$ for $j\in S_1^\prime$ are either identity, or $R_j$, which keeps $\sigma^3_j$ invariant. However, for qubit $j=4$ in $S_2^\prime$, $V_4=H_4$, which results in $H_4\sigma^3_4H_4^\dagger=\sigma^1_4$. Therefore, after the reduction of the graph according to Eq.~(\ref{eq:reduced_form}), the measurement setup in $S_2^\prime$ is not necessarily all $\sigma^3$.}
    \label{fig:regions}
\end{figure*}

\subsection{Localizable entanglement in graph states}
\label{subsec:le}

Let us now assume that the subsystem $S$ contains $n$ $(2\leq n \leq N-1)$ qubits labelled by $1,2,\cdots, n$, while the rest of the qubits $\{n+1,n+2,\cdots,N\}$ constitute the rest of the system $S^\prime$. Let us also assume that the state of the system is $\rho$. Localizable entanglement~\cite{verstraete2004,verstraete2004a,popp2005} over the subsystem $S$ is the maximum average entanglement that can be localized over $S$ by performing single-qubit projection measurements on all the qubits in $S^\prime$. Here, the maximization is performed over the complete set of single-qubit projection measurements. We denote the LE by $E_S$, given by
\begin{eqnarray}
E_S &=& \underset{\{M_k\}}{\max}\sum_{k}p_k E(\varrho_{S}^k),
\label{eq:le}
\end{eqnarray}
where $E$ is a chosen bipartite or multipartite entanglement measure~\cite{horodecki2009,guhne2009} computed over the unmeasured qubits after measurement. The set $\{M_k\}$ is the complete set of single-qubit projection measurements $M_k$ on qubits in $S^\prime$, $p_k$ is the probability of obtaining the measurement outcome $k$, given by 
\begin{eqnarray}
p_k &=& \text{Tr}\left[(I\otimes M_k) \rho (I\otimes M_k)^\dagger\right],
\end{eqnarray}
$\varrho_{S}^k$ denotes the post-measured state on $S$ corresponding to the measurement outcome $k$, given by 
\begin{eqnarray}
\varrho_{S}^k=\text{Tr}_{S^\prime}\left[\varrho_k\right]
\label{eq:k_labeled_post_measured_state}
\end{eqnarray}
with 
\begin{eqnarray}
\varrho_k=\frac{1}{p_k}\left[(I\otimes M_k) \rho (I\otimes M_k)^\dagger\right], 
\end{eqnarray}
and $I$ is the identity operator on the Hilbert space of $S$. For brevity, we write 
\begin{eqnarray}
\varrho_k=\frac{1}{p_k}\left[M_k \rho M_k^\dagger\right].
\label{eq:post_measured_state_full}
\end{eqnarray}

We consider rank-$1$ single qubit projective measurements on the qubits in $S^\prime$, such that the measurement operator $M_k$ takes the form 
\begin{eqnarray}
M_k=\otimes_{i\in S^\prime} P_{k_{i}},
\label{eq:measurement_operator}
\end{eqnarray}
where $P_{k_{i}}=\ket{k_{i}}\bra{k_{i}}$ are projectors corresponding to  $\ket{k_{i}}\in\{\ket{\mathbf{0}_{i}},\ket{\mathbf{1}_{i}}\}$, with 
\begin{eqnarray}
\ket{\mathbf{0}_{i}} &=& \cos\frac{\theta_{i}}{2}\ket{0_{i}}+\text{e}^{\text{i}\phi_{i}}\sin\frac{\theta_{i}}{2}\ket{1_{i}},\nonumber\\
\ket{\mathbf{1}_{i}} &=& \sin\frac{\theta_{i}}{2}\ket{0_{i}}-\text{e}^{\text{i}\phi_{i}}\cos\frac{\theta_i}{2}\ket{1_{i}}.
\label{eq:measurement_basis}
\end{eqnarray}
$\{\ket{0_{i}},\ket{1_{i}}\}$ being the computational basis in the Hilbert space of qubit $i$. The  measurement outcome-index $k$ can be identified as a multi-index $k\equiv k_{{n+1}}k_{{n+2}}\cdots k_{N}$. Therefore, the optimization involved in Eq.~(\ref{eq:le}) reduces to a $2(N-n)$-parameter optimization problem over real parameters $(\theta_i,\phi_i)$, $i\in S^\prime$, with $0\leq \theta_{i}\leq \pi$, $0\leq\phi_{i}\leq 2\pi$.

Note that for systems with large $N$, computation of $E_S$ can, in general, be computationally demanding. In this paper, we focus on the LE over a subsystem $S$ of a graph state $\ket{G}$ via only single-qubit Pauli measurements. 
A Pauli measurement-setup (PMS) in $S^\prime$ is defined by a configuration $\sigma^{\alpha_{n+1}}_{n+1}\sigma^{\alpha_{n+2}}_{n+2}\dots\sigma^{\alpha_{N}}_N$ of single-qubit Pauli measurements on the qubits in $S^\prime$, which is denoted by the label $\alpha\equiv\alpha_{n+1}\alpha_{n+2}\cdots\alpha_{N}$. 
%\begin{eqnarray}
%\bm{\sigma}\equiv\sigma^{\alpha_{n+1}}_{n+1}\sigma^{\alpha_{n+2}}_{n+2}\cdots\sigma^{\alpha_{N}}_N
%\label{eq:pauli_sequence}
%\end{eqnarray} 
There can be a total of $3^{N-n}$ such setups, since $\alpha_{i}=1,2,3$, $i=n+1,n+2,\cdots,N$\footnote{Note here that we discard $\alpha_i=0$ in order to ensure measurement on all qubits in $S^\prime$}. 
Let us denote the maximum LE obtained from all possible PMSs (i.e., all possible $\alpha$) by $E^{\mathcal{P}}_S$, where~\cite{Banerjee2020,Banerjee2022,amaro2018,amaro2020a} 
\begin{eqnarray}
    E_S^\mathcal{P}=\max_{\alpha}\sum_{k_\alpha=0}^{2^{N-n}-1}p_{k_\alpha}E(\varrho^{k_\alpha}_S),
    \label{eq:rle}
\end{eqnarray}
$k_\alpha$ being the label for the measurement outcomes corresponding to the PMS $\alpha$. We further discard the subscript $\alpha$ from the measurement outcome $k$ to keep the text uncluttered. Due to the restriction on the set of measurements, $E_S^{\mathcal{P}}$ is referred to as the restricted LE (RLE)~\cite{Banerjee2020,Banerjee2022,amaro2018,amaro2020a}, and the corresponding LGME the \emph{restricted} LGME (RLGME). Note that by definition of LE, 
\begin{eqnarray} 
0\leq E_{S}^\mathcal{P}\leq E_S.
\label{eq:lower_bound_1}
\end{eqnarray} 
While it is known that the Pauli measurements optimizes $E_S$ localized for subsystems $S$ constituted of only two qubits~\cite{Briegel2001,hein2006}, to the best of our knowledge, a similar result for subsystems constituted of multiple qubits is yet to be proved. In our numerical investigation, for an arbitrary graph state (also for an arbitrary stabilizer state, see Sec.~\ref{sec:topological}) in the noiseless scenario (i.e., for pure states), we always find $E_S=E_S^\mathcal{P}$. Motivated by this,  in the following, we discuss the computation of $E_S^{\mathcal{P}}$ in arbitrary graph states.

\subsubsection{Localization via Pauli measurements}
\label{subsubsec:pauli_measurement_le}

To analytically compute $E_S^{\mathcal{P}}$ for an arbitrary graph state $\rho=\ket{G}\bra{G}$ corresponding to a graph $G$, for ease of calculation, let us first re-index the qubits in $S^\prime$ as $j=i-n$, where $j=1,2,3,\cdots,N-n$. We denote the projectors $P_{k_{j}}^{\alpha_j}$ corresponding to a Pauli measurement on the qubit $j\in S^\prime$ as  
\begin{eqnarray}
P_{k_{j}}^{\alpha_j}=\frac{1}{2} \left[\sigma^0_j+(-1)^{k_{j}}\sigma^{\alpha_j}_j\right],
\label{eq:pauli_projectors}
\end{eqnarray} 
with $k_j=\pm 1$ being the measurement outcomes, and $\ket{k_j}$ is the eigenstate of $\sigma^{\alpha_j}_j$ corresponding to the eigenvalue $k_j$.  Note that a single-qubit $\sigma^{\alpha_j}_j$ ($\alpha_j=1,2$) measurement on a given graph state $\ket{G}$ is equivalent to a $\sigma^3_j$ measurement on the graph state $U^{\alpha_j\dagger}_j\rho_G U^{\alpha_j}_j$, where $U^{\alpha_j}_j$ is a Clifford unitary operator from the set $\{\sigma^3_j,H_j,R_j\}$, $H_j$ being the Hadamard operator, and $R_j=\sqrt{\sigma^3_j}$. Explicitly, 
\begin{eqnarray}
U^{1}_j=H_j,\;\; U^2_j = R_jH_j,
\label{eq:definition_of_U}
\end{eqnarray}
such that 
\begin{eqnarray}
\sigma^1_j &=& H_j\sigma^3_jH_j,\nonumber \\
\sigma^2_j &=& R_j H_j\sigma^3_jH_j R_j^\dagger.
\end{eqnarray}
Therefore, for a specific PMS $\alpha$, the measurement operator on $S^\prime$ having the form  
\begin{eqnarray}
M_{k}^\alpha=\otimes_{j\in S^\prime} P_{k_{j}}^{\alpha_j} 
\end{eqnarray}
would result in an overall Clifford unitary operator of the form 
\begin{eqnarray}
\mathbf{U}_\alpha=\otimes_{j\in S^\prime}U^{\alpha_j}_j, \alpha_j=1,2,
\label{eq:product_unitary}
\end{eqnarray}
on the subsystem $S^\prime$ (see Fig.~\ref{fig:regions}), such that $\mathbf{U}^\dagger_\alpha M_k^\alpha\mathbf{U}_\alpha\rightarrow M^\alpha_k$ with $\alpha_j=3$ $\forall j\in S^\prime$ after the transformation.  
Note that $\mathbf{U}_\alpha$ is specific to the chosen PMS $\alpha$. This further implies (from Eq.~(\ref{eq:post_measured_state_full})) 
\begin{eqnarray}
\varrho_k
&=&\mathbf{U}_\alpha M^\alpha_k\rho_{\alpha} M^{\alpha\dagger}_{k}\mathbf{U}^\dagger_\alpha
\label{eq:only_z_measurements}
\end{eqnarray}
with 
\begin{eqnarray}
\rho_{\alpha}&=&\mathbf{U}^\dagger_\alpha\rho \mathbf{U}_\alpha
\label{eq:primed_graph}
\end{eqnarray}
being specific to the original choice of $\alpha$. Fig.~\ref{fig:regions}(a) depicts an $8$-qubit graph state for demonstration, four of which are measured in Pauli basis, and the corresponding $U_i^{\alpha_i}$ for transforming all Pauli measurements to $\sigma^3$ measurements are shown in Fig.~\ref{fig:regions}(b).

To further simplify, note that the state $\mathbf{U}_\alpha\ket{G}$ can be written as~\cite{elliott2008,elliott2009} 
\begin{eqnarray}
\mathbf{U}^\dagger_\alpha\ket{G}=\text{e}^{\pm\text{i}\phi}\mathbf{V}_\alpha\ket{G^\prime_\alpha},
\label{eq:reduced_form}
\end{eqnarray}
where $\text{e}^{\pm\text{i}\phi}$ is an irrelevant global phase, $G^\prime_\alpha$ is a \emph{reduced graph} with modified connectivity that depends explicitly on $\alpha$ and is  defined on the nodes of the graph $G$. Here, 
\begin{eqnarray}
\mathbf{V}_\alpha=\otimes_{i\in G^\prime_\alpha}V_i
\label{eq:product_unitary_V}
\end{eqnarray}
is a unitary operator such that $V_i$ are local Clifford unitaries, and  
\begin{eqnarray}
V_j^\dagger P_{k_j}^3V_j=P_{k_j}^3 \forall j \in S^\prime_1,
\label{eq:V_rules}
\end{eqnarray}
where $k_j=\pm1$ and $S^\prime_1$ being the collection of qubits that construct the neighborhood of the qubits belonging to $S$. We denote the set of the rest of the qubits in $S^\prime$ by $S^\prime_2$, where $S^\prime=S^\prime_1\cup S^\prime_2$ and $S_1^\prime\cap S_2^\prime=\emptyset$. The transformation 
\begin{eqnarray}
\rho_\alpha\rightarrow\rho^{\prime}_{\alpha}=\mathbf{V}_\alpha\ket{G^\prime_\alpha}\bra{G^\prime_\alpha}\mathbf{V}^\dagger_\alpha,
\end{eqnarray} 
corresponding to Eq.~(\ref{eq:reduced_form}), can be performed via a series of local transformations on the underlying graph $G$ using a graphical representation of the states $\rho_\alpha$ and $\rho^{\prime}_\alpha$~\cite{elliott2008,elliott2009}, from which the graph $G^\prime_\alpha$ can also be extracted. This transformation is a rather technical one that involves  (a) redefining the attributes of the nodes of the graphs according to the single-qubit unitary operators $U_j^\dagger$ and $V_j$ applied to them, and (b) transforming the underlying graph along with the local unitary operators using a specific set of local graph operations. It ensures that
\begin{enumerate}
\item[\textbf{P1.}] $V_j$ corresponding to $j\in S^\prime_2$ are from the set $\{I_j, \sigma^3_j,H_j,H_j\sigma^3_j,R_j,R_j\sigma^3_j\}$,
\item[\textbf{P2.}] $V_i$ corresponding to a node $i$ that is (a) either in the neighborhood of a node $j\in S^\prime_2$ with $V_j\in\{H_j,H_j\sigma^3_j\}$, or (b) in $S^\prime_1$, are from the set $\{I_i,\sigma^3_i,R_i,R_i\sigma^3_i\}$, and 
\item[\textbf{P3.}] Any two nodes $j$ and $j^\prime$ $\in S^\prime_2$ having $V_j=H_j$ or $V_{j^\prime}=H_j\sigma^3_j$ does not have a link connecting them, 
\end{enumerate}
thereby satisfying Eq.~(\ref{eq:V_rules}). For the interested readers, details on this graph transformation can be found in Appendix~\ref{app:graph_operations}, while readers not interested in technical details can continue with this section. The transformation $\rho_\alpha\rightarrow\rho^\prime_\alpha$, according to Eq.~(\ref{eq:reduced_form}), is depicted in Fig.~\ref{fig:regions}(c), along with the sets $S_1^\prime$ and $S_2^\prime$, and the unitary operators $V_i$. The same transformation has also been depicted using the graph transformation notation in the Fig.~\ref{fig:grep} in  Appendix~\ref{app:graph_operations}. The overall protocol for obtaining the reduced graph $G^\prime_\alpha$ scales polynomially with the number of measured nodes, as discussed in Appendix~\ref{app:N_scaling}. 

Given a successful reduction of the graph $G\rightarrow G^\prime_\alpha$ for a specific PMS $\alpha$, the subgraph $G^\alpha_S$ on the chosen subsystem $S$ in $G^\prime_\alpha$ can be obtained by deleting all links between the nodes in $S$ and all other nodes in $S^\prime$. 
Using the above protocol, the following proposition can be proved for $E(\varrho_S^k)$ and the state $\ket{G^\alpha_S}$  (see Appendix~\ref{app:proof_proposition_1} for the proof).

\noindent\textbf{$\blacksquare$ Proposition 1.} \emph{For a given Pauli measurement setup on a subsystem $S^\prime$ of an arbitrary graph state, the average entanglement post-measurement on the subsystem $S$, where $S\cup S^\prime$ constitutes the entire system, is (a) independent of the measurement outcomes, (b) depends only on the chosen Pauli measurement setup, and (c) is given by}
\begin{eqnarray}
E(\varrho_S^k)=E\left(\ket{G^{\alpha}_S}\bra{G^{\alpha}_S}\right),
\end{eqnarray}
\emph{where $E$ is the chosen entanglement measure.}

\noindent Consequently, using the definition of $E^{\mathcal{P}}_S$ (see Eq.~(\ref{eq:rle})), the following corollary can also be written. 

\noindent\textbf{$\square$ Corollary 1.1} \emph{The value of $E_S^\mathcal{P}$ on the subsystem $S$ can be determined by performing an optimization over all the reduced graphs obtained from all possible Pauli measurement setups on $S^\prime$, and is given by}
\begin{eqnarray}
E_S^{\mathcal{P}}=\underset{\alpha}{\max}\left[E\left(\ket{G^{\alpha}_S}\bra{G^{\alpha}_S}\right)\right].\label{eq:measurement_outcome_independent_LE}
\end{eqnarray}

\subsubsection{Allowed and forbidden sets of outcomes}
\label{subsubsec:forbidden_sets}

A discussion on the different measurement outcomes $k$ corresponding to the measurement operators $M^\alpha_k$ is in order here. By design of the graph transformation (see Appendix~\ref{app:graph_operations}) ensuring \textbf{P1.}-\textbf{P3.}, $\sigma^3_i$ measurements over the set of nodes $\mathcal{Z}$ that form the combined neighborhood of the nodes in $S^\prime_2$ with $V_j=H_j,H_j\sigma^3_j$, and the nodes in $S$ remains unchanged due to the unitary transformations $\mathbf{V}_\alpha$\footnote{Clearly, $L\leq |\mathcal{Z}|\leq N-n$, $|\mathcal{Z}|$ being the size of $\mathcal{Z}$, and $L$ being the size of $S_1^\prime$.}. Therefore, following Sec.~\ref{subsubsec:pauli_measurement_le}, for a specific $\alpha$, we write the transformed $M^\alpha_k$ as $M^\alpha_{lm}$,
where we now re-label the measurement outcomes corresponding to $j\in \mathcal{Z}$ as $l$, and the same corresponding to nodes outside $\mathcal{Z}$ as $m$. Note that now the index $m$ corresponds to the nodes $j\in S^\prime_2$, for which $V_j=H_j,H_j\sigma^3_j$, i.e., for which a projection in the basis of $\sigma^1_j$ has to be applied as part of $M^\alpha_{lm}$. The projections on the nodes in $\mathcal{Z}$ leaves each of these node completely decoupled from the rest of the graph (since $\alpha_j=3$ $\forall j\in\mathcal{Z}$), and applies either $I$ or  $\sigma^3$ unitaries on the node depending on the outcomes $l$ corresponding to its neighborhood\footnote{See discussions related to Eqs.~(\ref{eq:gt_z}) and (\ref{eq:unitary_z_0})-(\ref{eq:unitary_z_1}) in Appendix~\ref{app:proof_proposition_1} for clarity.}. An isolated node $j$ in a graph is in the state $\ket{+_j}$, and is product with the rest of the graph (see Sec.~\ref{subsec:graph}). However, depending on the outcome $l$ on its neighborhood, after projecting all nodes in $\mathcal{Z}$, the isolated node $j\in S^\prime_2$ with $V_j=H_j,H_j\sigma^3_j$ either (a) remain in the state $\ket{+_j}$ (if $I$ applies to node $j$), or (b) changes to the state $\ket{-_j}$ (if $\sigma^3$ operator applies on $j$). Therefore, further projection on such nodes in $\sigma^1$ basis may result in $\langle m_j|+_j\rangle=0$, or $\langle m_j|-_j\rangle=0$, which leads to a set of outcomes $lm$, and consequently $lm\equiv k$ being \emph{forbidden}\footnote{The probabilities for finding these outcomes vanish.} for a given Pauli measurement setup $\alpha$.

\begin{figure}
    \centering
    \includegraphics[width=\linewidth]{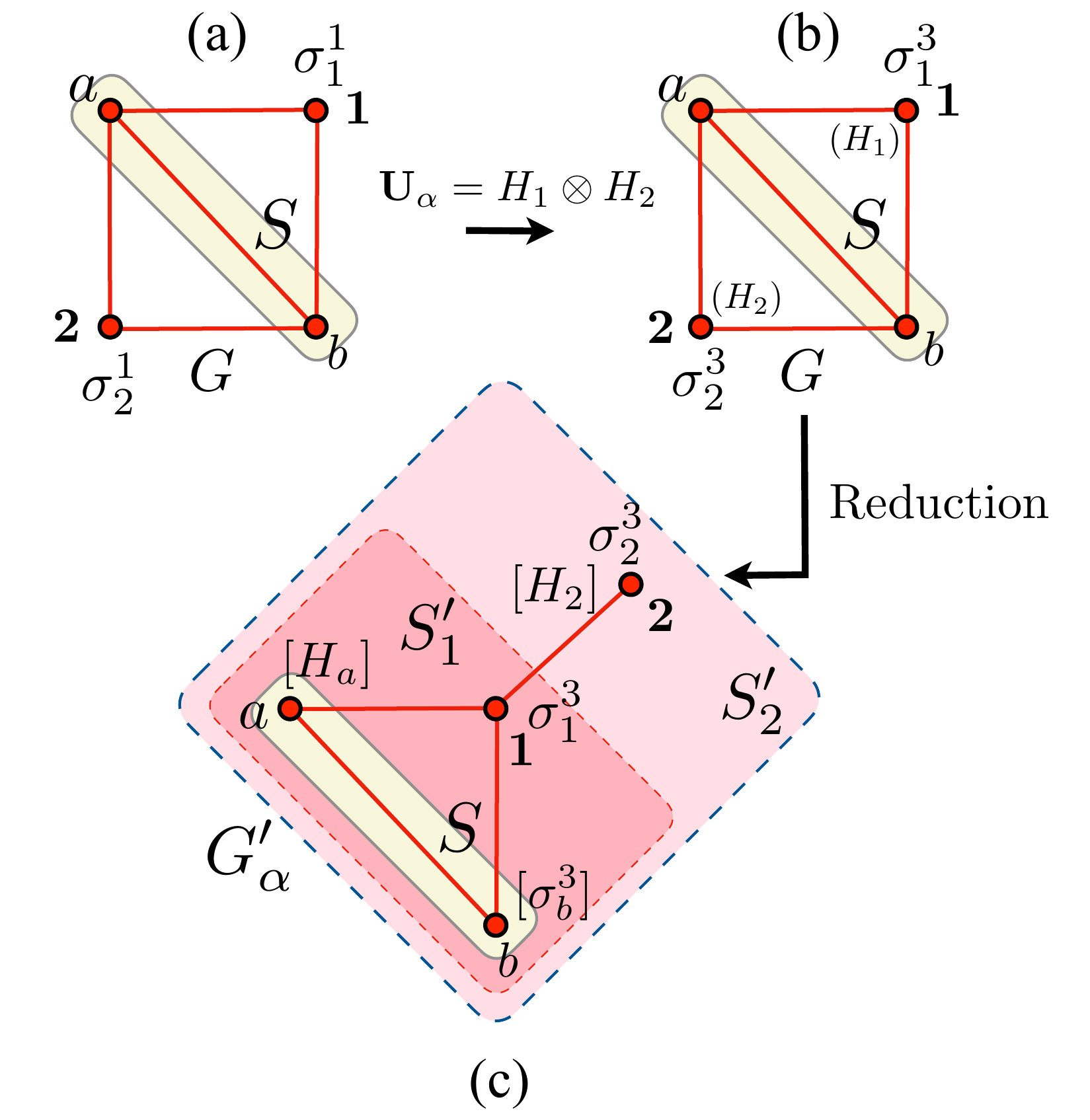}
    \caption{(a) A four-qubit graph state constituted of the qubits $a$, $b$, $1$, and $2$, where $\{1,2\}$ constitute the measured subsystems $A^\prime$. (b) A PMS given by $\alpha=11$ $(\alpha_1=1,\alpha_2=1)$ can be converted to a setup of $\sigma^3$ measurements on both qubits $1$ and $2$ via unitary operation $\mathbf{U}_\alpha=H_1\otimes H_2$, where $H_1$ and $H_2$ are shown in parenthesis. (c) The state $\rho_\alpha$ can be converted to $\rho_\alpha^\prime$ with the unitary operators $\mathbf{V}_\alpha=H_a\otimes\sigma^3_b\otimes I_3\otimes H_2$, where these unitary operators are depicted in square parenthesis on the reduced graph $G^\prime_\alpha$. On $G^\prime_\alpha$, qubit $1$ ($2$) constitutes the region $S^\prime_1$ $(S^\prime_2)$.}
    \label{fig:forbid}
\end{figure}

We demonstrate the occurrence of the forbidden and allowed set of measurement outcomes with an example of a single four-qubit connected graph depicted in Fig.~\ref{fig:forbid}(a). The region $S$ is constituted of the qubits $a$ and $b$, while the measured subsystem $S^\prime\equiv\{1,2\}$. For demonstration, we choose the PMS given by $\alpha\equiv 11$ ($\alpha_1=1,\alpha_2=1$). The four possible measurement outcomes corresponding to these measurements would be $k\equiv k_1k_2=(+1)(+1)$, $(+1)(-1)$, $(-1)(+1)$, and $(-1)(-1)$. The PMS can be converted to a $\sigma^3$ measurement setup on qubits $1$ and $2$ by the unitary operator $U_\alpha=H_1\otimes H_2$, which changes the measurement basis ($\alpha_1:1\rightarrow 3$, $\alpha_2:1\rightarrow 3$), but not the measurement outcomes (see Fig.~\ref{fig:forbid}(b)). Following the discussion in Sec.~\ref{subsubsec:pauli_measurement_le}, we now label the outcomes corresponding to qubit $1$ $(2)$ as $l_1$ $(m_2)$. The state $\rho_\alpha$ can now be transformed as $\rho^\prime_\alpha$ with $\mathbf{V}_\alpha=H_a\otimes \sigma^3_b\otimes I_1\otimes H_2$, and a transformed graph $G^\prime_\alpha$ (see Fig.~\ref{fig:forbid}(c)). Clearly, the $\sigma^3$ measurement on qubit $2$ is now transformed to a $\sigma^1$ measurement, implying $\alpha_2=3\rightarrow \alpha_2=1$, while $\alpha_1=3$ on qubit $1$ remains unchanged, and $l_1=\pm 1$, $m_2=\pm 1$. Application of $M_{l_1=+1}^{\alpha_1=3}$ $\left(M_{l_1=-1}^{\alpha_1=3}\right)$ on qubit $1$ decouples qubits $1$ and $2$ from the rest of the graph, and applies $I$ $(\sigma^3)$ on qubit $2$, leaving it in the state $\ket{+_j}$ $\left(\ket{-_j}\right)$. Therefore, further application of $M_{m_2=-1}^{\alpha_2=1}$  $\left(M_{m_2=+1}^{\alpha_2=1}\right)$ on qubit $2$ yields $0$, implying that the $k_1k_2\equiv l_1m_2=(+1)(-1),(-1)(+1)$ outcomes will never occur. Similar situation arises in the case of the example shown in Fig.~\ref{fig:regions}, where due to $V_{j=4}$ being a Hadamard operator, half of the set of $16$ possible outcomes $k\equiv k_1k_2k_3k_4$ are forbidden, given by the set 
\begin{eqnarray}
(+1)(+1)(-1)(+1)&,& (+1)(+1)(+1)(-1) \nonumber\\
(+1)(-1)(+1)(+1)&,& (+1)(-1)(-1)(-1)\nonumber \\
(-1)(+1)(-1)(+1)&,& (-1)(+1)(+1)(-1)\nonumber \\
(-1)(-1)(+1)(+1)&,& (-1)(-1)(-1)(-1)\nonumber
\end{eqnarray}

We point out here that the occurrence of such forbidden sets of measurement outcomes depend completely on the choice of the specific PMS $\alpha$. Therefore, one can divide the PMSs into two categories, (a) the ones forming the set $\Gamma$, for which all the measurement outcomes are allowed, and (b) the ones constituting the set $\overline{\Gamma}$, for which occurrence of such forbidden set is possible. For instance, in the example presented in Fig.~\ref{fig:forbid}, $\alpha\equiv 33$, i.e., $\alpha_1=3,\alpha_2=3$ represents a PMS belonging to $\Gamma$. Note also that the subgraph $G^\alpha_{S}$ after the application of $M^{\alpha_1=3}_{l_1=\pm 1}$ (in Fig.~\ref{fig:forbid}, this corresponds to the nodes $a$ and $b$, and the link connecting these two nodes), being fully decoupled from the nodes in $S^\prime_2$, is not affected by the application of $M^{\alpha_2=1}_{m_2}$ on the nodes in $S^\prime_2$. Therefore, in the case of pure graph states, this does not change the value of $E(\varrho^k_{S})$, and hence $E^{\mathcal{P}}_S$. 
%However, the situation is far more complex in the presence of local noise on qubits, which we demonstrate in Appendix.~\ref{app:demo}. 

\begin{figure}
    \centering
    \includegraphics[width=\linewidth]{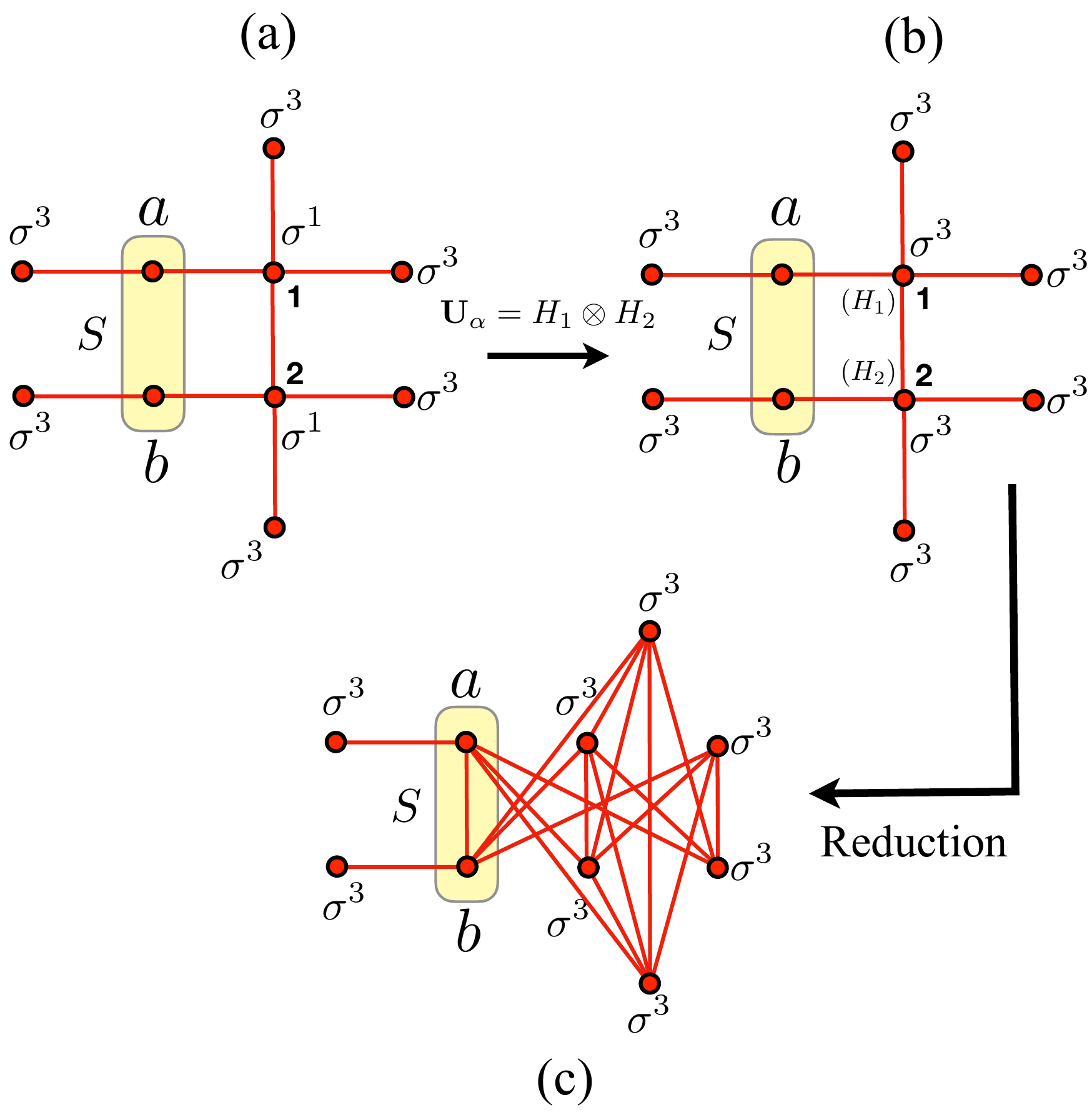}
    \caption{Transformation of a graph as per the discussion in Sec.~\ref{subsubsec:pauli_measurement_le} and Appendix~\ref{app:graph_operations} when $\sigma^1$ measurements are performed on qubits $1$ and $2$ situated on the path connecting the qubits $a$ and $b$ in subsystem $S$ (denoted by the shaded region). In the reduced graph, a link exists between the qubits $a$ and $b$, such that decoupling of $S$ from the rest of the qubits leads to a connected qubit-pair on $S$, resulting in a maximally entangled Bell-pair. The components of the local unitary transformations $\mathbf{U}_\alpha^\dagger$ ($\mathbf{V}_\alpha$ is identity here) are shown in parenthesis.}
    \label{fig:examples_two}
\end{figure}

\subsection{Examples}
\label{subsec:gme_examples_pure_graphs}

By virtue of Proposition 1 and Corollary 1.1, it is now possible to localize entanglement over a subsystem $S$ of a multi-qubit system, via an optimization over $3^{N-n}$ reduced graph states $\{\ket{G^\prime_\alpha}\}$, resulting from all possible PMSs, labelled by $\alpha$, where $N-n$ is the size of $S^\prime$. The localized entanglement over $S$ is \emph{genuinely multipartite} in nature if the subgraph $G^\alpha_S$, corresponding to the state $\ket{G^\alpha_S}$, in $G^\prime_\alpha$ is a connected one. Therefore, to obtain RLGME, one can perform the optimization in Eq.~(\ref{eq:measurement_outcome_independent_LE}) over only those $G^\prime_\alpha$ for which $G^\alpha_S$ is connected. While this restriction over the optimization may still yield a large number of connected graphs over $S$ making the optimization difficult, in the case of \emph{typical} regular graphs, eg. the linear graph, the graph with a ladder structure, and the square graph, the number of $G^\prime_\alpha$ resulting in a connected subgraph $G^\alpha_S$ is considerably low, and therefore advantageous. Moreover, the connected subgraphs $G^\alpha_S$ can be further classified into different \emph{orbits} such that the members of individual orbits are connected to each other via local complementation operations and graph isomorphism~\cite{hein2004,hein2005,hein2006}, thereby having identical GME. Therefore, the number of connected subgraphs that one needs to consider for performing the optimization in Eq.~(\ref{eq:measurement_outcome_independent_LE}) can be reduced to the number of orbits, as we demonstrate in the following examples.  Note also that in order to \emph{quantify} the LGME, one needs to compute a GME measure over $G^\alpha_S$, which can still be a non-trivial task due to the possible optimizations involved in the calculation. In this paper, we use the \emph{Schmidt measure}~\cite{hein2004,hein2006}  and the \emph{generalized geometric measure} (GGM)~\cite{aditi2010,biswas2014,sadhukhan2017} for this purpose in the noiseless scenario, the definition of which can be found in Appendix.~\ref{app:entanglement_in_graph}.

\subsubsection{Two-qubit subsystems in arbitrary graph}
\label{subsubsec:two_qubit}

We first re-visit the extensively investigated case of the two-qubit subsystem on an arbitrary graph. It is well-known~\cite{hein2006} that the optimal PMS corresponding to localizing maximum entanglement on the chosen two-qubit subsystem $S$ in an arbitrary graph $G$ is given by either (a) $\sigma^3$ measurement on all qubits except the chosen subsystem if a direct link exists between the selected qubits in the graph, or (b) $\sigma^1$ measurements on all qubits on a chosen path between the two qubits belonging to $S$, and $\sigma^3$ measurements everywhere else. The former involves a decoupling of $S$ from the rest of the qubits in $G$, thereby leaving only a connected pair of qubits on $S$. On the other hand, in the latter prescription, the $\sigma^1$  measurements create a direct link between the two chosen qubits, while the $\sigma^3$ measurement decouples $S$ from the rest of the qubits. The purpose of maximizing the localizable entanglement is fulfilled due to the fact that a connected graph of two qubits is equivalent to a maximally entangled Bell state, or its local unitary equivalents, which are also maximally genuinely multiparty entangled. It is straightforward to see that for an arbitrary graph $G$, the PMS described in the prescription (b), via the the graph transformations introduced in Sec.~\ref{subsubsec:pauli_measurement_le}, leads to a reduced graph where the selected nodes are connected by a direct link, thereby leading to maximal entanglement over $S$. See Fig.~\ref{fig:examples_two} for a demonstration. It is worthwhile to note that there may be a number of paths between the chosen qubits in $G$, each of which results in a specific optimal PMS for the LE, and subsequently a specific reduced graph with a link between the qubits in $S$. However, each such PMS guarantees the creation of a Bell-pair or its local unitary equivalent on $S$.    

\begin{figure}
    \centering
    \includegraphics[width=\linewidth]{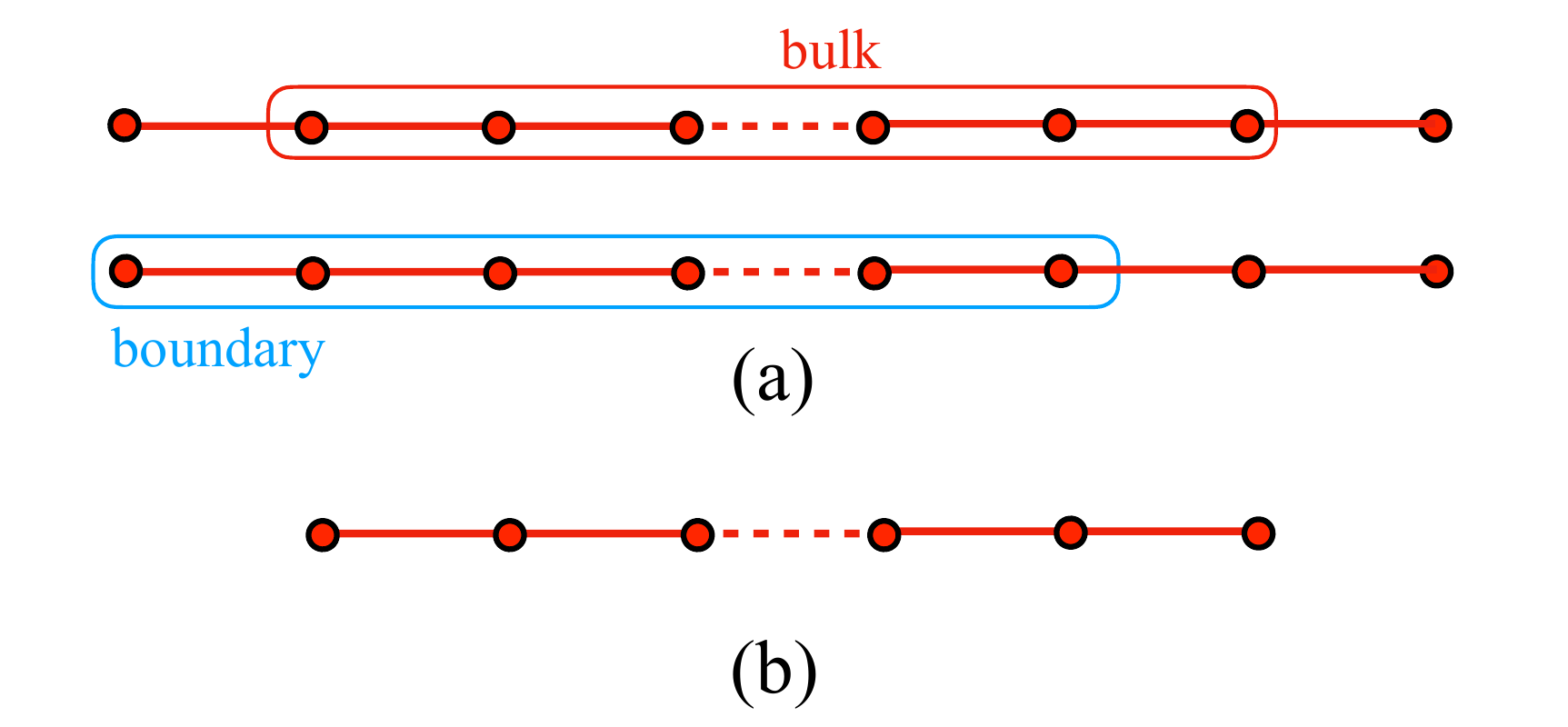}
    \caption{(a) Connected subsystem of qubits situated at the boundary or bulk of a linear graph. (b) The only connected graph $G^\alpha_S$ that may occur on such a subsystem $S$ in a linear graph as a result of the application of the PMSs, denoted by $\alpha$, on the graph is a linear graph on $S$.}
    \label{fig:example_linear}
\end{figure}

\begin{figure*}
    \centering
    \includegraphics[width=0.8\textwidth]{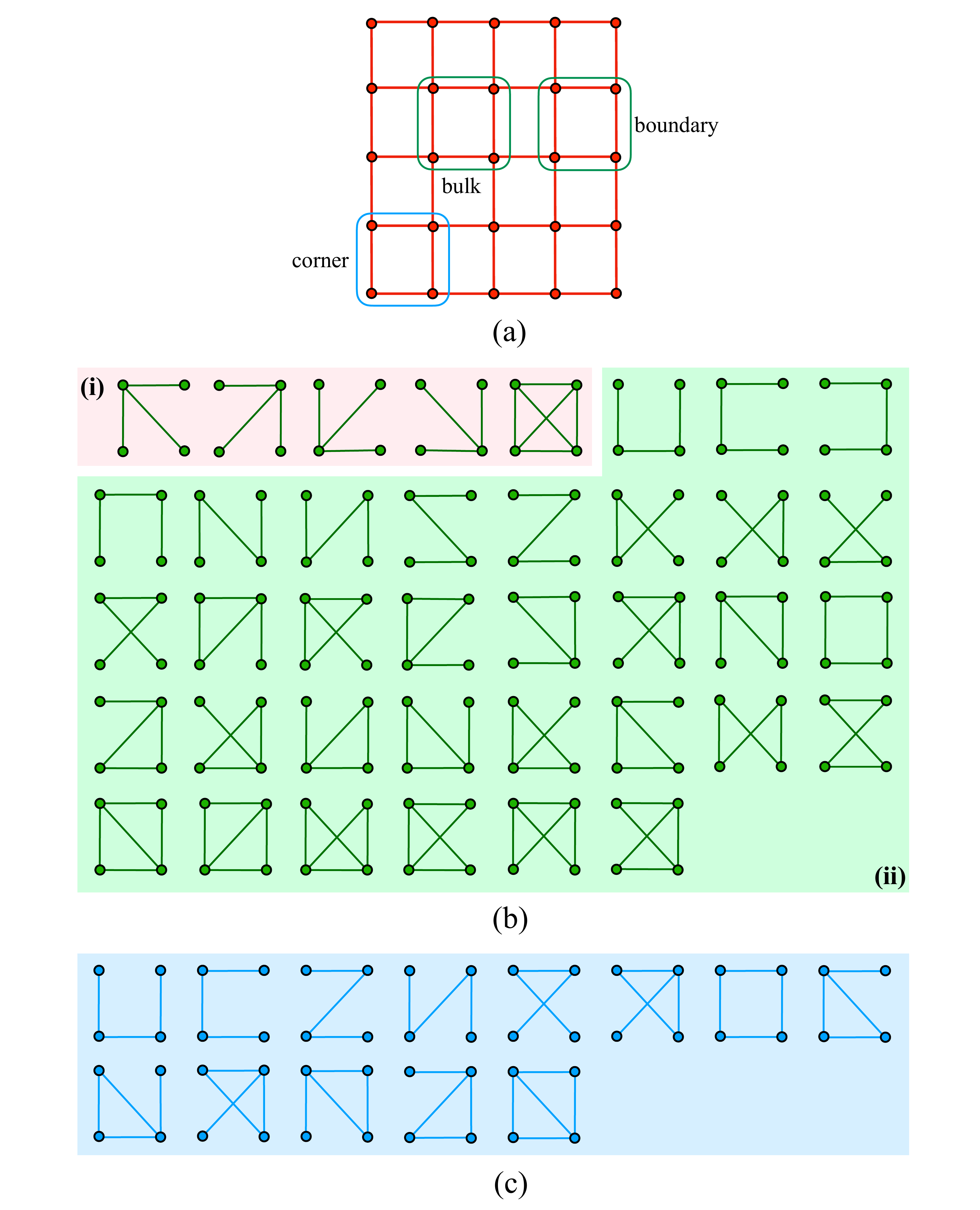}
    \caption{(a) A plaquette of four qubits as the chosen subsystem $S$ on a square graph, situated at the corner, bulk, or boundary of the graph. (b) If the plaquette $S$ is located at the bulk or the boundary, there are $38$ possibilities for the four-qubit connected subgraph $G^\alpha_S$ in the reduced graphs $\{G^\prime_\alpha\}$. These $38$ four-qubit graphs can be further divided into two orbits, shown as two separate shaded blocks (i) and (ii), containing $5$ and $33$ subgraphs respectively. (c) If the four-qubit plaquette is located at one corner of the graph, then the number of possibilities for $G^\alpha_S$ is reduced to $13$, all of which belong to the same orbit.}
    \label{fig:example_square}
\end{figure*}

\subsubsection{Connected subsystems in linear graphs}
\label{subsubsec:linear}

We next consider a subsystem of qubits, $S$, in a linear graph. Here, $S$ forms a \emph{connected patch}, such that each of the qubits in this patch  is connected via a link with at least one other qubit in the same patch (see Fig.~\ref{fig:example_linear}). We further assume that $S$ contains more than two qubits. As discussed above, we are interested only in the PMSs that provide a connected $G^\alpha_S$ on $S$. We find that the application of the protocol discussed in Sec.~\ref{subsubsec:pauli_measurement_le} leads to only \emph{linear connected graphs} on $S$ (see Fig.~\ref{fig:example_linear}(b)), irrespective of whether $S$ is situated in the \emph{bulk}, or at the \emph{boundary} of the linear graph (see Fig.~\ref{fig:example_linear}(a)), for all PMSs that lead to a connected $G^\alpha_S$. Therefore, RLGME over a connected subsystem $S$ on a linear graph equals to the GME in a linear graph of size $|S|=n$, which is known to be $\lfloor n/2\rfloor$~\cite{hein2004} when Schmidt measure is used for quantification.

\subsubsection{Connected subsystems in square graph}
\label{subsubsec:square}

We next consider a square graph, and take one of the plaquettes as the chosen subsystem $S$. The plaquette can be located in the bulk, or at the corner, or at one of the boundaries of the graph (see Fig.~\ref{fig:example_square}(a)). In the situations where it is located in the bulk, or at the boundary, application of the methodology described in Sec.~\ref{subsubsec:pauli_measurement_le} and Appendix.~\ref{app:graph_operations} leads to a total of $38$ possible connected $G^\alpha_S$ (see Fig.~\ref{fig:example_square}(b)) in the reduced graphs $\{G^\prime_\alpha\}$, obtained from all possible PMSs, labelled by $\alpha$. The 38 connected subgraphs can further be divided into two orbits -- (i) one containing the star subgraphs and the fully connected subgraph on $S$ (see Fig.~\ref{fig:example_square}(b)), and (ii) the other containing the rest $33$ connected subgraphs. Since the members of individual orbits are connected to each other via local complementation (see Appendix~\ref{app:graph_operations} for definition) and graph isomorphism, they have identical entanglement properties, and it is therefore sufficient to consider a representative from each of the orbits to perform the optimization in Eq.~(\ref{eq:measurement_outcome_independent_LE}). The lower and the upper bound of the Schmidt measure for $G^\alpha_S$ in (i) are found to be equal to $1$, while for $G^\alpha_S$ in (ii), it is $2$. This implies that the values of the Schmidt measure in these two cases are $1$ and $2$ respectively, leading to $E_S^\mathcal{P}=1$ and $E_S^\mathcal{P}=2$ respectively. 

On the other hand, if the chosen four-qubit plaquette is located at one of the corners of the square graph, the number of connected  subgraphs $G^\alpha_S$ reduces to $13$ (see Fig.~\ref{fig:example_square}(c)), all of which belong to the same orbit. Therefore, it is sufficient to compute the Schmidt measure of any one of these $13$ subgraphs. Similar to the former case, here also the upper and lower bounds of the Schmidt measure are found to be equal, having a value $2$, which leads to $E_S^\mathcal{P}=2$.

\begin{figure*}
    \centering
    \includegraphics[width=0.8\textwidth]{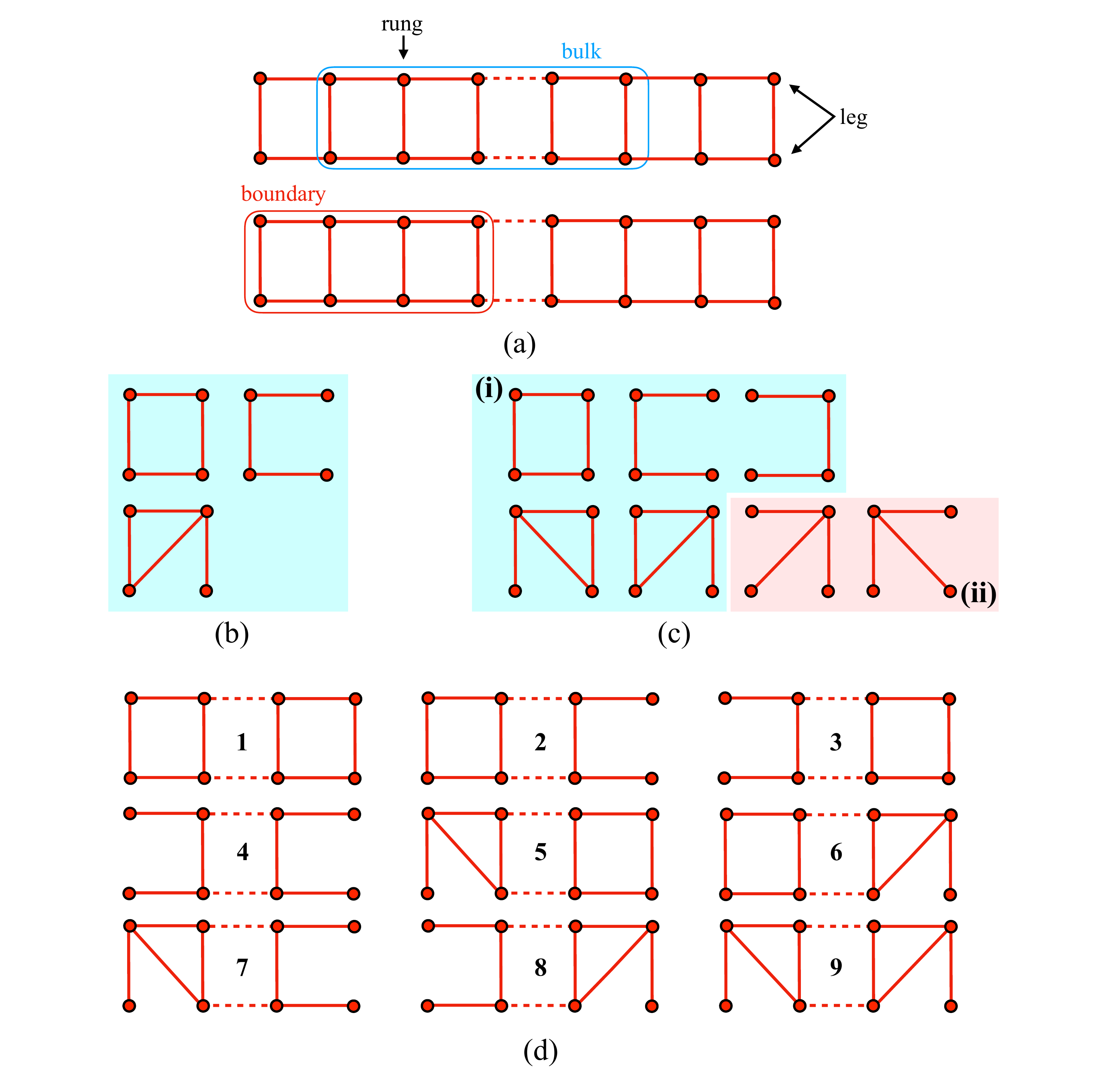}
    \caption{(a) A set of $2R$ connected qubits, where $R(<N/2)$ is the number of rungs, as the chosen subsystem $S$ on a graph with a ladder structure with two \emph{legs} and $N/2$ \emph{rungs}, $N$ being the total number of qubits. The subsystem $S$ can be situated at the bulk, or the boundary of the graph. (b) For $R=2$ and $S$ is connected and is located at the boundary, only three distinct connected subgraphs $G^\alpha_S$ are possible on $S$, all of which belong to a single orbit. (c) However, if the connected subsystem $S$ with $R=2$ is in the bulk of the system, there are $7$ possibilities for the four-qubit connected subgraph $G^\alpha_S$, including the four-qubit star graph, which can be divided into two orbits, (i) and (ii), shown in two different shaded blocks. (d) If the subsystem $S$ is connected, and is constituted of $R>2$ rungs, then a total of $9$ possibilities for the connected subgraphs $G^\alpha_S$ arise.}
    \label{fig:example_ladder}
\end{figure*}

\subsubsection{Connected subsystems in Ladders}
\label{subsubsec:ladders}

We now consider a graph having a ladder structure (see Fig.~\ref{fig:example_ladder}(a)) where two legs are connected by rungs and the nodes of the graph are the points where the rungs meet the legs. We choose a connected subsystem consisted of a number, say, $R$, of rungs of the ladder as $S$, and apply the protocol described in Sec.~\ref{subsubsec:pauli_measurement_le} and Appendix~\ref{app:graph_operations}. For $R=2$, and $S$ situated at the boundary, the set of connected subgraphs $\{G^\alpha_S\}$ consists of only three four-qubit graphs, given in Fig.~\ref{fig:example_ladder}(b), and $E_S^\mathcal{P}=2$ as in the cases described earlier. On the other hand, of the two-rung subsystem is situated in the bulk of the ladder, then the four-qubit star graphs also appear in the set of connected subgraphs $\{G^\alpha_S\}$, containing a total of $7$ distinct graph structures that can be divided into two orbits (see Fig.~\ref{fig:example_ladder}(c), with two orbits marked by (i) and (ii)). Here, we have reduced the number of graphs in an orbit by discarding the four-qubit graphs that are connected directly by local complementations with any one of the members of the orbit. In this case also, the RLGME, as quantified by the Schmidt measure, is obtained as $E_S^\mathcal{P}=2$.

We now consider the scenario of a connected subsystem $S$ with $R>2$ rungs. We find that the set of connected $G^\alpha_S$ on $S$ in $G^\prime_\alpha$, obtained from the PMSs denoted by $\alpha$, contains only $9$ distinct connected graph structures, shown in Fig.~\ref{fig:example_ladder}(d), labelled as $1-9$, where graphs that are connected to the structures 5-9 via local complementation are discarded. Moreover, each of these $9$ connected graphs on $G^\alpha_S$ has the same value for the lower and upper bounds of the Schmidt measure, implying that the value of the measure is equal to either of the bounds. In the case of $R=3$ $(n=6)$, the value of the Schmidt measure for the structure $4$ is $2$, while all of the other graphs in Fig.~\ref{fig:example_ladder}(d) has the value of the Schmidt measure equals to $3$, implying that $E_S^\mathcal{P}=3$. On the other hand, for all cases with $R>3$ $(n>6)$, for each of the structures $1-9$, the Schmidt measure is  given by $n/2$, implying that the RLGME, as quantified by the Schmidt measure, is $n/2$. 

\begin{figure}
    \centering
    \includegraphics[width=\linewidth]{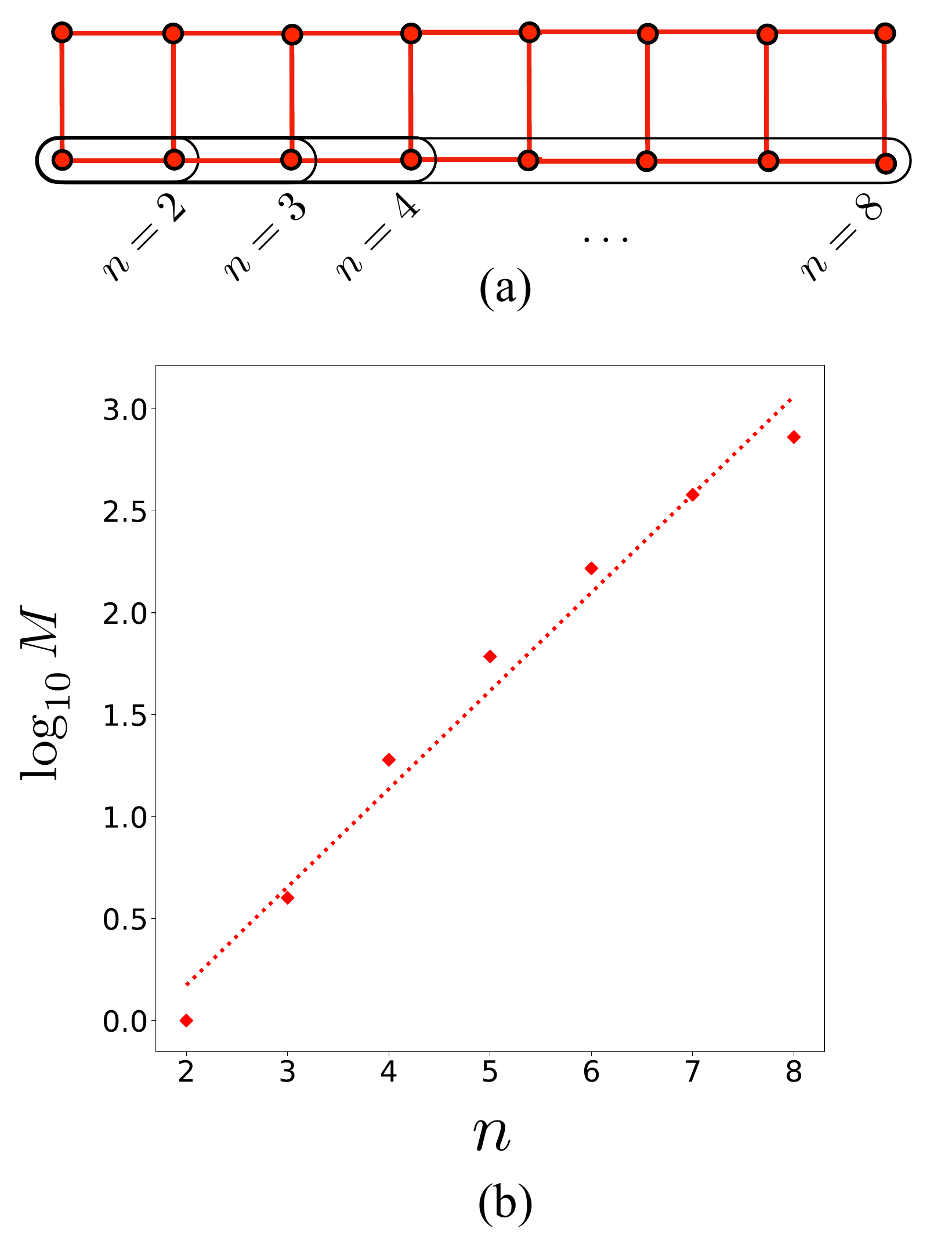}
    \caption{(a) Connected subsystem of qubits of size $n$ ($2\leq n\leq 8$) in the form of a linear subgraph on one of the legs of the two-leg ladder graph. (b) Variation of $\log_{10}M$ with $n$ for $N=16$, where $M$ is the total number of connected subgraphs $\{G^\alpha_S\}$ resulting from the reduced graphs $\{G^\prime_\alpha\}$, obtained by the application of all PMSs on $S^\prime$. The data is fitted to a straight line of slope $0.481$ and intercept $-0.788$. See Eq.~(\ref{eq:M_fitted})}
    \label{fig:example_ladders_linear}
\end{figure}

We further consider a connected subsystem $S$ in the form of a linear graph on one of the legs of the ladder (see Fig.~\ref{fig:example_ladders_linear}(a)), where the size, $n$, of $S$ is increasing. In this situation, with increasing $n$, the number, $M$, of connected subgraphs $\{G^\alpha_S\}$ increases exponentially (see Fig.~\ref{fig:example_ladders_linear}(b)) with $n$ as 
\begin{eqnarray}
    M=10^{\alpha+\beta n},
    \label{eq:M_fitted}
\end{eqnarray}
where $\alpha$ and $\beta$ depends on $N$, and can be determined from fitting the data to Eq.~(\ref{eq:M_fitted})\footnote{For example, with $N=16$, $\alpha=-0.781$ and $\beta=0.481$.}. However, the RLGME, as quantified by the Schmidt measure, has the value $\lfloor n/2\rfloor$.

\noindent\textbf{Note.} We also compute the RLGME in terms of the GGM, which is a measure for the GME present in a quantum state, for each of the above cases, and find it to be $1/2$ for all the examples. This implies that genuine multipartite entanglement is localized on all the chosen subsystems $S$ in all the examples. We comment on the loss of LGME with introduction of noise in Sec.~\ref{sec:le_noisy_graph}.      

\subsection{Localizable bipartite entanglement}

It is worthwhile to point out here that one can also compute localizable bipartite entanglement over a chosen bipartition of the subsystem $S$ using the same methodology. However, the optimization in Eq.~(\ref{eq:measurement_outcome_independent_LE}), in this case, has to be performed over all possible subgraphs $G^\alpha_S$, connected or otherwise, obtained from all possible reduced graphs $G^\prime_\alpha$, resulting from the PMSs. Therefore, for an arbitrary graph $G$, the optimization can be more demanding compared to the calculation of RLGME. However, a few conclusions can be drawn without performing the optimization. For example, 
\begin{enumerate}
\item[(a)] if at least one connected $G^\alpha_S$ is obtained for the subsystem $S$, and 
\item[(b)] if the chosen bipartition $A:B$ of $S$ is such that $A$ is made of only $1$ qubit, and $B$ consists of the rest of the qubits in $S$, 
\end{enumerate}
then maximum bipartite entanglement will be localized between $A$ and $B$ since all qubits in a connected graph has maximally mixed marginals~\cite{hein2006,Verstraete2003} (see Appendix~\ref{app:bipartite_entanglement}). Note that this is the case for all the examples discussed above. Further, if $n=2,3$ in addition to the conditions (a) and (b), where $|S|=n$, then maximum bipartite entanglement can be localized over all bipartitions in $S$. For $n>3$, relaxing condition (b) and assuming $A$ to be the smaller subsystem constituted of at least $2$ qubits $\{i,j\}\in G^\alpha_S$,  $\rho_{A}=\text{Tr}_B\left(\ket{G^\alpha_S}\bra{G^\alpha_S}\right)$ is either maximally mixed, or a rank-$2$ mixed state~\cite{hein2006}. In the case of the former, $\ket{G^\alpha_S}$ is maximally entangled in the bipartition $A:B$, whereas in the case of the latter, it is not. Similar conclusions on the localizable bipartite entanglement over the partition $A:B$ in $S$ hold.

%\subsubsection{Dumbbell graphs}
%\label{subsubsec:dumbbell}

\section{LGME in noisy graph states}
\label{sec:le_noisy_graph}

We now provide a brief description of the noise models considered in this paper, and discuss the calculation of $E_S^\mathcal{P}$, and one of its lower bounds corresponding to a specific PMS in graph states under Markovian and non-Markovian noisy channels. 

\subsection{Single-qubit Pauli noise on graph states}
\label{subsec:noise}

To describe the effect of single-qubit Pauli noise on the graph states, we use the Kraus operator formalism as 
\begin{eqnarray}
\rho=\Lambda(\rho_0)=\sum_{s}K_s\rho_0 K^\dagger_s,
\label{eq:evolve}
\end{eqnarray}
where $\{K_s\}$ are the Kraus operators satisfying
\begin{eqnarray}
\sum_{s}K_s^\dagger K_s=I, 
\end{eqnarray}
with $I$ being the identity operator in the Hilbert space of the $N$-qubit graph. The states $\rho_0=\ket{G}\bra{G}$, and $\rho$ denote respectively the pure graph state, and the mixed state of the system under noise. In the case of Pauli noise, individual Kraus operators are $K_s=\sqrt{q_s}J_s$, where 
\begin{eqnarray}
\label{eq:kraus_pauli}
J_s&=&\otimes_{i=1}^N\sigma^{s_i}, \\ q_s&=&\prod_{i=1}^Nq_{s_i},
\label{eq:kraus_prob}
\end{eqnarray}
with $s_i\in\{0,1,2,3\}\forall i\in G$, and $\sum_{s_i=0}^3q_{s_i}=1$. Note here that the index $s$ on the left hand side can be interpreted as the multi-index  $s\equiv s_1s_2\dots s_N$. The factorized form of $q_s$ implies that no spatial correlation exists between the qubits in the graph. In this paper, we focus on non-dissipative Pauli channels.  More specifically, we consider Markovian bit-flip (BF), bit-phase-flip (BPF), phase-damping (PD), and depolarizing (DP) noise channels. 
%The Markovianity of these channels is reflected in  $q$ having specific dependence on time, $t$, arising out of the Markovian approximation. For instance, $q$ is often a monotonically increasing function of $t$, such as $q=1-\text{e}^{-\Gamma t}$~\cite{Yu2006}, where $\Gamma$ is a decay-rate. 
In terms of the probabilities associated to different Pauli operators in the maps, the BF, BPF, PD, and DP channels are given by   
\begin{eqnarray}
q_{0}&=&1-\frac{q}{2},q_{1}=\frac{q}{2},q_{2},q_{3}=0,
\label{eq:mbf_channel_prob}
\end{eqnarray}
\begin{eqnarray}
q_{0}&=&1-\frac{q}{2},q_{2}=\frac{q}{2},q_{1},q_{3}=0,
\label{eq:mbpf_channel_prob}
\end{eqnarray}
\begin{eqnarray}
q_{0}&=&1-\frac{q}{2},q_{3}=\frac{q}{2},q_1,q_2=0,
\label{eq:mpf_channel_prob}
\end{eqnarray}
and 
\begin{eqnarray}
q_{0}=1-\frac{3q}{4},q_{1}=q_2=q_3=\frac{q}{4},
\label{eq:mdp_channel_prob}
\end{eqnarray}
respectively, where $0\leq q\leq 1$, and we assume $q$ to be identical for all qubits. 

%In the case of a non-Markovian noise, $q$ may be an oscillatory function of time with decaying amplitude~\cite{Daffer2004}. 
There exists non-Markovian versions of the single-qubit Pauli noise, such as the non-Markovian PD\footnote{The non-Markovian version of the BF and the BPF channels can also be defined in a similar fashion.} and DP channels~\cite{Daffer2004,Shrikant2018,Gupta2020}, given by 
\begin{eqnarray}
q_0&=&\left(1-\frac{q}{2}\right)\left[1-\epsilon \frac{q}{2}\right], \nonumber \\
q_{3}&=& \frac{q}{2}\left[1+\epsilon\left(1-\frac{q}{2}\right)\right],\nonumber \\
q_1&=&q_2=0,
\label{eq:nmpd_channel_prob}
\end{eqnarray}
and 
\begin{eqnarray}
q_0&=&\left(1-\frac{3q}{4}\right)\left[1-\frac{9\epsilon q}{4}\right], \nonumber \\
q_{1}&=& q_2=q_3= \frac{q}{4}\left[1+3\epsilon\left(1-\frac{3q}{4}\right)\right],
\label{eq:nmdp_channel_prob}
\end{eqnarray}
respectively, with $0\leq q\leq 1$, and $0\leq\epsilon\leq 1$. The Markovian PD and the DP channels are recovered from their non-Markovian counterparts for $\epsilon=0$.

Note here that the stabilizer description of graph states implies that application of $\sigma_i^{s_i}$, $s_i=1,2$, on the qubit $i$ in the graph state leads to
\begin{eqnarray}\label{eq:pauli_map_identities}
\sigma_i^1\ket{G} &=&\otimes_{j\in\mathcal{N}_i}\sigma_i^3\ket{G},\nonumber\\
\sigma_i^2\ket{G}&=&\sigma_i^3\otimes_{j\in\mathcal{N}_i}\sigma_i^3\ket{G}.
\end{eqnarray}
Therefore, all Pauli noise described in Eqs.~(\ref{eq:mbf_channel_prob}), (\ref{eq:mdp_channel_prob}), and (\ref{eq:nmdp_channel_prob}) are equivalent to PD noise (Eq.~\ref{eq:mpf_channel_prob}), with a modified probability distribution $\{\tilde{q}_s\}$, which is not necessarily factorized (i.e., can not be written in a form similar to Eq.~(\ref{eq:kraus_prob})). Note also that a PD noise on a connected graph state results in a GD state (Eq.~(\ref{eq:GDstate})), where all the diagonal elements of the density matrix are not-necessarily non-zero.

\begin{table}
    \centering
    \begin{tabular}{|c|c|c|}
    \hline
    $U$ & $\sigma^{s}$ & $\sigma^{s^\prime}=U^\dagger \sigma^{s}U$  \\
    \hline 
    $H$ & $\sigma^1$ & $\sigma^3$ \\
    \hline 
    $H$ & $\sigma^2$ & $\sigma^2$ \\
    \hline 
    $H$ & $\sigma^3$ & $\sigma^1$ \\

    \hline
    $R$ & $\sigma^1$ & $\sigma^2$\\
    \hline
    $R$ & $\sigma^2$ & $\sigma^1$\\
    \hline
    $R$ & $\sigma^3$ & $\sigma^3$\\
    \hline    
    $RH$ & $\sigma^1$ & $\sigma^2$\\
    \hline
    $RH$ & $\sigma^2$ & $\sigma^3$\\
    \hline
    $RH$ & $\sigma^3$ & $\sigma^1$\\
    \hline    
    \end{tabular}
    \caption{Transformation of Pauli matrices up to a phase factor under the Clifford operation $U$, where $U$ represents the unitary operations $U_i$ and $V_i$ corresponding to the transformation of the graph state $\rho\rightarrow\rho_\alpha\rightarrow\rho^\prime_\alpha$ (see Sec.~\ref{subsubsec:pauli_measurement_le}). The transformation of the Pauli noise, as described in Sec.~\ref{subsec:noise}, can also be determined using these transformations.}
    \label{tab:pauli_transformation}
\end{table}

\begin{figure*}
    \centering
    \includegraphics[width=0.9\textwidth]{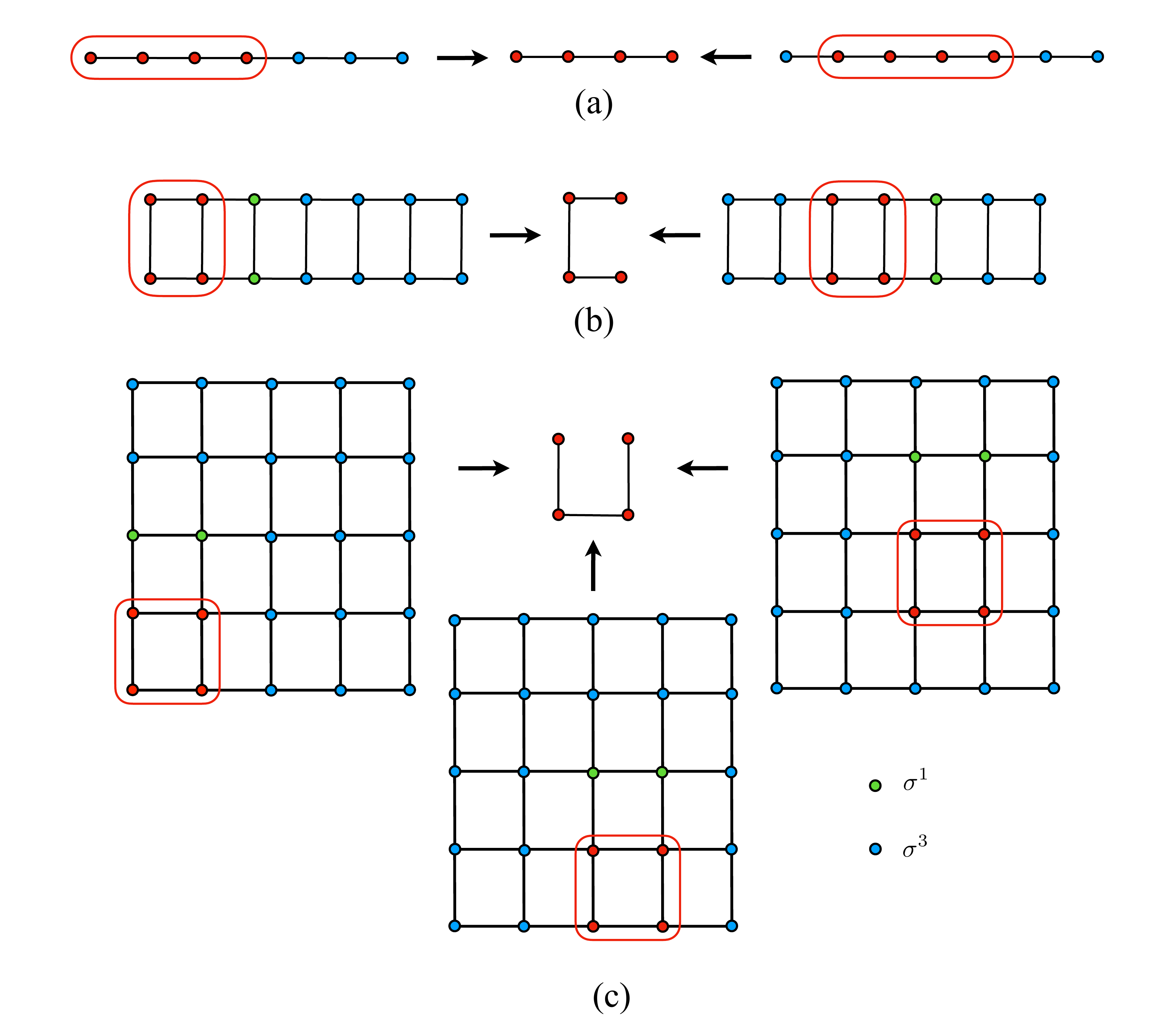}
    \caption{The PMS used for the lower bound calculation in noisy (a) linear (boundary, bulk), (b) ladder (boundary, bulk) and (c) square (corner, boundary, bulk) graph structures. The red nodes indicates the unmeasured four nodes on which a GD state of linear graph will be localized. The blue(green) nodes indicates $\sigma^3(\sigma^1)$ measurements. Translation to any higher system size only requires additional $\sigma^3$ measurements on the additional nodes maintaining the $\sigma^1$ measurement pattern w.r.t. localizing four nodes.}
    \label{fig:demo_PMS}
\end{figure*}

\subsection{Localizable entanglement via Pauli measurements}
\label{subsec:pauli_noise_pauli_measurement_le}

We now discuss the calculation of $E^{\mathcal{P}}_S$. For a chosen $\alpha$ representing a specific PMS, we proceed in a fashion similar to the pure graph states (see Sec.~\ref{subsubsec:pauli_measurement_le}), and transform $\alpha_j\rightarrow 3$ $\forall j\in S^\prime$ via the application of the local Clifford operation $\mathbf{U}_\alpha$. Explicitly, we write 
\begin{eqnarray}
\varrho_k&=& M_k^\alpha\left[\sum_{s}q_sJ_s\rho_0 J_s^\dagger\right]M^{\alpha\dagger}_k, 
\nonumber\\ 
&=&\mathbf{U}_\alpha\left[\sum_s q_s M^\alpha_k J_{s^\prime} \rho_\alpha J_{s^\prime}^\dagger M^{\alpha\dagger}_k\right]\mathbf{U}^\dagger_\alpha, 
\label{eq:noisy_system_1}
\end{eqnarray} 
where definitions of $M^\alpha_{k}$ and $\rho_\alpha$ are as in Sec.~\ref{subsubsec:pauli_measurement_le} respectively, and 
\begin{eqnarray}
J_{s^\prime}&=&\mathbf{U}^\dagger_\alpha J_s \mathbf{U}_\alpha.
\label{eq:noise_transform_in_S_prime}
\end{eqnarray}
Further transformation of $\rho_\alpha\rightarrow\rho^{\prime}_\alpha$ modifies  Eq.~(\ref{eq:noisy_system_1}) as 
\begin{eqnarray}
\varrho_k &=& \mathbf{U}_\alpha\mathbf{V}_\alpha \tilde{\varrho}_k\mathbf{V}^\dagger_\alpha\mathbf{U}^\dagger_\alpha,
\end{eqnarray}
with
\begin{eqnarray}
\tilde{\varrho}_k
&=&\sum_s q_s M^\alpha_{lm} J_{s^{\prime\prime}} \ket{G_\alpha^\prime}\bra{G^\prime_\alpha} J_{s^{\prime\prime}}^\dagger M^{\alpha\dagger}_{lm}, 
\end{eqnarray}
where we have followed the notations introduced in Sec.~\ref{subsubsec:pauli_measurement_le}, and 
\begin{eqnarray}
J_{s^{\prime\prime}}=\mathbf{V}_\alpha^\dagger J_{s^\prime}\mathbf{V}_\alpha.
\label{eq:pauli_noise_transformation}
\end{eqnarray}
Note here that since both $\mathbf{U}$ and $\mathbf{V}$ are Clifford unitaries, Pauli noise is mapped to Pauli noise (see Table~\ref{tab:pauli_transformation}), and the transformation $s\rightarrow s^\prime\rightarrow s^{\prime\prime}$ leads to the multi-index $s^{\prime\prime}\equiv s^{\prime\prime}_1s^{\prime\prime}_2\dots s^{\prime\prime}_{N-n}$ with $s^{\prime\prime}_j=1,2,3$ for $j\in S^\prime$.  
Note further that 
\begin{eqnarray}
\sigma^{\alpha^\prime_j}_jP_{l_j,m_j}^{\alpha_j}\sigma^{\alpha^\prime_j}_j\rightarrow P_{l^{\prime}_j,m_j^{\prime}}^{\alpha_j},
\label{eq:transformed_projector}
\end{eqnarray} 
where $l_j^{\prime}=l_j$ and $m^{\prime}_j=m_j$ ($l_j^{\prime}\neq l_j$ and $m^{\prime}_j\neq m_j$) if $\alpha_j=\alpha_j^\prime$ ($\alpha_j\neq\alpha_j^\prime$), $\alpha_j,\alpha_j^\prime=1,2,3$, leading to 
\begin{eqnarray}
\tilde{\varrho}_k
&=&\sum_s q_s J_{s^{\prime\prime}} M^{\alpha}_{l^\prime m^\prime} \ket{G_\alpha^\prime}\bra{G^\prime_\alpha} M^{\alpha\dagger}_{l^\prime m^\prime} J_{s^{\prime\prime}}^\dagger. 
\label{eq:measurement_outcomes_changed}
\end{eqnarray}

\begin{figure*}
\includegraphics[width=0.8\textwidth]{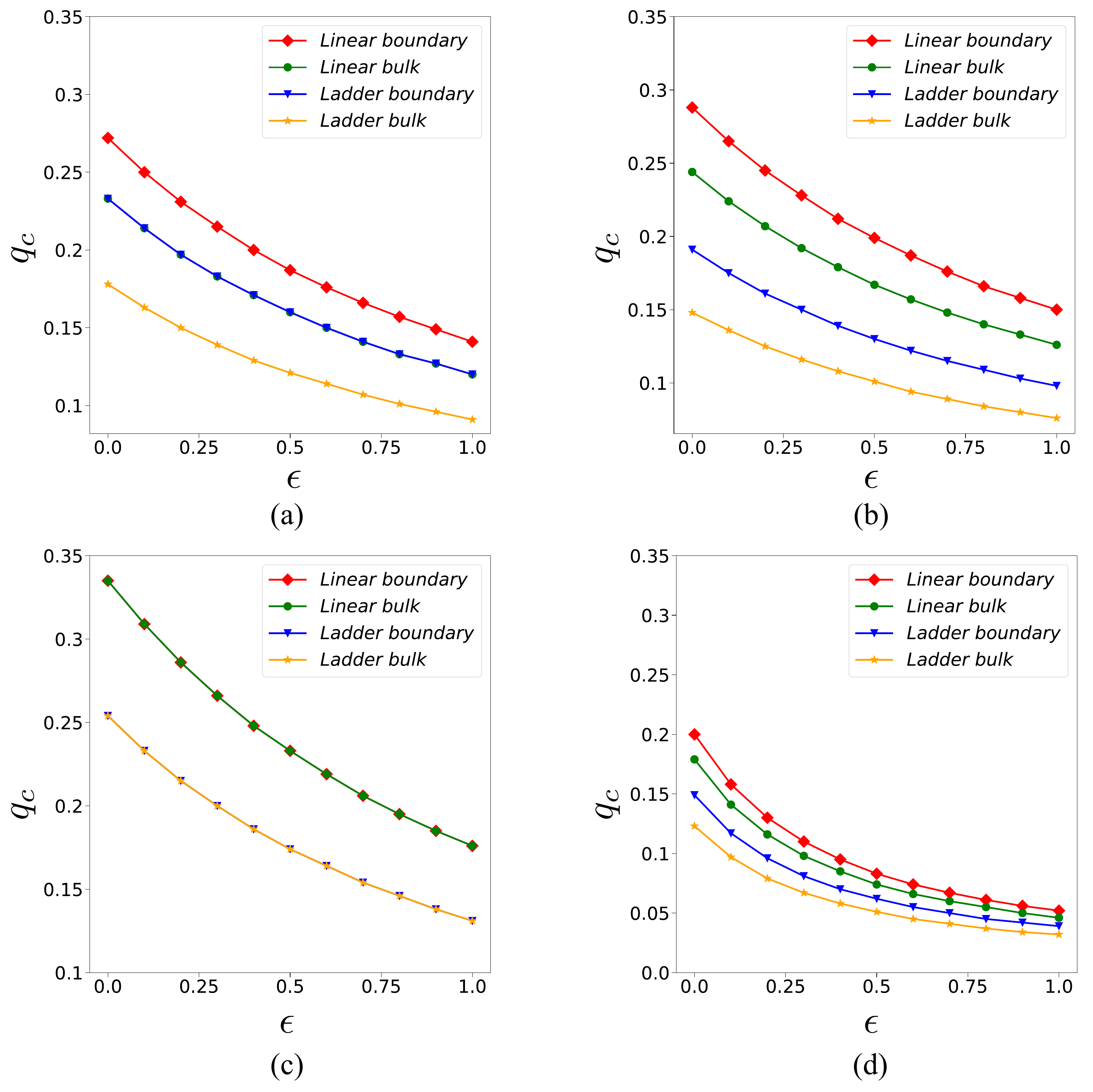}
\caption{Variations of $q_c$ as a function of $\epsilon$ in the case of the (a) BF, (b) BPF, (c) PD, and (d) DP noise, for a connected subsystem $S$ of four qubits (see Figs.~\ref{fig:demo_PMS}(a) and (b)) located at the boundary, or bulk of a linear graph, and a graph in the shape of a ladder. All the axes in all figures are dimensionless.}
\label{fig:noise_linear_ladder}
\end{figure*}

We now separately consider the two types of PMSs, constituting the sets $\Gamma$ and $\overline{\Gamma}$, where $\Gamma\cup\overline{\Gamma}$ is the full set of all possible PMSs, and $\Gamma\cap\overline{\Gamma}=\emptyset$. The set $\Gamma$ is defined such that for a specific PMS $\alpha\in\Gamma$, all outcomes $k$ are equally probable with probability $2^{-(N-n)}$, which is not true for a PMS belonging to $\overline{\Gamma}$. Under this consideration, the optimization of RLGME can be considered as $E_S^{\mathcal{P}}=\max\{E_S^{\Gamma},E_{S}^{\overline{\Gamma}}\}$, where (see Eq.~(\ref{eq:rle}))
\begin{eqnarray}
\label{eq:rle_separate_1}
E_S^{\Gamma}&=&\underset{\alpha\in\Gamma}{\max}\sum_{k_{\alpha}=0}^{2^{N-n}-1}p_{k_\alpha}E\left(\varrho^{k_\alpha}_S\right), \nonumber\\
\label{eq:rle_separate_2}
E_S^{\overline{\Gamma}}&=&\underset{\alpha\in\overline{\Gamma}}{\max}\sum_{k_{\alpha}=0}^{2^{N-n}-1}p_{k_\alpha}E\left(\varrho^{k_\alpha}_S\right),
\end{eqnarray}
where we have re-instated the index $\alpha$ in $k$ to make the equations comprehensive.   Starting with the set $\Gamma$, the following proposition can be written (see Appendix~\ref{app:proof_proposition_2} for the proof.)

\noindent\textbf{$\blacksquare$ Proposition 2.} \emph{For a subsystem $S$ of an arbitrary graph state subjected to single-qubit Pauli noise on all qubits, if the chosen Pauli measurement setup on $S^\prime$ with $S\cup S^\prime$ constituting the entire system is such that $\alpha\in\Gamma$, then the average entanglement post-measurement on the subsystem $S$ is independent of the measurement outcomes, and depends only on $\alpha$.}

Using Proposition 2 in Eq.~(\ref{eq:rle_separate_1}), Corollary 2.1 is straightforward. 

\noindent\textbf{$\square$ Corollary 2.1.} \emph{The value of $E^{\Gamma}_S$ on the subsystem $S$ is given by}
\begin{eqnarray}
E_S^{\Gamma}=\underset{\alpha\in\Gamma}{\max}\left[E\left(\varrho^k_S\right)\right].
\label{eq:final_eq_noise_outcome_independence}
\end{eqnarray}
See Eq.~(\ref{eq:unitary_equivalence_under_noise}) for the definition of $\varrho^k_S$. Note that Eq.~(\ref{eq:final_eq_noise_outcome_independence}) reduces to Eq.~(\ref{eq:measurement_outcome_independent_LE}) in the absence of noise, and in a situation when $\overline{\Gamma}$ is a null set. 

Similar to the proof of Proposition 1 and Corollary 1.1 (see Appendix~\ref{app:proof_proposition_1}), the proof of Proposition 2 and Corollary 2.1 is also established on the fact that for $\alpha\in\Gamma$, the post-measured states $\varrho^k_S$ on $S$ corresponding to different $k$ are local-unitarily connected to each other, where the unitary operators are constructed by single-qubit unitary operators. Therefore $\varrho_S^k$ for different $k$ have the same entanglement content (see Appendix~\ref{app:proof_proposition_2}). In contrast, In the situation where $\overline{\Gamma}$ has a non-zero cardinality, and for a PMS $\alpha\in\overline{\Gamma}$, $\varrho_S^k$ corresponding to different $k$ are, in general, not connected by local unitary operators. In this case, explicit calculation depends on the fact that the change $lm\rightarrow l^\prime m^\prime$ may result in a transition between allowed and  forbidden sets of outcomes, making the computation for the post-measured states $\varrho^k_S$ difficult. For interested readers, we demonstrate the calculation step-by-step in Appendix~\ref{app:demo} in both the cases of $\alpha\in\Gamma$ and $\alpha\in\overline{\Gamma}$, using the example of the graph in Fig.~\ref{fig:forbid}. However, in all the specific examples considered in this paper, we restrict ourselves to specific PMSs belonging to $\Gamma$, as discussed in Sec.~\ref{subsec:gme_examples_mixed_graphs} and \ref{sec:topological}, where Proposition 2 and Corollary 2.1 are applicable.

\subsection{Examples}
\label{subsec:gme_examples_mixed_graphs}

We now revisit the examples considered in Sec.~\ref{subsec:gme_examples_pure_graphs}, but in the presence of single-qubit Pauli noise of different types, which adds obstacles in quantifying the LGME. On one hand, understanding how to characterize mixed states in terms of GME is far from complete, and computable measures of multipartite entanglement in mixed states are scarce~\cite{horodecki2009}. On the other hand, in the case of the mixed states, the manifestation of the effect of the measurement outcomes is different compared to the pure graph states, indicating the necessity of rigorous optimization over all $3^{N-n}$ PMSs. To overcome these challenges, we focus on a lower bound of the RLGME (and in consequence, of LGME), which can be obtained by judiciously choosing a PMS, labelled by $\alpha_c$. Since the definition of RLGME involves a maximization over all possible $\alpha$, the average entanglement $E_{S}^{\alpha_c}$ obtained for the chosen PMS $\alpha_c$ provides a lower bound of $E_S^{\mathcal{P}}$, and therefore of $E_S$ i.e., $E_{S}^{\alpha_c}\leq E_S^{\mathcal{P}}\leq E_S$. The existence of LGME on $S$ is guaranteed by an $\alpha_c$ such that $E_S^{\alpha_c}>0$. We refer to  $E_S^{\alpha_c}$ as the \emph{floor} of LGME (FLGME). 

\begin{figure*}
\includegraphics[width=0.8\textwidth]{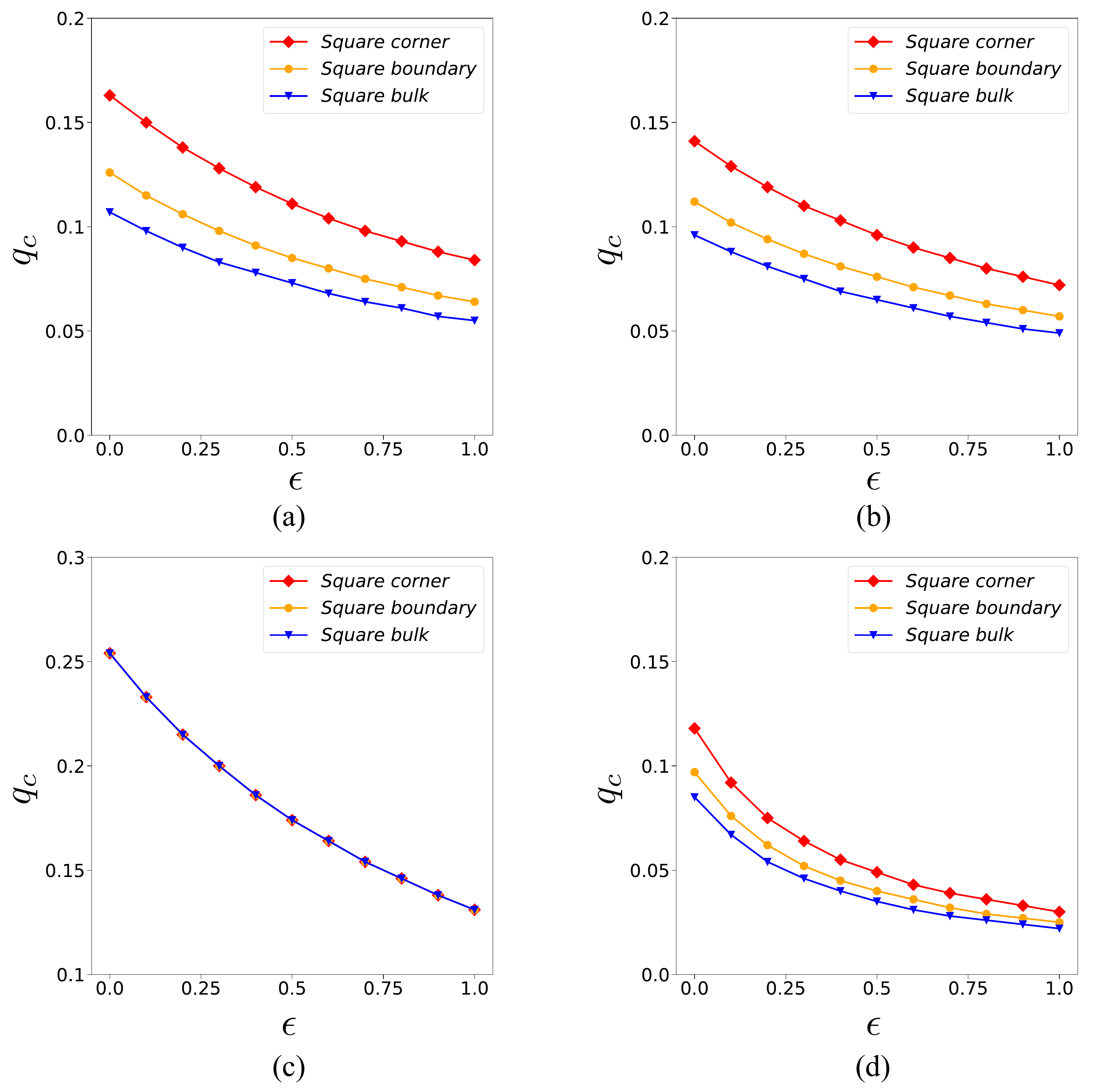}
\caption{Variations of $q_c$ as a function of $\epsilon$ in the case of the (a) BF, (b) BPF, (c) PD, and (d) DP noise, for a connected subsystem $S$ of four qubits (see Fig.~\ref{fig:demo_PMS}(c)) located at the boundary, corner, or bulk of a square graph.  All the axes in all figures are dimensionless.}
\label{fig:noise_square}
\end{figure*}

A word on the choice of the PMS labelled by $\alpha_c$ is in order here. We choose $\alpha_c$ such that 
\begin{enumerate}
    \item[(a)] $\alpha_c\in\Gamma$, i.e., all measurement outcomes corresponding to the PMS $\alpha_c$ are equally probable with probability $1/2^{N-n}$,  
    \item[(b)] the subgraph $G^{\alpha_c}_S$ on $S$ in the reduced graph $G^\prime_{\alpha_c}$ is a connected graph,  and 
    \item[(c)] the pattern of the PMS can be generalized to any system size.  
\end{enumerate}
We find that such a PMS $\alpha_c$ can be constructed in terms of only $\sigma^1$ and $\sigma^3$ measurements for all the examples discussed in Sec.~\ref{subsec:gme_examples_pure_graphs}. See Fig.~\ref{fig:demo_PMS} for a demonstration.

\begin{figure}
    \centering
    \includegraphics[width=0.8\linewidth]{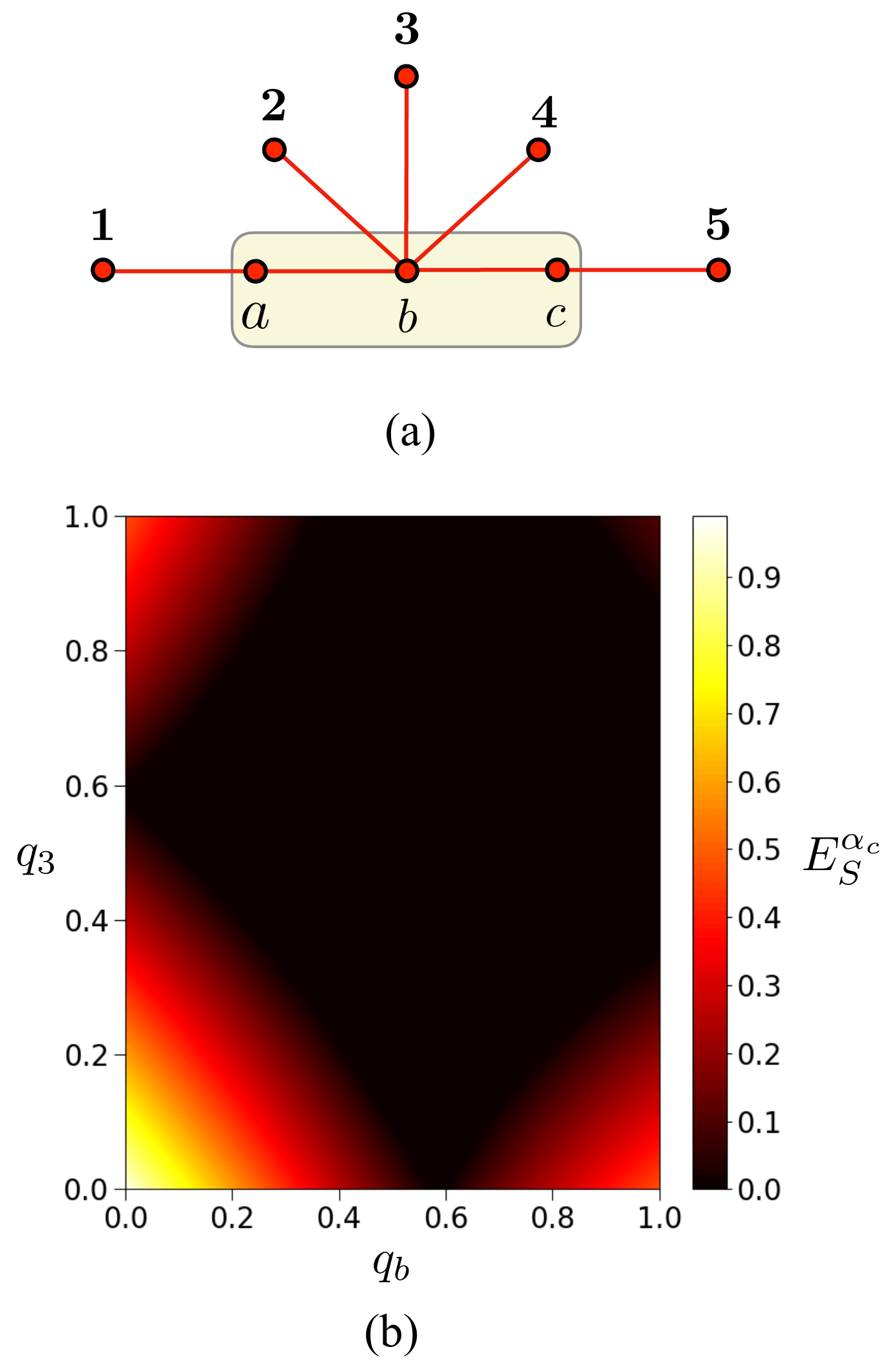}
    \caption{(a) Example of an $8$-qubit graph, where the noise applied on different qubits are as follows: $1$: BF ($5\times 10^{-3},10^{-1}$), $2$: BPF ($10^{-3},1.0$), $3$: BF ($q_3,1.0$), $4$: BF ($10^{-3},7\times 10^{-1}$), $5$: BPF ($3\times 10^{-3},9\times 10^{-2}$), $a$: BF ($4\times 10^{-3},8\times 10^{-1}$), $b$: BPF ($q_b,1.0$), $c$: PD ($8\times 10^{-3},9\times 10^{-1}$), where the numbers in the parenthesis are the values of $(q,\epsilon)$. The FLGME is computed over the subsystem $S$ constituted of the three-qubits labelled by $a$, $b$, and $c$. (b) Variation of FLGME (quantified by genuine multiparty concurrence, see Appendix~\ref{app:entanglement_in_gd_states}) as a function of $q_b$ and $q_3$, where the PMS $\alpha$ corresponds to $\sigma^3$ measurements on qubits $1-5$. All quantities plotted are dimensionless.}
    \label{fig:comment2}
\end{figure}

In this section, we particularly focus on the evolution of a quantum state from having GME localized on a chosen subsystem to a state where the post-measured states on $S$ are biseparable. Noticing that the post-measured states on $S$ after Pauli measurements are GD states in the case of Pauli noise (see Sec.~\ref{subsec:pauli_noise_pauli_measurement_le}), we choose four-qubit subsystems as $S$, for which the criteria for the GD state being genuine multipartite entangled is known~\cite{Guhne_2011} (see Appendix~\ref{app:multipartite_entanglement_gd} for details). Using this criteria, for each PMS labelled by $\alpha$, a \emph{critical} value of $q$, denoted by $q_c$, can be determined such that for $q_{c}\leq q\leq 1$, all post-measured states corresponding to the PMS $\alpha$ are biseparable. For the choice of $\alpha=\alpha_c$, the values of $q_{c}$ are plotted (see Figs.~\ref{fig:noise_linear_ladder} and \ref{fig:noise_square}) against the non-Markovianity parameter $\epsilon$ corresponding to the different types of Pauli noise (see Sec.~\ref{subsec:noise}) for the different cases discussed in Sec.~\ref{subsec:gme_examples_pure_graphs}, where the subsystem $S$ is a four-qubit one.  From the figures, the following observations can be made. 
\begin{enumerate}
\item[(a)]  For the PMS denoted by $\alpha_c$ and for a specific type of Pauli noise,  the values of $q_c$ depend on the effective neighbourhood $S_1^{\prime\prime}$ of $S$ (see the discussion in Appendix~\ref{app:GDstate} for the definition of $S_1^{\prime\prime}$). As one moves from the periphery (corner or boundary) to the bulk of the graph, $|S_1^{\prime\prime}|$ increases, and $q_c$ decreases. 

\item[(b)] Note, however, that in the case of PD noise (Figs.~\ref{fig:noise_linear_ladder}(c) and \ref{fig:noise_square}(c)), $|S_1^{\prime\prime}|$ remains the same in the case of the bulk, the boundary, and the corner, which manifests in $q_c$ remaining the same for all these cases. 

\item[(c)] Moreover, for all types of Pauli noise and for a specific category of subsystem $S$ (bulk, boundary, or corner), the value of $q_c$ decreases with increasing $\epsilon$, implying a quicker loss of FLGME with increasing non-Markovianity in the noise.

\item[(d)] In the case of the BF, BPF, and PD noise, $q_c$ has a quadratic dependence on $\epsilon$ ($q_c\sim a_2 \epsilon^{2}+a_1\epsilon+a_0$), while in the case of the DP noise, a cubic dependence ($q_c\sim a_3\epsilon^3+ a_2 \epsilon^{2}+a_1\epsilon+a_0$) is observed, The values of the parameters $a_i$s $(i=1,2,3)$ depends on the type of noise, the choice of the graph, and the location of the subsystem $S$ in the graph, while $a_0$ is the value of $q_c$ for $\epsilon=0$. 
\end{enumerate} 
We point out here that for a specific type of Pauli noise, the values of $q_c$ are independent of system size, $N$, for the chosen PMS $\alpha_c$ (cf.~\cite{dur2004}).  The values of $q_{c}$ in the Markovian cases can be extracted from the data for $\epsilon=0$ (see Sec.~\ref{subsec:noise}). Due to the maximization involved in the definition of LGME, the actual critical noise strength beyond which the LGME perishes is greater than, or equal to $q_c$. Note further that in the case of noisy systems with even noise models as simple as Pauli noise, it is not guaranteed whether $E_S=E_S^\mathcal{P}$. While the equality can occur for specific noise types and for specific values of the noise strength,  in general, $E_S^\mathcal{P}<E_S$. This calls for a comment on the performance of $E_S^{\alpha_c}$  and $E_S^{\mathcal{P}}$ as lower bounds of $E_S$. We revisit this in Sec.~\ref{sec:topological}, with an example of a stabilizer state from Kitaev's toric code~\cite{kitaev2001,kitaev2006}. 

Note that in the above examples, $(q_i,\epsilon_i)$s on different qubits are assumed to the identical. One can also consider a scenario where different noise of different strengths and different non-Markovianity parameters apply to the qubits (see Appendix~\ref{app:proof_proposition_2} and \ref{app:GDstate}), and similar methodologies apply. For a demonstration,  consider the example of the $8$-qubit graph shown in Fig.~\ref{fig:comment2}(a), where different types of noises with different strengths $q$ as well as different values of the non-Markovianity parameter $\epsilon$ are applied on different qubits. For the graph shown in Fig.~\ref{fig:comment2}(a), Eq.~(\ref{eq:new_eqn}) of Appendix~\ref{app:GDstate} representing the probability associated to the subclass $(r,m)=(2,1)$ constituted of qubits $2$, $3$, and $4$, takes the form 
\begin{eqnarray}
    P_1(2,1) &=& p_2+p_3+p_4-2(p_2p_3+p_3p_4+p_2p_a)\nonumber\\&&+p_2p_3p_4 \nonumber\\ 
    P_0(2,1) &=& 1-P_1(2,1).
\end{eqnarray}
where $p_i=\frac{q_i}{2}\left[1+\epsilon_i\left(1-\frac{q_i}{2}\right)\right]$ for $i=2,3,4$ are the probability for occurrence of noise on $i$th qubit. In Fig.~\ref{fig:comment2}(b), we plot the FLGME as a function of the noise parameters $q_b$ and $q_3$. We point out here that this example deals with different types of Pauli noise, different noise strengths, as well as different non-Markovianity parameters for different qubits, and thereby is a typical case of the most general scenario covered by the noise model and the methodology considered in this paper.

\begin{figure}
    \centering
    \includegraphics[width=0.7\linewidth]{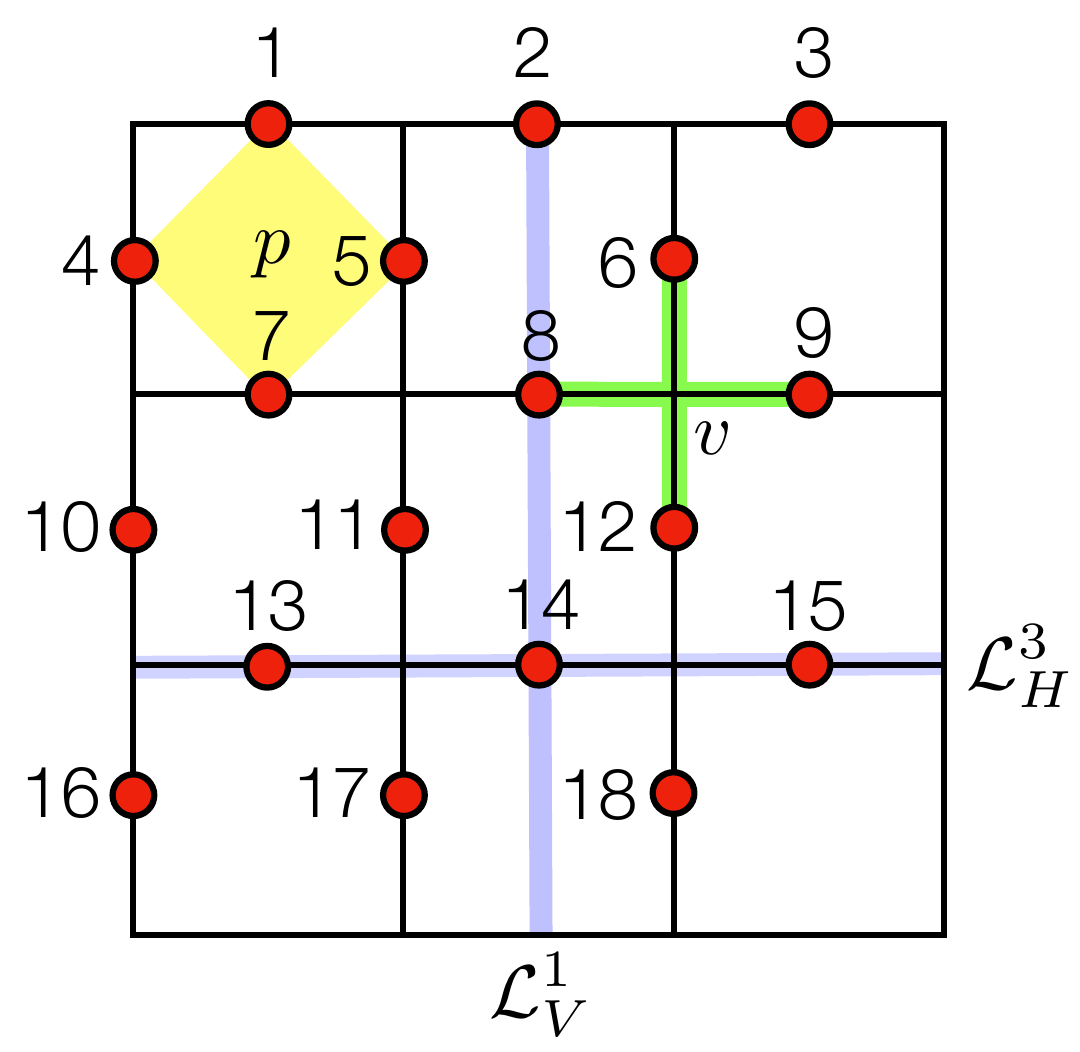}
    \caption{A toric code of $18$ qubits on a square lattice, where each of the plaquettes $p$ and the vertices $v$ are constituted of four qubits. Periodic boundary condition is assumed along both the horizontal and the vertical directions on the lattice.}
    \label{fig:toric_code}
\end{figure}

\section{Application in topological quantum codes}
\label{sec:topological}

Since any stabilizer state can be mapped to a graph state via local unitary transformations belonging to the Clifford group~\cite{van-den-nest2004}, the methodology discussed in Secs.~\ref{sec:le_noiseless_graph}-\ref{sec:le_noisy_graph} can also be applied to any stabilizer state, as long as the structure of the graph underlying the local unitarily connected graph state is known. In this section, we demonstrate this by applying the protocol to the toric code~\cite{kitaev2001,kitaev2006} defined on a square lattice, where each qubit in the system is placed on one of the edges of the lattice (see Fig.~\ref{fig:toric_code}). Two types of stabilizer operators, namely, the plaquette operators $\mathcal{S}_p=\otimes_{i\in p}\sigma^3_i$, and the vertex operators $\mathcal{S}_v=\otimes_{i\in v}\sigma^1_i$ are defined on the toric code, where $p$ and $v$ are respectively the plaquette and the vertex index. The stabilizer state $\ket{\mathcal{S}}$ on the toric code is given by the common eigenstate of all the stabilizer operators corresponding to the $(+1)$ eigenvalue, i.e.,
\begin{eqnarray}
    \mathcal{S}_p\ket{\mathcal{S}}=(+1)\ket{\mathcal{S}}, \quad  \mathcal{S}_v\ket{\mathcal{S}}=(+1)\ket{\mathcal{S}}.
\end{eqnarray}
Under periodic boundary condition assumed along both the horizontal and the vertical directions, the square lattice can  embedded on a \emph{genus-$1$} torus, hosting two non-trivial loops. We denote the sets of nodes, in the horizontal
and vertical directions, on which these non-trivial loops are defined, by $H$ and $V$, respectively (see Fig.~\ref{fig:toric_code}), and represent four non-trivial loop operators as
\begin{eqnarray}
    \mathcal{L}_\alpha^1=\otimes_{i\in\alpha}\sigma_i^1,\quad \mathcal{L}_\alpha^3=\otimes_{i\in\alpha}\sigma_i^3,
\end{eqnarray}
with $\alpha=H,V$.

\begin{figure*}
    \centering
    \includegraphics[width=\textwidth]{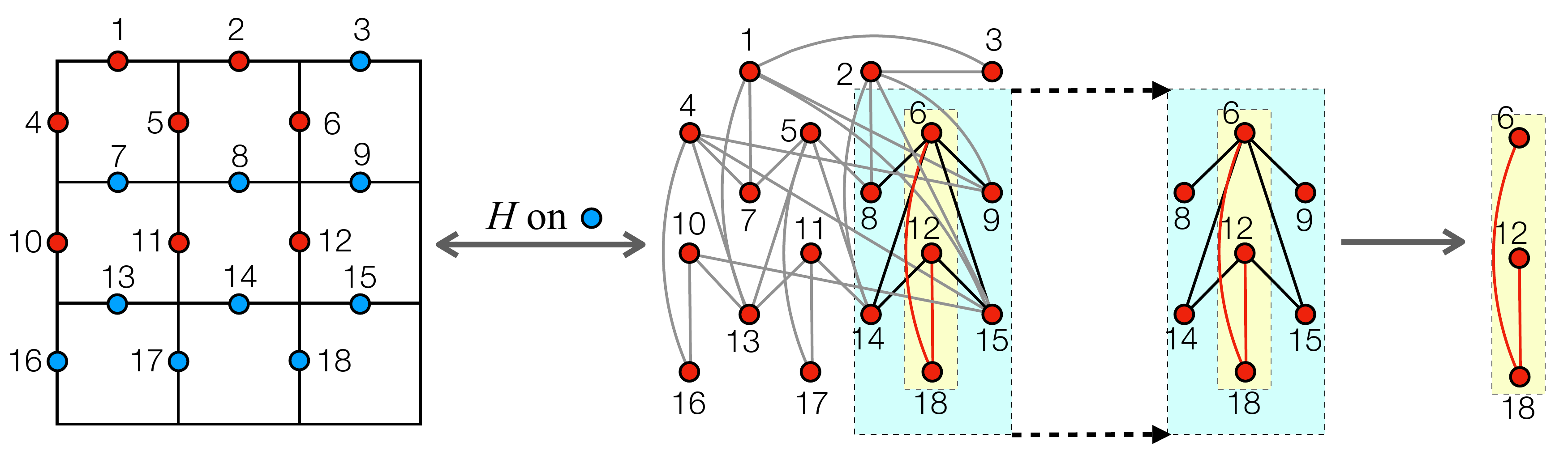}
    \caption{Transformation of a stabiliser state $\ket{\mathcal{S}}$ corresponding to a toric code defined on a $N_P=3$ square lattice via application of Hadamard operators on the control qubits. The control qubits are denoted by the blue nodes. The chosen subsystem $S$ is a non-trivial vertical loop of size $n=N_P$, on which a star graph can be obtained via a careful choice of the set of control qubits.}
    \label{fig:toric_loop}
\end{figure*}

In this section, we apply the methodology discussed in Secs.~\ref{sec:le_noiseless_graph}-\ref{sec:le_noisy_graph} on the stabilizer states $\ket{\mathcal{S}}$, and localize GME on the non-trivial loops in the horizontal or the vertical directions. Note that in the case of a square lattice of $N_P\times N_P$-architecture hosting a Kitaev model, where $N_P$ is the number of plaquettes in the horizontal or vertical direction, the number of qubits in the system grows as $N=2N_P^2$, while the size of a non-trivial loop is $N_P$. Therefore, the protocol discussed in Secs.~\ref{sec:le_noiseless_graph}-\ref{sec:le_noisy_graph} requires an optimization over all the reduced graphs having a connected subgraph $G^\alpha_S$ on $S$, which are obtained from  $3^{N_P(2N_P-1)}$ PMSs. While computation for pure states is still possible for moderately large system size, in the case of noisy toric codes, this is practically impossible to compute when $N_P$ (and subsequently $N$) is large.  However, following the same approach as in the case of the noisy graph states, one may compute FLGME, as we discuss below. 

\begin{figure}
    \centering
    \includegraphics[width=0.9\linewidth]{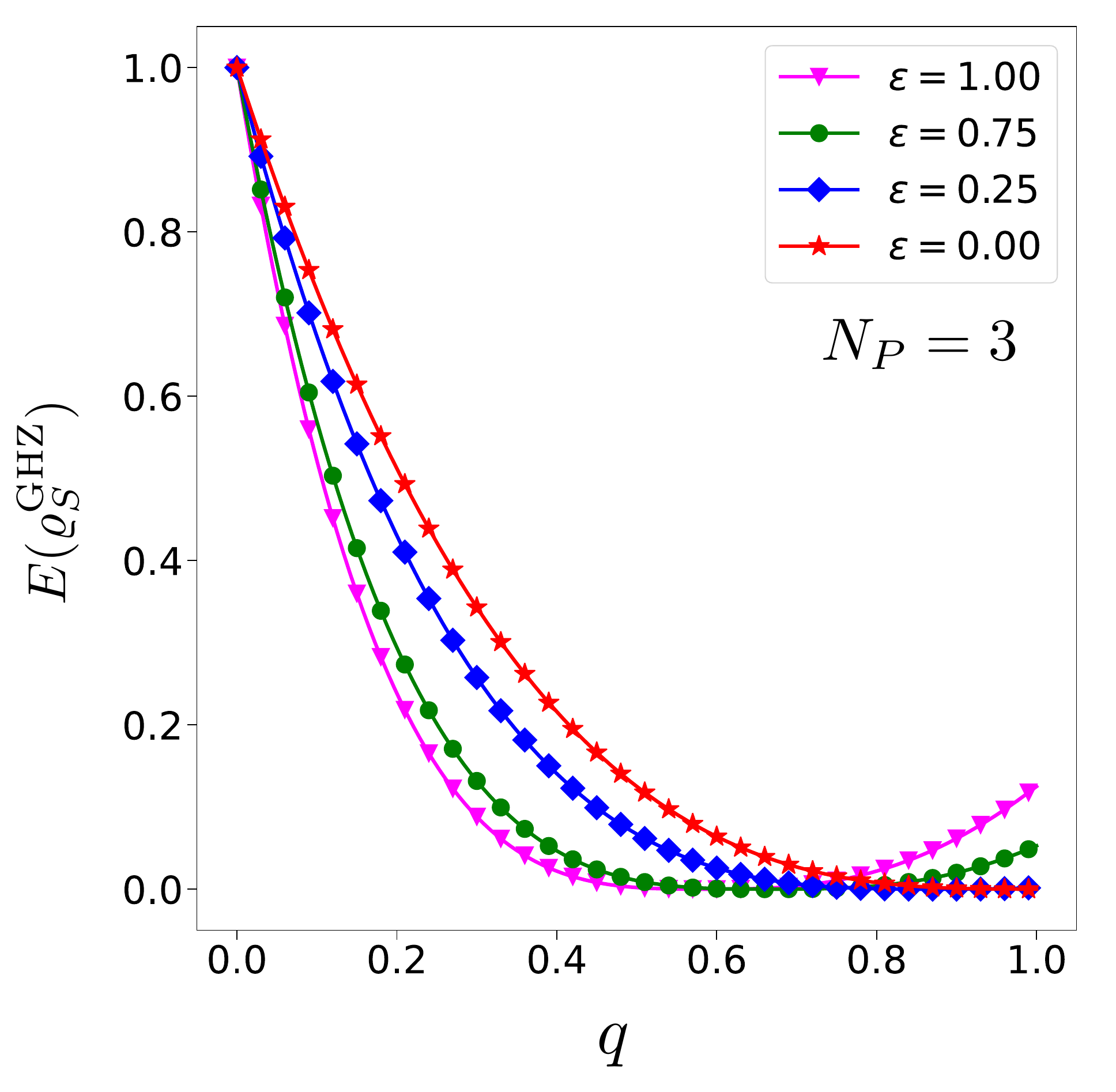}
    \caption{Variations of $E(\varrho_S^{\text{GHZ}})$ as a function of $q$ in the case of the BF noise applied to all qubits in a toric code defined on a $N_P=3$ square lattice, where $S$ is a non-trivial loop representing $\mathcal{L}^3_{H,V}$, and the non-Markovianity parameter $\epsilon$ takes values $\epsilon=0,0.25,0.75$, and $1$. Genuine multiparty concurrence is chosen as the entanglement measure. All the axes in all figures are dimensionless.}
    \label{fig:toric_loop_qc_1}
\end{figure}

The stabilizer state $\ket{\mathcal{S}}$ can be mapped to a graph state via Hadamard operations on judiciously chosen qubits~\cite{lang2012}, referred to as the \emph{control qubits}, such that a star subgraph can be obtained on $S$. In Fig.~\ref{fig:toric_loop}, an example of the mapping is presented for a toric code defined on an $N_P=3$ square lattice, where $S$ is chosen to be a non-trivial loop of $\mathcal{L}_{H,V}^3$ type. We choose a PMS $\alpha_c$ that, throughout the protocol discussed in Sec.~\ref{subsubsec:pauli_measurement_le} and Appendix~\ref{app:graph_operations}, keeps the star structure of the graph $G^{\alpha_c}_S$ unchanged. This can be achieved by 
\begin{enumerate}
\item[(a)] measuring $\sigma^1$ on the control qubits, and 
\item[(b)] measuring $\sigma^3$ on the rest of the qubits, 
\end{enumerate}
which we refer to as the \emph{star measurement setup} (SMS). Note that for a star subgraph $G^{\alpha_c}_S$ is local unitary equivalent to an $n$-qubit GHZ state, which is a genuinely multiparty entangled state, leading us to the following proposition. 

\noindent\textbf{$\blacksquare$ Proposition 3.} \emph{It is always possible to localize GME on a non-trivial loop of a toric code via single-qubit Pauli measurements on the rest of the system.} 

\noindent Computation of FLGME reveals that on the non-trivial loop of the toric code,  LGME is bounded from below by $E_S^{\alpha_c}=1$, as long as Schmidt measure is used for quantifying GME over $S$. If GGM is used, $E_S^{\alpha_c}=1/2$.

\begin{figure*}
    \centering
    \includegraphics[width=\textwidth]{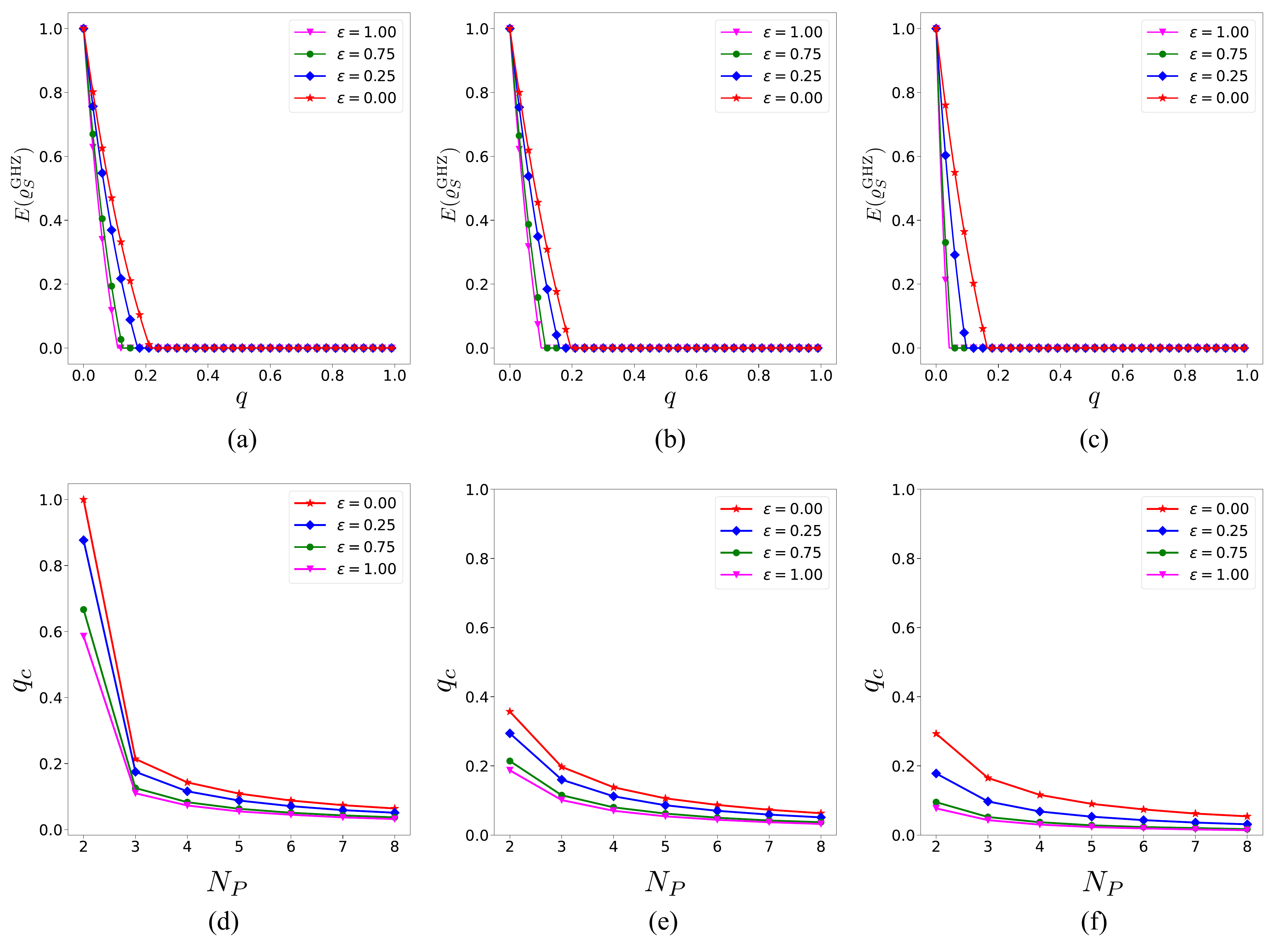}
    \caption{In (a)-(c), we depict variations of $E(\varrho_S^{\text{GHZ}})$ as a function of $q$ in the case of the (a) PD, (b) BPF, and (c) DP noise, applied to all qubits in a toric code defined on a $N_P=3$ square lattice, where $S$ is a non-trivial loop representing $\mathcal{L}^3_{H,V}$, and the non-Markovianity parameter $\epsilon$ takes values $\epsilon=0,0.25,0.75$, and $1$. Genuine multiparty concurrence is chosen as the entanglement measure. 
    In (d)-(f), we show variations of $q_c$ as a function of $N_P$ in the case of the (d) PD, (e) BPF, and (f)  DP noise, for a non-trivial loop of a toric code defined on a $N_P\times N_P$ square lattice, for $\epsilon=0,0.25,0.75$, and $1$. In all figures, $\epsilon=0$ stands for the case of Markovian noise. All the axes in all figures are dimensionless.}
    \label{fig:toric_loop_qc_2}
\end{figure*}

In order to determine FLGME under noise, we follow the methodology described in Sec.~\ref{subsec:no_forbidden_set}. We consider the toric code to be subjected to the Markovian and non-Markovian single-qubit Pauli noise discussed in Sec.~\ref{subsec:noise}, and investigate the evolution of the FLGME, which we now discuss. We first consider the case of BF noise applied to all qubits in the toric code. Our choice of $\alpha_c$ ensures that $|S^{\prime\prime}_1|=0$, further implying that the corresponding FLGME $E_S^{\alpha_c}$ is given by $E_S^{\alpha_c}=E(\varrho_S)$, where $\varrho_S$ is a GD state obtained when the hub of the star subgraph on $S$ in $G^\prime_{\alpha_c}$ is subjected to PD noise, and the rest of the qubits in $S$ are subjected to BF noise\footnote{This can be seen easily by following the modification of the noise through the protocol described in Sec.~\ref{subsubsec:pauli_measurement_le} and Appendix~\ref{app:graph_operations}.}. We also note that the star subgraph on $S$ is equivalent to the $N_P$-qubit GHZ state via Hadamard operations on the qubits with BF noise, implying that $E(\varrho_S)=E(\varrho_S^{\text{GHZ}})$, $\varrho_S^{\text{GHZ}}$ being the mixed state obtained when PD noise is applied to all qubits on the GHZ state. In the case of non-Markovian phase damping noise~\cite{krishnan2022},
\begin{eqnarray}
\varrho_S^{\text{GHZ}} &=& \frac{1}{2}\Big[ (\ket{0}\bra{0})^{\otimes N_P}+(\ket{1}\bra{1})^{\otimes N_P}\nonumber\\ 
&&+(1-f)^{N_P}\left((\ket{0}\bra{1})^{\otimes {N_P}}+(\ket{1}\bra{0})^{\otimes {N_P}}\right)\Big],\nonumber\\
\end{eqnarray}
with 
\begin{eqnarray}
f &=& q\left[1+\epsilon\left(1-\frac{q}{2}\right)\right]. 
\end{eqnarray}
Using this, the FLGME, as quantified by the genuine multiparty concurrence~\cite{Hashemi2012} (see Appendix~\ref{app:entanglement_in_gd_states}) as the GME measure, $E$, is given by $2\max[0,\lambda]$ with 
\begin{eqnarray}
\lambda &=& \left|\frac{(1-f)^{N_P}}{2}\right|. 
\end{eqnarray}
It is now straightforward to see that $E(\varrho_S^{\text{GHZ}})=0$ at a \emph{critical} value of $q$ given by 
\begin{eqnarray}
q_c=\frac{1+\epsilon-\sqrt{1+\epsilon^2}}{\epsilon}, 
\end{eqnarray}
while for all other values of $0\leq q< q_c$ and $q_c<q<1$, $E(\varrho_S^{\text{GHZ}})>0$ (see Fig.~\ref{fig:toric_loop_qc_1}). Note also that $q_c$ is independent of the system size $N_P$. 

\begin{figure*}
    \includegraphics[width=0.7\textwidth]{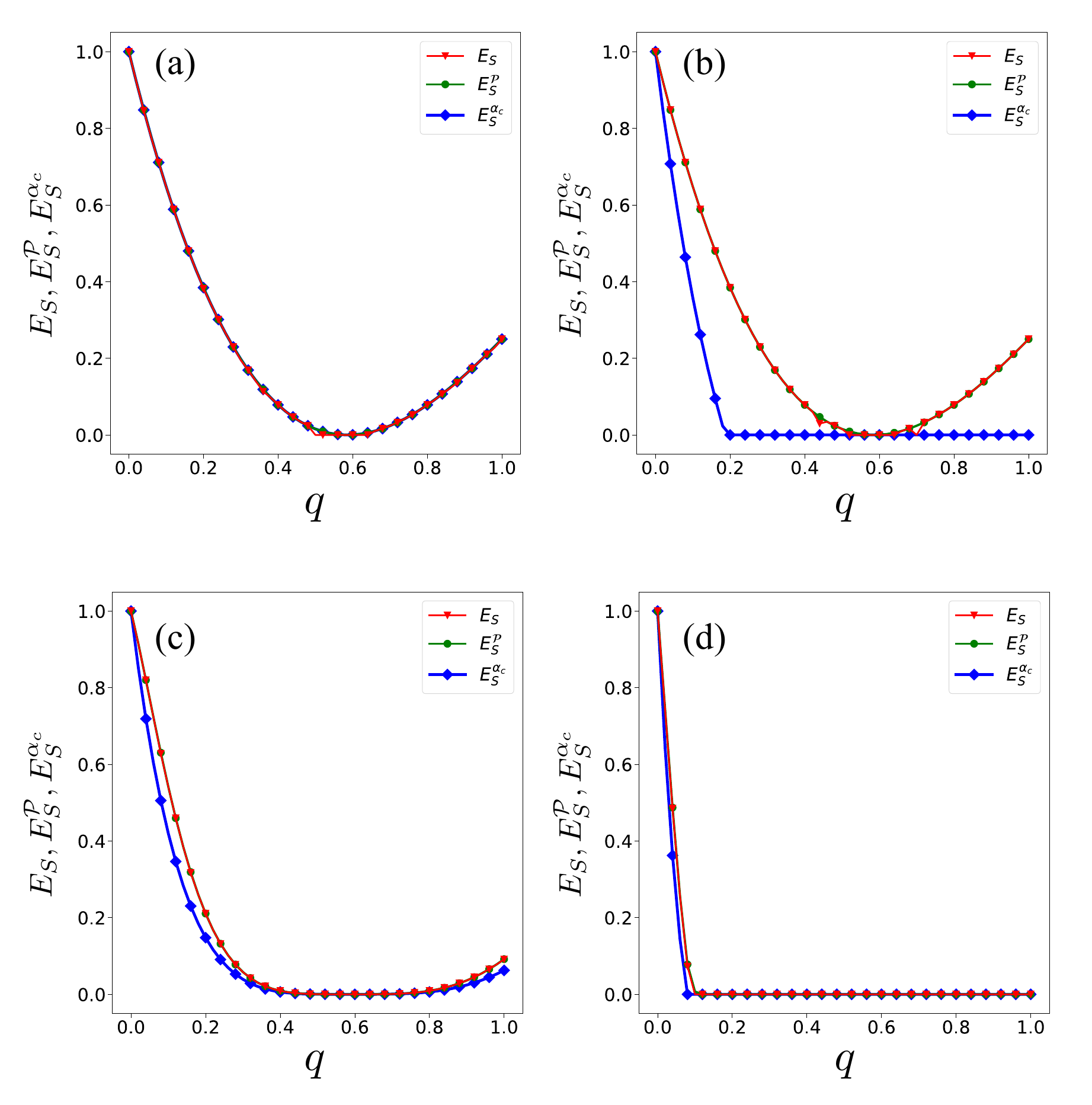}
    \caption{Variations of $E_S$, $E_S^{\mathcal{P}}$, and $E_S^{\alpha_c}$ as a function of $q$ in the case of the stabilizer state corresponding to a $2\times 2$ ($N_P=2$) toric code subjected to (a) BF, (b) BPF, (c) PD, and (d) DP channels on all qubits with $\epsilon=1.0$. All quantities plotted are dimensionless.}
    \label{fig:comment1}
\end{figure*}

\begin{figure}
    \centering
    \includegraphics[width=0.9\linewidth]{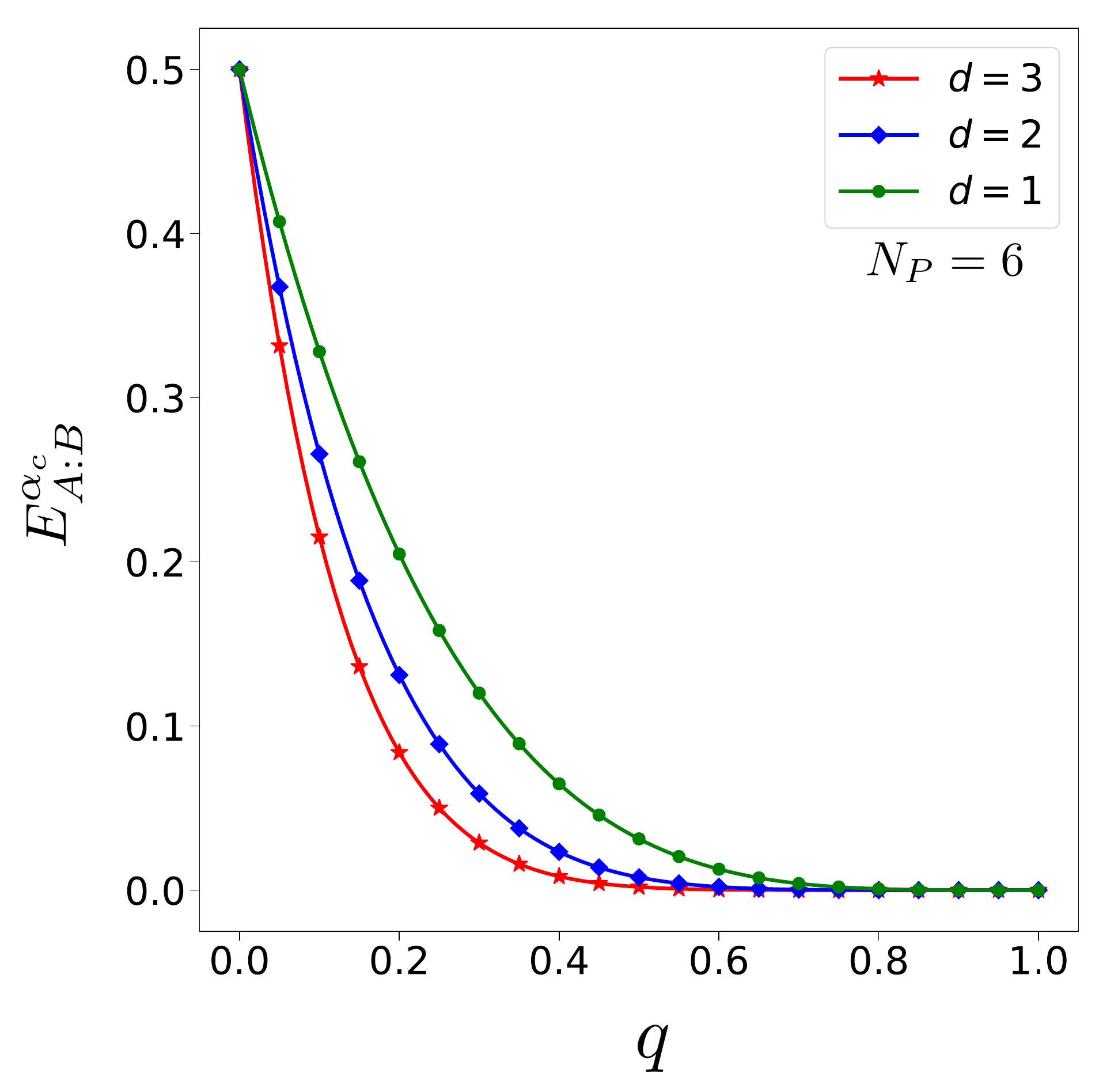}
    \caption{Variations of $E^{\alpha_c}_{A:B}$ as a function of $q$ in the case of the BF noise applied to all qubits in a toric code defined on a $N_P=6$ square lattice for various distances, $d$, between the non trivial loops. Both the axes in the figure are dimensionless. See Appendix \ref{app:xy_measurement_protocol} for the description of $\alpha_c$.}
    \label{fig:toric_loop_qc_6}
\end{figure}

We also perform similar investigations in the case of the PD, BPF, and DP noise. For each of these noise, using the results discussed in Appendix~\ref{app:entanglement_in_gd_states}, we determine the values of $q_c$ beyond which all post-measured states corresponding to the SMS $\alpha_c$ becomes biseparable\footnote{For $\epsilon\neq 0$, there may be revival of the FLGME after it decays to zero. In such cases, we consider $q_c$ to be the first instance at which FLGME vanishes. For example, see $2\times 2$ toric code under non-Markovian PD noise (Fig.~\ref{fig:comment1}(c))}. In contrast to the BF noise, $q_c$ is found to decrease monotonically with increasing $N_P$, implying a faster loss of FLGME for larger system. The variations of $q_c$ as a function of the system size $N_P$, for different values of $\epsilon$, is given in Fig.~\ref{fig:toric_loop_qc_2}.   Note also that for a fixed system size $N_P$, $q_c$ decreases monotonically with increasing non-Markovianity parameter $\epsilon$.

We now estimate the performance of $E_S^{\alpha_c}$ and $E_S^{\mathcal{P}}$ as  lower bounds  of $E_S$. For demonstration, we consider Kitaev's toric code on a square lattice of dimension $2\times 2$, subjected to single-qubit Pauli noise on all qubits, where $E_S$ and $E_S^\mathcal{P}$ as a function of the noise strength $q$ are computed, taking $S$ to be the two-qubit non-trivial loop on the code. In all cases of bit-flip, bit-phase-flip, phase-damping, and depolarizing channels, we find $E_S=E_S^\mathcal{P}$ for all noise strengths $0\leq q\leq 1$ and for all non-Markovianity parameters $0\leq \epsilon\leq 1$, with examples corresponding to $\epsilon=1.0$ for different types of noise depicted in Fig.~\ref{fig:comment1}. The performance of the bound $E_S^{\alpha_c}$, on the other hand, depends on the judicious choice of $\alpha_c$, and $E_{S}^{\mathcal{P}}=E_S^{\alpha_c}$ iff an optimal Pauli measurement setup is chosen as $\alpha_c$. For the choice of the SMS as $\alpha_c$, in Fig.~\ref{fig:comment1}, we plot $E_S^{\alpha_c}$ as functions of $q$ for  $\epsilon=1.0$.  It is clear from the figures that along with the choice of $\alpha_c$, the performance of $E_S^{\alpha_c}$ as a lower bound of $E_S^\mathcal{P}$ depends also on the type of noise as well as noise strength. For example, in the case of the bit-flip noise, $E_S^\mathcal{P}=E_S^{\alpha_c}$ for all values of $q$, while for the bit-phase-flip noise,  there exists range of $q$ over which $E_S^{\mathcal{P}}$ ($=E_S$) is non-zero, while $E_S^{\alpha_c}$ is not.

We point out here that the protocol discussed in Sec.~\ref{subsubsec:pauli_measurement_le} and Appendix~\ref{app:graph_operations}, and the methodology for obtaining a lower bound of localizable entanglement in the presence of noise works as long as a suitable $\alpha_c$ can be constructed. For example, one may also aim to compute the lower bound for the localizable bipartite entanglement over a specific bipartiton $AB$ of the subsystem $S$. We demonstrate this by computing a lower bound $E^{\alpha_c}_{A:B}$ of the localizable bipartite entanglement between two non-trivial loops in the toric code, where the two loops combined form the subsystem $S$. The appropriate PMS $\alpha_c$ used for this calculation is discussed in Appendix~\ref{app:xy_measurement_protocol}, while negativity~\cite{peres1996,horodecki1996,Karol1998,vidal2002,Plenio2005,plenio2005a} (see Appendix~\ref{app:entanglement_in_gd_states}) is used as the bipartite entanglement measure $E$. Variations of  $E^{\alpha_c}_{A:B}$ as a function of $q$ is depicted for an $N_P=6$ square lattice, and for increasing distances between the two chosen non-trivial loops of $\mathcal{L}_{H,V}^{3}$ type in Fig.~\ref{fig:toric_loop_qc_6}. Note that with increasing distance, $E^{\alpha_c}_{A:B}$ between two non-trivial loops decreases when $q$ is fixed.

\section{Conclusion and Outlook}
\label{sec:conclusion}

In this paper, we investigate LGME on subsystems of arbitrary large stabilizer states in noiseless and noisy scenarios. We demonstrate the calculation of lower bounds of LGME for pure stabilizer states using multi-qubit graph states as their representatives, and adopting a graph-based technique for performing single-qubit Pauli measurements on graphs. We also show that the calculation of a lower bound of LGME using the graphical technique has a polynomial scaling with the system size. We calculate LGME over subsystems of linear, ladder, and square graphs of arbitrary sizes. We further expand the calculation in the case of arbitrary graph states in the presence of single-qubit Pauli noise of Markovian and non-Markovian types. Using the linear, ladder, and square graphs of arbitrary sizes as examples, we demonstrate the existence of a critical noise strength corresponding to a lower bound of LGME for a specific Pauli measurement setup, beyond which all post-measured states corresponding to the chosen measurement setup become biseparable. The results for the graph states can be translated directly to arbitrary stabilizer states due to their local unitary connection with graph states. We demonstrate this by considering the stabilizer state corresponding to a toric code defined on a square lattice. We also provide a specific Pauli measurement setup to determine a lower bound of the localizable bipartite entanglement between two non-trivial loops situated a distance apart in the code.  

We conclude with a brief overview of the future research directions arising out of this paper. The graph-based methodology for computing appropriate lower bounds of LGME over chosen subsystems of arbitrary stabilizer states open up the possibility of thoroughly studying and characterizing LGME on subsystems of topological quantum codes, including the toric code~\cite{kitaev2001,kitaev2006} and the color code~\cite{bombin2006,bombin2007} in the absence and presence of naturally occurring local Pauli noise in experiments~\cite{Schindler_2013,Bermudez2017}. It also provides avenues to explore localizable bipartite entanglement between two non-trivial loops representing different logical operators in the codes, and provides the motivation to look for appropriate witness operators to construct witness-based lower bounds of LGME, in the same vein as in~\cite{amaro2018,amaro2020a}. Moreover, our paper provides a way of characterizing parts of a large quantum network build out of stabilizer states~\cite{Englbrecht2022}.  

In view of the importance of devising methodologies that are applicable in noisy intermediate-scale quantum (NISQ) devices~\cite{Bharti2022}, which are ideal test-beds for dynamics of open quantum systems, we point out that determination of LGME using our proposed methodology requires (a) Pauli measurements on subsets of qubits in a multi-qubit system, and (b) determination of entanglement of a quantum state post measurement. Regarding the former, measuring Pauli operators in NISQ devices is possible, and is studied extensively~(see~\cite{Bharti2022} and the references therein).  With respect to the latter, we point out that while we focus on genuine multipartite entanglement measures, one can also perform a similar study with other types of entanglement, eg. bipartite entanglement. This requires detecting and quantifying entanglement of arbitrary states in NISQ devices, which has been an active area of research in recent times.  Specific algorithms towards this goal already exist using the positive map criterion~\cite{Wang2020}, for computing tangle~\cite{Salinas2020} for three-qubit systems, and for computing entanglement spectrum~\cite{LaRose2019,Cerezo2022}. Moreover, we point out here that the localizable entanglement can be connected to entanglement witness operators~\cite{amaro2018,amaro2020a,Hk1}, the expectation value of which, computed in the post-measured states, can provide a lower bound of localizable entanglement -- bipartite, or multipartite. In the case of stabilizer states, these witness operators, are typically designed with stabilizer operators, which are constituted of Pauli matrices, and whose expectation values can be accessed in NISQ devices~(\cite{hamilton2022}, see also~\cite{Bharti2022}, and the references therein).  Therefore, our methodology can potentially be applied to NISQ devices also, while the specific algorithm remains to be worked out.

\acknowledgements 

We acknowledge the support from the Science and Engineering Research Board (SERB), India through the Start-Up Research Grant (SRG) (File No. -  SRG/2020/000468 Date: 11 November 2020), and the use of \href{https://github.com/titaschanda/QIClib}{QIClib} -- a modern C++ library for general purpose quantum information processing and quantum computing.

\appendix 

\section{Operations on graph}
\label{app:graph_operations}

\begin{figure}
\includegraphics[width=\linewidth]{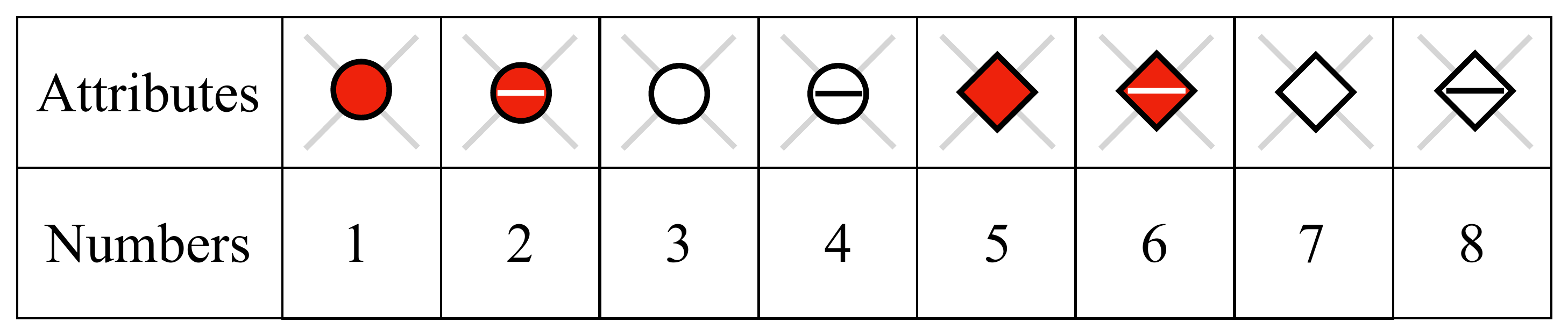}
\caption{Attributes for the nodes of a graph according to the application of (1) $I_i$, (2) $\sigma^3_i$, (3) $H_i$, (4) $H_i\sigma_i^3$, (5) $R_i$, (6) $R_i\sigma^3_i$, (7) $H_iR_i$, and (8) $H_iR_i\sigma^3_i$ on a node in the graph. We denote the different attributes by the corresponding numbers in the  text.} 
\label{fig:nodes}
\end{figure}

In this Section, we discuss a number of graph operations, and the related local unitary operations on the corresponding graph states. In order to distinguish between different unitary operations on a specific node of a graph, we provide different attributes, $a$, to each node. We particularly focus on the unitary operations $\{\sigma^3, H,R=\sqrt{\sigma^3}\}$, where the node attributes corresponding to different combinations of these unitary operations are shown in Fig.~\ref{fig:nodes}. The importance of these unitaries is discussed in Sec.~\ref{subsubsec:pauli_measurement_le}. Note that the attributes depicted in Fig.~\ref{fig:nodes} can be considered to be different combinations of three basic binary attributes -- (a) shape $(s)$ (circle $(c)$,  or diamond $(d)$), (b) fill $(f)$ (red $(r)$, or white $(w)$), and (c) sign $(sg)$ (plus $(+)$ or minus $(-)$, where we represent the plus sign by the absence of the sign to keep the figures uncluttered). At the graph level, we often describe the operations in terms of changing one or more of these attributes, while each such operation corresponds to a local unitary operation on the corresponding graph state.

\begin{figure}
\includegraphics[width=\linewidth]{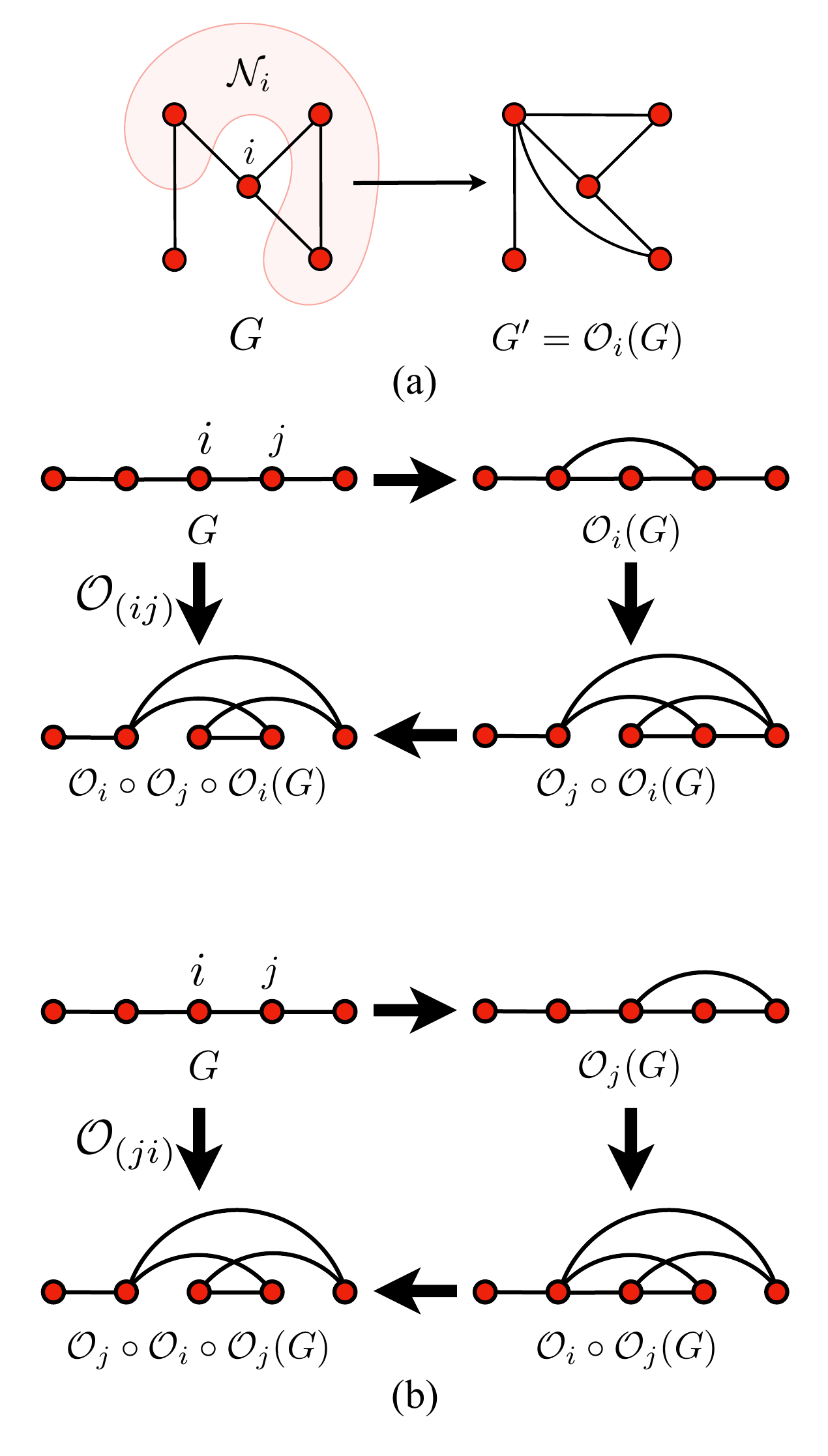}
\caption{(a) Local complementation operation on the graph $G$ w.r.t. the node $i$. The shaded region contains the nodes that constitute the neighborhood $\mathcal{N}_i$ of the node $i$. (b) Equivalence of $\mathcal{O}_{ij}$ and $\mathcal{O}_{ji}$ demonstrated on a linear graph of $5$ nodes.} 
\label{fig:lc}
\end{figure}

\begin{figure}
    \centering
    \includegraphics[width=0.9\linewidth]{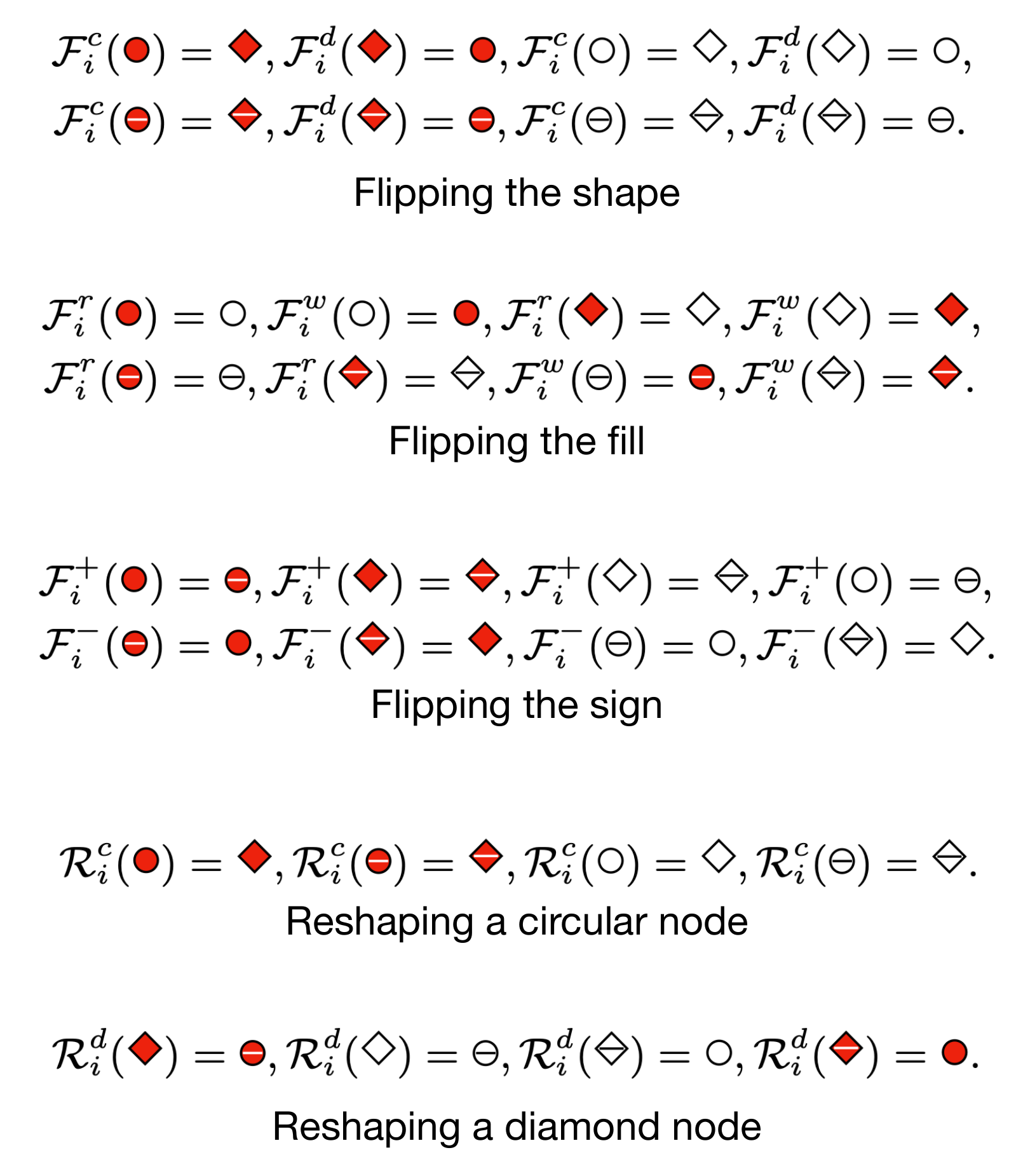}
    \caption{Fill and reshape operations on the node attributes.}
    \label{fig:graph_equations}
\end{figure}

\paragraph{Local complementation} Local complementation~\cite{VandenNest2004,VandenNest2004a,hein2006,Bouchet1991,Bouchet1993} of a connected graph $G$ w. r. t. a node $i\in G$ is a simple local graph operation, performed by deleting (creating) all the links $(j,k)$ between the nodes $j$ and $k$  $\in\mathcal{N}_i$, if $(j,k)$ is present (absent) in the graph. Here, $\mathcal{N}_i$ is the neighborhood of the node $i$, constituted of all the nodes in the graph that are connected to the node $i$ via a link. An example of the local complementation operation is demonstrated in Fig.~\ref{fig:lc}(a) for a graph with $5$ nodes. Let $G^\prime=\mathcal{O}_i(G)$ represents the graph resulting from the local complementation on the node $i$ in $G$. The corresponding graph states, $\ket{G^\prime}$ and $\ket{G}$, are connected by an LC operation on the node $i$ and its neighboring nodes, as~\cite{van-den-nest2004} $\ket{G^\prime}=U_{\text{LC}}^{i}\ket{G}$, with 
\begin{eqnarray}
U_{\text{LC}}^{i} &=& \text{e}^{-\text{i}\frac{\pi}{4}\sigma^1_{i}}\otimes_{j\in\mathcal{N}_{i}} \text{e}^{\text{i}\frac{\pi}{4}\sigma^3_{j}}
\label{eq:LC-LU}
\end{eqnarray}

\begin{figure*}
    \centering
    \includegraphics[width=\textwidth]{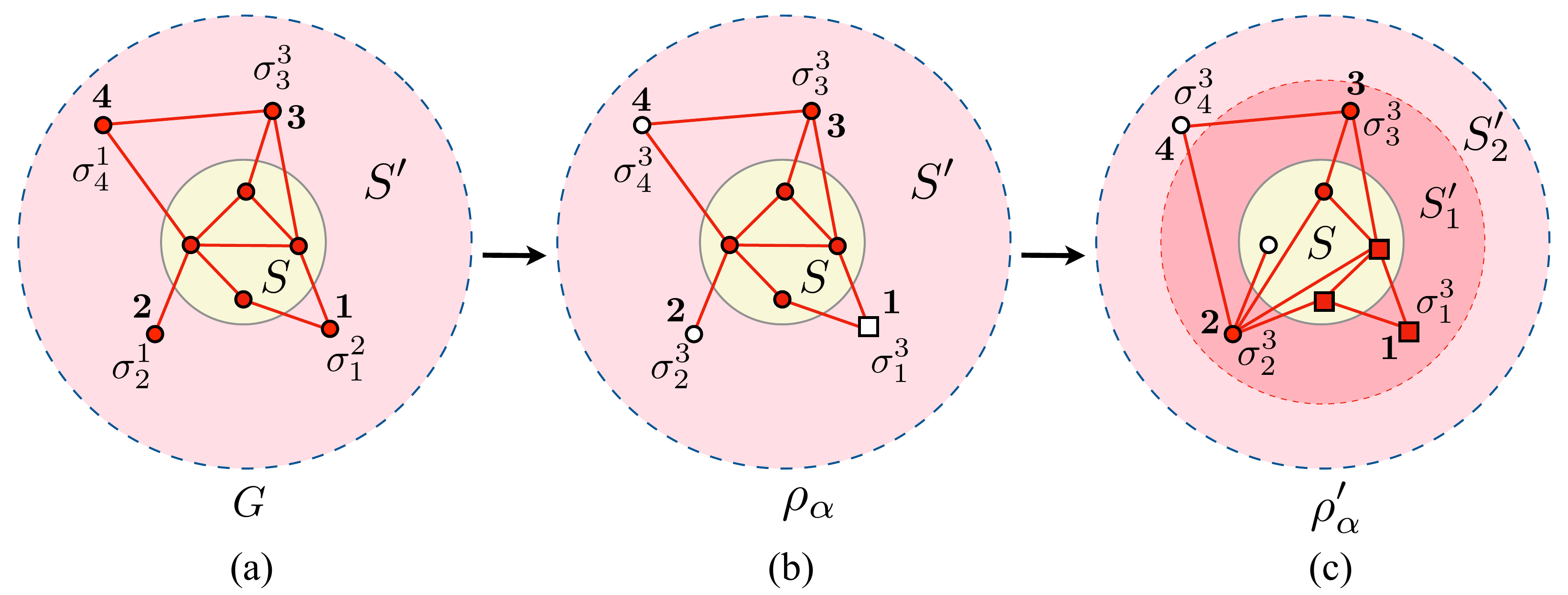}
    \caption{Transformation of the graph $G\rightarrow G_R$ shown in Fig.~\ref{fig:regions} in the graphical representation described in Appendix~\ref{app:graph_operations}.}
    \label{fig:grep}
\end{figure*}

\paragraph{Local complementation along an edge} Local complementation along an edge $(i,j)\in G$ is a sequence of three local complementations on $G$ w.r.t the nodes $i$ and $j$, given by $\mathcal{O}_{(ij)}=\mathcal{O}_i\circ\mathcal{O}_j\circ\mathcal{O}_i$, where $\mathcal{O}_j\circ\mathcal{O}_i(G)=\mathcal{O}_j\left(\mathcal{O}_i(G)\right)$. Local complementation along an edge is symmetric with an interchange of $i$ with $j$, i.e., $\mathcal{O}_{(ij)}\equiv\mathcal{O}_{(ji)}$. This is demonstrated in Fig.~\ref{fig:lc}(b) in the case of a linear graph with $5$ nodes. Evidently, local complementation along an edge represents a local unitary operation on the corresponding graph state, which is constituted by the unitaries of the form $U^i_{\text{LC}}$, as given in Eq.~(\ref{eq:LC-LU}). 

\paragraph{Flip} The flip operation $\mathcal{F}_{i}^a$ on the node $i$ in $G$ reverses one of the binary attributes, shape, fill, and sign of the node, where the superscript $a$ in $\mathcal{F}_{i}^a$ represents the value of the flipped binary attribute, with $a=s,f,sg$. The flip operations are consolidated in Fig.~\ref{fig:graph_equations}.

\paragraph{Reshape} Reshaping is an operation that is specific to the shape attribute of a node. We denote this operation by $\mathcal{R}_i^{s}$, where the superscript $s$ is the shape attribute of the node prior to applying  $\mathcal{R}^s_i$, on which the action of $\mathcal{R}^s_i$ depends. If the shape of a node is a circle, then $\mathcal{R}_i^c$ results in only a \emph{flip of the shape} that changes the shape attribute from circle to diamond, i.e., $\mathcal{R}_i^c\equiv\mathcal{F}_i^c$. On the other hand, if the shape of a node is a diamond, then $\mathcal{R}_i^d$ is equivalent to (a) a \emph{flip of the shape} that changes the shape attribute from diamond to circle, and (b) a \emph{flip of the sign} from $+$ ($-$) to $-$ ($+$). These rules are also consolidated in Fig.~\ref{fig:graph_equations}.

\subsection{Obtaining the reduced graph}
\label{app:obtaining_GR}

We now discuss the graph transformation $G\rightarrow G^\prime_\alpha$ for a specific choice of the PMS $\alpha$, which is a key ingredient of the protocol for computing $E^\mathcal{P}_S$ over an arbitrary $S$ in an arbitrary $G$, as discussed in Sec.~\ref{subsubsec:pauli_measurement_le}. Starting from $\rho$ corresponding to an arbitrary graph $G$, we achieve this in two steps, \textbf{Step A} and \textbf{Step B}. In \textbf{Step A}, we graphically represent the state $\rho_\alpha=\mathbf{U}^\dagger_\alpha\rho\mathbf{U}_\alpha$ (see Eq.~(\ref{eq:primed_graph})). Next, we perform \textbf{Step B} to obtain the graphical representation of the state $\rho^\prime_\alpha=\mathbf{V}_\alpha\ket{G^\prime_\alpha}\bra{G^\prime_\alpha}\mathbf{V}^\dagger_\alpha$, from which the graph $G^\prime_\alpha$ can be straightforwardly extracted.

\noindent\textbf{Step A.} As stated in Sec.~\ref{subsubsec:pauli_measurement_le}, we are specifically interested in the single-qubit unitary operations $\{\sigma^3_i,H_i,R_i\}$ on the qubit~$i$. Based on the graph operations described above, these unitaries can be represented as follows~\cite{elliott2008,elliott2009}. 
\begin{enumerate}
    \item[A.1] $H_i$ on any node: $\mathcal{F}_i^{f}$ ($\mathcal{F}_i^r(G)$ or $\mathcal{F}_i^w(G)$, depending on whether $f=r$ or $w$ for the node $i$).
    \item[A.2] $R_i$ on a node with $f=r$: $\mathcal{R}_i^s$ ($\mathcal{R}_i^c(G)$ or $\mathcal{R}_i^d(G)$, depending on whether $s=c$ or $d$ for the node $i$).
    \item[A.3] $R_i$ on a node with $f=w$, $s=c$, $sg=+$:~$\mathcal{R}^{s}_{\mathcal{N}_i}\circ\mathcal{O}_i$, where $\mathcal{R}^{s}_{\mathcal{N}_i}$ represents $\mathcal{R}^s_j$ operations on all nodes $j\in\mathcal{N}_i$.  
    \item[A.4] $R_i$ on a node with $f=w$, $s=c$, $sg=-$: $\mathcal{F}^{sg}_{\mathcal{N}_i}\circ\mathcal{R}^{s}_{\mathcal{N}_i}\circ\mathcal{O}_i$, where $\mathcal{R}^{s}_{\mathcal{N}_i}$ represents $\mathcal{R}^s_j$ operations on all nodes $j\in\mathcal{N}_i$, and $\mathcal{F}^{sg}_{\mathcal{N}_i}$ stands for $\mathcal{F}^{sg}_j$ $\forall j\in\mathcal{N}_i$. 
    \item[A.5] $R_i$ on a node with $f=w$, $s=d$, $sg=+$: $\mathcal{F}^{sg}_{\mathcal{N}_i} \circ\mathcal{R}^{s}_{\mathcal{N}_i}\circ\mathcal{O}_i\circ\mathcal{F}^{d}_{i}\circ\mathcal{F}^{w}_{i}$ where $\mathcal{F}^{sg}_{\mathcal{N}_i} ,\mathcal{R}^{s}_{\mathcal{N}_i}$ denotes flip and reshape operations respectively on entire neighbourhood of the node.
    \item[A.6] $R_i$ on a node with $f=w$, $s=d$, $sg=-$: $\mathcal{R}^{s}_{\mathcal{N}_i}\circ\mathcal{O}_i\circ\mathcal{F}^{d}_{i}\circ\mathcal{F}^{w}_{i}$
    \item[A.7] $\sigma^3_i$ on a node with $f=r$: $\mathcal{F}^{sg}_i$ ($\mathcal{R}_i^+(G)$ or $\mathcal{R}_i^-(G)$, depending on whether $sg=\pm$ for the node $i$)
    \item[A.8] $\sigma^3_i$ on a node with $f=w$, $s=c$: $\mathcal{F}_{\mathcal{N}_i}^{sg}$. Where $\mathcal{F}^{sg}_{\mathcal{N}_i}$ denotes flip of sign on all neighbourhood nodes. 
    \item[A.9] $\sigma^3_i$ on a node with $f=w$, $s=d$: $\mathcal{F}^{sg}_{i} \circ\mathcal{F}_{\mathcal{N}_i}^{sg}$
\end{enumerate}
For ease of representation, we denote the graph operations corresponding to the rules $A.1,\cdots, A.9$ as $\mathcal{A}_1,\cdots,\mathcal{A}_9$, where, for example, $\mathcal{A}_3\equiv\mathcal{R}_{\mathcal{N}_i}^s\circ\mathcal{O}_i$. Note that a subset of the transformations $\{\mathcal{A}_1,\cdots,\mathcal{A}_9\}$ according to a specific unitary operator $\mathbf{U}_\alpha$ corresponding to a specific PMS $\alpha$ provides a graphical representation of the state $\rho_\alpha$ (see Eq.~(\ref{eq:primed_graph}) and the corresponding discussions).   

\noindent\textbf{Step B.} To achieve this step,  we additionally use two graph transformations, referred to as the \emph{equivalence transformations}, that correspond to Eq.~(\ref{eq:reduced_form}) in Sec.~\ref{subsubsec:pauli_measurement_le}. These graph transformations, denoted by $\mathcal{B}_1$ and $\mathcal{B}_2$, are described below. 
\begin{enumerate}
    \item[B.1] The graph operations constituting $\mathcal{B}_1$ on a node $i$ with $s=d$  are as follows, in the specified order: (1) flip the fill of the node $i$, (2) perform local complementation on the node $i$, (3) reshape the neighbors of the node $i$ in the transformed graph, (4) flip the sign of the node $i$, and (5) if $sg=-$ for the node $i$ in the resulting graph, then flip the signs of the neighbors of the node $i$ in the transformed graph also. 
    \item[B.2] The graph operations constituting $\mathcal{B}_2$ on a link $(i,j)\in G$, such that both $i$ and $j$ have $s=c$, are as follows, in the specified order: (1) flip the fills of the nodes $i$ and $j$, (2) perform a local complementation along the edge $(i,j)$, (3) flip the signs of the nodes that are connected to both of $i$ and $j$ in the transformed graph, and (4) if either of the nodes $i$ and $j$ in the transformed graph has $sg=-$, flip the sign of that node as well as the signs of its neighbors in the transformed graph.  
\end{enumerate}

We now present the algorithm for obtaining $G^\prime_\alpha$ in the form of a pseudo code, using the above equivalence transformations. 

\noindent\hrulefill

\noindent\textbf{input} (1) The graphical representation of $\rho_\alpha$, as obtained from \textbf{Step A}, with nodes of eight different attributes and the links, and (2) sets $S^\prime$ and $S$ of measured and unmeasured nodes.
\begin{enumerate}
\item \emph{obtain the set $\mathcal{D}_{dw}$ of all nodes with $s=d$, $f=w$ (i.e., type $7$ and type $8$ nodes (see Fig.~\ref{fig:nodes})),  and the set $\mathcal{L}_{cw}$ of all pairs $(i,j)$ of nodes with $s=c$, $f=w$, (i.e., (type $3$, type $3$), (type $3$, type $4$), (type $4$, type $3$), and (type $4$, type $4$) links) such that the link $(i,j)$ exists. The sizes of these sets are given by $|\mathcal{D}_{dw}|$ and $|\mathcal{L}_{cw}|$, respectively. }
    \begin{enumerate} 
        \item \textbf{while} $|\mathcal{L}_{cw}|>0$ \emph{or} $|\mathcal{D}_{dw}|>0$, \textbf{do}
        \begin{enumerate}[label=(\Roman*)]
            \item \emph{operate $\mathcal{B}_1$ on all nodes $i\in\mathcal{D}_{dw}$}
            \item \emph{operate $\mathcal{B}_2$ on all $(i,j)\in\mathcal{L}_{cw}$}
        \end{enumerate}
        \item \emph{determine $|\mathcal{L}_{cw}|$, $|\mathcal{D}_{dw}|$}.
        \item \textbf{if} $|\mathcal{L}_{cw}|>0$ \emph{or} $|\mathcal{D}_{dw}|>0$, \textbf{repeat} (a)-(c).
        \item[]\textbf{else} \textbf{exit} 
    \end{enumerate} 
    \item \emph{obtain the set $\mathcal{E}_{B}$ of all boundary links $(i,j)$ having node $i\in S^\prime$, and node $j\in S$, such that $i$ is either of type $3$, type $4$ nodes (see Fig.~\ref{fig:nodes}). Let us denote the set of these nodes $i\in S^\prime$ as $\mathcal{N}_{S}$.}   
    \begin{enumerate}
        \item \textbf{for} \emph{all $(i,j)\in \mathcal{E}_B$ such that $i\in\mathcal{N}_{S}$,} \textbf{do}
        \begin{enumerate}[label=(\Roman*)]
            \item \textbf{if} \emph{$j$ has $s=c$, apply $\mathcal{B}_2$ on $(i,j)$}
            \item[]\textbf{else} \emph{first apply $\mathcal{B}_1$ to node $j$, and then to node $i$}
        \end{enumerate}
    \end{enumerate}
    \item \emph{extract the unitary operation $\mathbf{V}$ from the node attributes}
    \item \emph{replace all nodes by type 1 nodes (see Fig.~\ref{fig:nodes}) 
    converting it to a normal graph, and extract the connectivity of $G^\prime_\alpha$.} 
\end{enumerate}
\noindent\textbf{output} (1) The reduced graph $G^\prime_\alpha$, and (2) the unitary $\mathbf{V}$ in terms of the set of unitaries $\{V_i,i\in G^\prime_\alpha\}$. 

\noindent\hrulefill

Implementation of the algorithm presented in the pseudo code upto step 2 ensures the following for $\rho^\prime_\alpha$.
\begin{enumerate}
    \item[(a)] The neighborhood $S_1^\prime$ of $S$ has at least one node, and the nodes in $S_1^\prime$ can be only of the types $1$, $2$, $5$, or $6$ (see Fig.~\ref{fig:nodes}), 
    so that $\sigma^3$ measurement on these nodes remain unchanged.
    \item[(b)] The nodes in $S_2^\prime$, in addition to being of the types $1$, $2$, $5$, or $6$,   
    can also be of the types $3$ or $4$ (see Fig.~\ref{fig:nodes}). This ensures that only $\sigma^3$ and $\sigma^1$ measurements are allowed in $S_2^\prime$.
    \item[(c)] If a type $3$ node, or a type $4$ node occurs in $S_2^\prime$, then its neighborhood will be constituted of types $1$, $2$, $5$, and $6$ nodes only. 
\end{enumerate}
Because of these characteristics of $\rho^\prime_\alpha$, it is  always possible to fully disconnect types $3$ and $4$ nodes of $S_2^\prime$ by measuring $\sigma^3$ only on the nodes of types $1$, $2$, $5$, and $6$. Note that when $\sigma^3$ is measured on a node $i\in\{\text{type }1, \text{type }2, \text{type }5, \text{type }6\}$, the outcomes $\pm 1$ are equiprobable, and the graph post-measurement can be obtained by
\begin{enumerate}
\item deleting all the edges between the node $i$ and its neighbours $\mathcal{N}_i$, 
\item replacing the node $i$ with type $3$ (type $4$) node if the outcome is  $+1$ ($-1$), and  
\item performing $\mathcal{F}_{\mathcal{N}_i}^{sg}$ if the outcome is $-1$.
\end{enumerate} 
If one now measures $\sigma^3$ on a node $i\in \{\text{type }3,\text{type }4\}$ belonging to $S_2^\prime$, due to the absence of connection between the node and the rest of the graph, the measurement leaves the graph unaltered, and the measurement outcome is $+1$ ($-1$) for the node of type $3$ (type $4$) with certainty.

\subsection{Scaling with system-size}
\label{app:N_scaling}

We now explore how the protocol discussed in Sec.~\ref{subsubsec:pauli_measurement_le} and Appendix~\ref{app:graph_operations} scales with the number of measured nodes $|S^\prime|=N-n=m$ in a graph $G$, which is given by the dependence of the total number of \emph{graph operations}, $C$, performed during the protocol on $m$. Since the graph $G$ changes during each step of the protocol, it is in general difficult to obtain the exact dependence of $C$ on $m$. However, to estimate an \emph{upper bound}, $C^\prime$, of $C$,  consider a fully connected graph of size $N$, where $\sigma^2$ measurement\footnote{This is the costliest Pauli measurement in terms of graph operations. See Appendix~\ref{app:graph_operations}.} on $m$ nodes are performed so that a maximum number of graph operations in terms of flipping of the node attributes and creation or deletion of the links in local complementations have to be performed. The  measurement transformation $\sigma^2 \rightarrow \sigma^3$ in \textbf{Step A} on $m$ nodes corresponds to $2m$ graph operations. On the other hand, the number of graph operations during the reduction in \textbf{Step B} is upper bounded by $m$ iteration of $\mathcal{B}_1$, each of which requires \emph{at most} $N^2-N+4$  graph operations. Therefore, the dependence of $C$ on $m$ is overall upper bounded by
\begin{equation}
    C^\prime=m(N^2-N+6).
\end{equation}
In the limit $N\approx m$\footnote{This is typically the case when the actual system is large enough compared to the subsystem on which entanglement is localized.}
\begin{equation}
    C^\prime=m^3-m^2+6m,
\end{equation}
which is a polynomial dependence on $m$.

\begin{figure}
    \centering
    \includegraphics[width=0.9\linewidth]{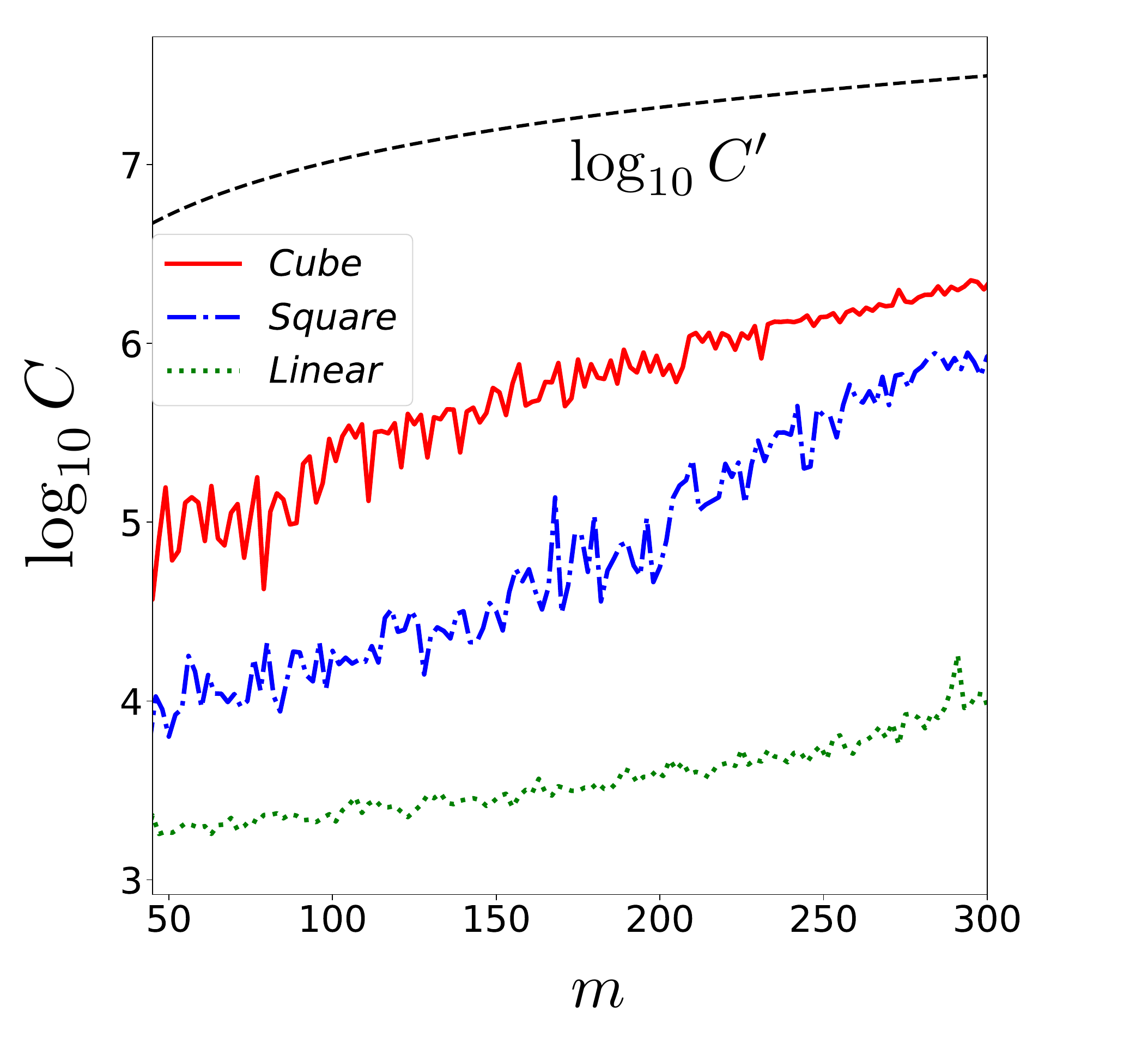}
    \caption{Variation of the total number of graph operations with the number of measured nodes, $m$, in cubic, square and linear graphs on respectively $N=343,324$, and $325$ nodes. To draw the variation of the analytically estimated upper bound, $C^\prime$, we use $N=324$.}
    \label{fig:complexity}
\end{figure}

In order to test this estimate, in Fig.~\ref{fig:complexity}, we present the numerical data for the variation of $C$ with $m$, for cubic, square, and linear graphs states on $N=8^3=343$, $N=18^2=324$, and $N=325$ nodes, respectively. We randomly choose $n$ nodes from the graph, and perform $\sigma^2$ measurement on $m=N-n$ nodes, followed by a  calculation of $C$. We see that the numerical data is  always upper bounded by $C^\prime$, as per our estimation. Note here that the actual values of the total graph operations depend on the structure of the chosen graph, as expected. 

\section{Proof of Proposition 1}
\label{app:proof_proposition_1}

\begin{proof} 
For reasons that will be clear shortly, we relabel the measurement outcomes corresponding to qubits in $S^\prime_1$ $(S^\prime_2)$ by $l$ $(m)$, such that $l_j=k_j$ ($m_j=k_j$) if $j\in S^\prime_1$ ($S^\prime_2$). We can therefore regroup the outcome-index $k$ as $k\equiv lm$ with $l\equiv l_{1}l_{2}\dots l_{L}$, and $m\equiv m_{L+1}m_{L+2}\dots m_{N}$, where we have assumed, without any loss in generality, that $L$ is the size of $S^\prime_1$ such that $S^\prime_2$ has $N-n-L$ qubits. 
In this notation, $M^\alpha_k\equiv M_{lm}^\alpha$, and using Eq.~(\ref{eq:reduced_form}) in Eq.~(\ref{eq:only_z_measurements}), we obtain 
\begin{eqnarray}
\varrho_k&=&\mathbf{U}_\alpha M_{lm}^{\alpha}\rho_\alpha M_{lm}^{\alpha\dagger}\mathbf{U}^\dagger_\alpha\nonumber\\
&=&\mathbf{U}_\alpha\mathbf{V}_\alpha M_{lm}^{\alpha}\ket{G^\prime_\alpha}\bra{G^\prime_\alpha}M_{lm}^{\alpha\dagger}\mathbf{V}^\dagger_\alpha\mathbf{U}^\dagger_\alpha,
\label{eq:long_equation}
\end{eqnarray} 
with $M_{lm}^{\alpha}\leftarrow \mathbf{V}_\alpha^\dagger M_{lm}^{\alpha} \mathbf{V}_\alpha$, such that after this transformation, $\alpha_j=3$ only $\forall j\in S^\prime_1$. Note that $\mathbf{V}_\alpha$ can change only the measurement basis, not the measurement outcome.  
The graph transformation algorithm discussed in Appendix~\ref{app:graph_operations} ensures that the measurement on any of the qubits in $S_2^\prime$ can be either a $\sigma^3$, or a $\sigma^1$, i.e., $\alpha_j=1,3$ if $j\in S^\prime_2$. Now, application of $M^\alpha_{lm}$ on the graph state $\ket{G^\prime_\alpha}$ results in the normalized state of the form  
\begin{eqnarray}
%M^{\beta}_{l}M^{\delta^\prime}_{m}\ket{G^\prime_\alpha}\bra{G^\prime_\alpha}M^{\beta\dagger}_{l}M^{\delta^\prime\dagger}_{m} =
M_{lm}^{\alpha}&&\ket{G^\prime_\alpha}\bra{G^\prime_\alpha}M_{lm}^{\alpha\dagger}=\mathbf{W}_{lm}\big[\left(\otimes_{j\in S^\prime_1}\ket{l_j}\bra{l_j}\right)\otimes \nonumber\\ 
&&\left(\otimes_{j\in S^\prime_2}\ket{m_j}\bra{m_j}\right)
\otimes\left(\ket{G^{\alpha}_S}\bra{G^{\alpha}_S}\right)\big]\mathbf{W}_{lm}^\dagger,
\label{eq:longest_eq}
\end{eqnarray}
where $G^{\alpha}_S$ is the subgraph on $S$ in $G^{\prime\prime}_{\alpha}$, which is obtained by performing the graph transformations corresponding to the single-qubit Pauli measurements on the graph $G^\prime_{\alpha}$ for all qubits in $S^\prime$. These transformations, for a measurement on a single qubit, are given by~\cite{hein2004,hein2006}  
\begin{eqnarray}
\label{eq:gt_x}
\sigma^1_j:\quad &G^{\prime}_{\alpha}& \rightarrow \mathcal{O}_{a}(\mathcal{O}_{j}(\mathcal{O}_a(G^\prime_\alpha))\backslash j), \\ 
%\label{eq:gt_y}
%\sigma^2_j:\quad &G^{\prime\prime}_{\beta\delta^\prime}& = \mathcal{O}_{j}(G^\prime_\alpha)\backslash j, \\
\label{eq:gt_z}
\sigma^3_j:\quad &G^{\prime}_{\alpha}& \rightarrow G^\prime_\alpha\backslash j,
\end{eqnarray}
with $a$ being any node in the neighborhood $\mathcal{N}_{j}$ of the node $j\in G^\prime_\alpha$, and $G^\prime_\alpha\backslash j$ is obtained from $G^\prime_\alpha$ by deleting all links connected to the node $j$. The graph operation $\mathcal{O}_j(G)$ w.r.t. a node $j$ represents a \emph{local complementation}~\cite{VandenNest2004,VandenNest2004a,hein2006,Bouchet1991,Bouchet1993} operation (see also Appendix~\ref{app:graph_operations} for the definition). These single-qubit Pauli measurements on all $j\in S^\prime$ also results in Clifford  unitary operators operating post-measurement on the qubits constituting the graph, leading to the unitary operations $\mathbf{W}_{lm}$, given by~\cite{hein2004,hein2006}
\begin{eqnarray}
\mathbf{W}_{lm} &=& \left(\otimes_{j\in S^\prime_1} W_{l_j}^3\right)\left( \otimes_{j\in S^\prime_2} W_{m_j}^{1,3}\right),
\end{eqnarray}
where for each $j$, 
\begin{eqnarray}
\label{eq:unitary_x_0}
W_{+1}^1 &=& \text{e}^{\text{i}\frac{\pi}{4}\sigma^2_a} \otimes_{b\in\mathcal{N}_{j}\backslash (\mathcal{N}_a\cup a)}\sigma^3_b,\\
\label{eq:unitary_x_1}
W_{-1}^1 &=& \text{e}^{-\text{i}\frac{\pi}{4}\sigma^2_a}\otimes_{b\in\mathcal{N}_{a}\backslash (\mathcal{N}_{j}\cup j)}\sigma^3_b,\\
\label{eq:unitary_z_0}
W^3_{+1} &=& I_j, \\
\label{eq:unitary_z_1}
W^3_{-1} &=& \otimes_{b\in\mathcal{N}_{j}}\sigma^3_b,
\end{eqnarray}
with $a$ being chosen as in Eq.~(\ref{eq:gt_x}). 

It is worthwhile to point out here that Eqs.~(\ref{eq:gt_x})-(\ref{eq:gt_z}) indicate that $G^{\prime\prime}_{\alpha}$, and consequently $G^{\alpha}_S$ is independent of the measurement outcomes $l$ and $m$. 
%However, we include the outcome indices in the subscript, the purpose of which will be clear in Sec.~\ref{sec:le_noisy_graph}.    
Using~(\ref{eq:longest_eq}) in Eq.~(\ref{eq:long_equation}), and tracing out the subsystem $S^\prime$, we obtain  
\begin{eqnarray}
\varrho_S^k=\mathbf{V}_{S,\alpha}\mathbf{W}_{S,l}\ket{G^{\alpha}_S}\bra{G^{\alpha}_S}\mathbf{W}_{S,l}^\dagger\mathbf{V}_{S,\alpha}^\dagger,
\label{eq:state_on_S}
\end{eqnarray}
where the subscript $S$ is introduced in the unitary operators to signify the components of the unitaries that have support only on $S$, and we have dropped the index $m$ from $\mathbf{W}_S$ as the components of $\mathbf{W}$ on $S$ are governed only by $l$. Therefore,
\begin{eqnarray}
E(\varrho_S^k)=E\left(\ket{G^{\alpha}_S}\bra{G^{\alpha}_S}\right),
\end{eqnarray}
which is independent of the measurement outcomes $l$ and $m$. This proves Proposition 1.
\end{proof}

\section{Entanglement in pure graph states} 
\label{app:entanglement_in_graph}

Here we discuss quantification and characterization of bipartite and multipartie entanglement in graph states. 

\subsection{Bipartite entanglement}
\label{app:bipartite_entanglement}

Let us consider a system of qubits $V\equiv\{1,2,\cdots,N\}$ forming a \emph{connected graph} $G$. Let us now consider a bipartition of the system $V$ as $A\cup B=V$, and $A\cap B=\emptyset$, where $d_A=\dim(\mathcal{H}_A)$ ($d_B=\mathcal{H}_B$), $\mathcal{H}_A$ ($\mathcal{H}_B$) being the Hilbert space associated to $A$ ($B$). For each pure state $\ket{G}$ in $\mathcal{H}_A\otimes\mathcal{H}_B$, orthonormal basis $\{b_A^i\}\in\mathcal{H}_A$ and  $\{b_B^j\}\in\mathcal{H}_B$ exist such that in Schmidt decomposed form,
\begin{eqnarray}
\ket{G} = \sum_{i=1}^d\sqrt{\lambda_i}\ket{b_A^ib_B^i},
\label{eq:sd}
\end{eqnarray}
with $d=\min\{d_A,d_B\}$, and non-negative Schmidt coefficients $\sqrt{\lambda_i}$ with $\lambda_i\geq 0$ and $\sum_{i=1}^d\lambda_i=1$. The marginals of $\ket{G}$, given by 
\begin{eqnarray}
\rho_A(B) &=&\text{Tr}_{B(A)}[\ket{G}\bra{G}]\nonumber \\ &=&\sum_{i=1}^d\lambda_i\ket{b_{A(B)}^i}\bra{b_{A(B)}^i},
\end{eqnarray}
has the spectrum $\{\lambda_i;i=1,\cdots,d\}$. The state $\ket{G}$ is \emph{maximally entangled} (ME)~\cite{horodecki2009,guhne2009} in the bipartition $A:B$ if $\lambda_i=1/d$ $\forall i$, which corresponds to the marginal of the subsystem of dimension $d$ to be maximally mixed, i.e., $I/d$, where $I$ is the identity matrix on the Hilbert space of dimension $d$. Otherwise, $\ket{G}$ is \emph{non-maximally entangled} (NME) in the partition $A:B$. In situations where $\ket{G}$ is maximally entangled across any bipartition (see also~\cite{Gisin1998,Higuchi2000}), they are referred to as the \emph{absolutely maximally entangled} (AME) states~\cite{Facchi2008,Gour2010,Cerf2013,Goyeneche2015,Enriquez2016}.  
To compute the bipartite entanglement $E_{A:B}(G)$ between the partitions $A$ and $B$ of $G$, one needs to compute the marginals $\rho_{A(B)}$ for $\ket{G}$, which can be obtained in terms of the generators $\tilde{g}_{i}^{A(B)}$ as~\cite{hein2004,hein2006}
\begin{eqnarray}
\rho_{A(B)} &=& \frac{1}{d_{A(B)}}\sum_{\left\{\tilde{g}_{i}^{A(B)}\right\}} \tilde{g}_{i}^{A(B)}, 
\end{eqnarray}
where 
\begin{eqnarray}
\left\{\tilde{g}_{i}^{A(B)}\right\} = \left\{g_{i}|\text{supp}\left(g_{i}\right)\in A(B)\right\}, 
\end{eqnarray} 
i.e., $\left\{\tilde{g}_{i}^{A(B)}\right\}$ is the subset of $\{g_{i}\}$ with support on $i\in A(B)$. For an arbitrary connected graph $G$, the single-qubit marginals corresponding to all qubits are maximally mixed~\cite{Verstraete2003,hein2006}, implying $\ket{G}$ is ME in all  bipartitions of a qubit and the rest. Also, note that for $N=2,3$, $d=2$ (see Eq.~(\ref{eq:sd})), and $\ket{G}$ is AME. On the other hand, in case of $G$ with arbitrary $N>3$, and assuming $A$ to be the smaller subsystem constituted of $2$ qubits $\{i,j\}\in G$ (i.e., $d=4$),  $\rho_{A}=\rho_{ij}$ is either maximally mixed, or a rank-$2$ mixed state~\cite{hein2006}. In the case of the former, $\ket{G}$ is ME in the bipartition $A:B$, whereas in the case of the latter, it is not.

Note that it is sufficient to investigate connected graphs. In the case of graphs that are not connected, the bipartite entanglement present in the graph will be decided by the bipartite entanglement present in the graph states corresponding to the connected segments of the full graph.

\subsection{Multipartite entanglement}
\label{app:multipartite_entanglement}

We now discuss the multipartite entanglement in graph states. A multi-qubit quantum state is genuinely multiparty entangled if it is not separable in any possible bipartition. Connected pure graph states are known to be genuinely multiparty entangled~\cite{hein2004}, where the degree of entanglement can be quantified using the \emph{Schmidt measure}~\cite{hein2004,hein2006}, and the generalized geometric measure~\cite{aditi2010,sadhukhan2017}.  

\subsubsection{Schmidt measure}
\label{app:schmidt_measure}

A pure state of $N$ qubits can be written as
\begin{equation}
    \ket{\psi}=\sum_{i=1}^R c_i \ket{\psi^{(1)}_i}\otimes ... \otimes \ket{\psi^{(N)}_i},
\end{equation}
where $c_i$s are complex numbers subject to normalization $\sum_i|c_i|^2=1$, $\ket{\psi^{(l)}_i}$ belongs to the Hilbert space of the qubit $l$, $l=1,2,\dots,N$, and $R$ is an integer $<2^N$. The Schmidt measure of $\ket{\psi}$ is given by~\cite{hein2004,eisert2001}
\begin{equation}
    E(\ket{\psi}\bra{\psi})=\log_2(r),
\end{equation}
where $r$ is the minimum possible value of $R$ over all possible linear decompositions of $\ket{\psi}$ into product states. In general, Schmidt measure is a difficult one to compute for an arbitrary pure multi-qubit state. However, there exist computable lower and upper bounds of the measure for certain multi-qubit states, including the graph states.  A lower bound to the Schmidt measure in the case of graph states can be obtained as follows~\cite{hein2004}. For a given graph $G$ and a bipartition $A:B$ such that $G=A \cup B$, the adjacency matrix $\Gamma_G$ can be written, without any loss in generality, as 
    \begin{eqnarray}
        \Gamma_{G}=\begin{pmatrix} \Gamma_{A} & \Gamma_{AB}^{T} \\ \Gamma_{AB} & \Gamma_{B} \end{pmatrix} 
    \end{eqnarray}
by rearranging the nodes, where $T$ denotes the transposition operation. The Schmidt measure of the graph state w.r.t. the partition $A:B$ is given by the rank of the matrix $\Gamma_{AB}$, i.e,
    \begin{equation}
        E(\ket{G}\bra{G})=\text{rank}[\Gamma_{AB}].
    \end{equation}
The lower bound to the Schmidt measure w.r.t the partitioning of the system into smallest possible subsystems\footnote{In the present case, each partition would hold one qubit.} is given by maximizing $E(\ket{G}\bra{G})$ over all possible bipartitions~\cite{hein2004}, as 
    \begin{equation}
        E_L(\ket{G}\bra{G})=\max_{\forall \{A:B\}} E(\ket{G}\bra{G}).
    \end{equation}

On the other hand, an upper bound to Schmidt measure in the case of graph states is given by the \emph{Pauli persistency}, defined as the minimum number of local Pauli measurements required to completely disentangle the state~\cite{hein2004}. We exploit the measurement protocol for the evaluation of this upper bound as follows. 
For a graph state $\ket{G}$ on $N$ qubits, and for a chosen Pauli measurement setup $\mathcal{P}$ on these $N$ qubits, let us denote the the reduced graph, obtained according to Eq.(\ref{eq:reduced_form}), by $G^\prime$. The {minimal vertex cover} of this reduced graph $G^\prime$ is defined as the minimum cardinality of the subset of nodes in $G^\prime$ such that upon deletion of edges on them leaves behind a completely disconnected graph on $N$ vertices and we denote it by $M(G^\prime)$. An upper bound to the Schmidt measure of $\ket{G}$ is given by minimizing $M(G^\prime)$ over all possible Pauli measurements i.e,
    \begin{equation}
        E_U(\ket{G}\bra{G})=\min_{\mathcal{P}} M(G^\prime). 
    \end{equation}

\subsubsection{Generalized geometric measure} 
\label{app:ggm}

Let us consider a \emph{$K$-separable} $N$-qubit quantum state $\ket{\phi_K}$, which can be divided into $K$ product state partitions, where $2\leq K\leq N$. The GME in an $N$-qubit state $\ket{\psi}$ 
is quantified by the $K$-geometric measure (K-GM) of entanglement, defined as the minimum distance of the state $\ket{\psi}$ from the set $\mathcal{S}_K$ of all possible $K$-separable states, i.e.,     
\begin{eqnarray}
 E(\ket{\psi})=1-\underset{\mathcal{S}_K}{\max}|\langle{\phi_K}|\psi\rangle|^2.
 \label{eq:gm_ksep}
\end{eqnarray}
The original definition of the geometric measure is recovered for $K=N$, while at the other extremum $K=2$, the measure is called the generalized geometric measure (GGM). The optimization in the definition of the GGM can be achieved via a maximization of the Schmidt coefficients across all possible bipartitions of $\ket{\psi}$\cite{biswas2014}, leading 
to 
\begin{eqnarray}
E=1-\underset{\mathcal{S}_{A:B}}{\max}\{\lambda^2_{A:B}\},
\label{eq:ggm_exp}
\end{eqnarray}
where $\lambda_{A:B}$ is the Schmidt 
coefficient of $\ket{\psi}$ maximized over the set $\mathcal{S}_{A:B}$ of all arbitrary $A:B$
bipartitions of the $N$-qubit system. The GGM of a pure state of arbitrary number of qubits, $N$, can be computed using Eq.~(\ref{eq:ggm_exp}).

\section{Proof of Proposition 2}
\label{app:proof_proposition_2}

\begin{proof}
Using
\begin{eqnarray}
J_{s^{\prime\prime}}&=&J_{S,s^{\prime\prime}}\otimes J_{S^\prime,s^{\prime\prime}}
\end{eqnarray} 
and 
\begin{eqnarray}
q_s=q_{S,s}\times q_{S^\prime,s}
\end{eqnarray}
with $q_{S(S^\prime),s}=\prod_{j\in S(S^\prime)}q_{s_j}$, and $\sum_{(S^\prime,s)}q_{S^\prime,s}=\sum_{(S,s)}q_{S,s}=1$, one can trace out $S^\prime$ from $\varrho_k$ to obtain 
\begin{eqnarray}
\varrho^k_S=\mathbf{V}_{S,\alpha}\tilde{\varrho}^k_S \mathbf{V}_{S,\alpha}^\dagger
\label{eq:unitary_equivalence_under_noise}
\end{eqnarray}
with  
\begin{eqnarray}
\tilde{\varrho}_S^k = \sum_{(S,s)} q_{S,s}  J_{S,s^{\prime\prime}} \tilde{\rho}_S J_{S,s^{\prime\prime}}^\dagger,
\label{eq:noisy_state_on_s}
\end{eqnarray} 
where $\tilde{\rho}_S$ is a GD state in the graph-state basis corresponding to the graph $G_S^\alpha$ on the subsystem $S$, obtained from $G^\prime_\alpha$ (see Eq.~(\ref{eq:state_on_S})). 
% Hereafter we are dropping the superscript $\alpha$ to avoid cluttering but the following discussion is unique for each $\alpha$.
We write $\tilde{\rho}_S$ as, 
\begin{eqnarray}\label{eq:tilde_rho_S}
\tilde{\rho}_S=\sum_{\psi} \lambda_{\psi}(\alpha,s)\ket{\psi}\bra{\psi}
\label{eq:equiv_eq.70}
\end{eqnarray}
$\{\ket{\psi}\}$ being the graph basis in the Hilbert space of $S$ where $\psi\in [0,2^{n}-1]$, and $\psi$ can be identified as $\psi\equiv\psi_1\psi_2\dots\psi_{n}$, where $\psi_i\in\{0,1\}\forall i\in S$. In this notation $\ket{0}\bra{0}\equiv\ket{G_S^\alpha}\bra{G_S^\alpha}$, and (see Sec.~\ref{subsec:graph})
\begin{eqnarray}
\ket{\psi}&=&\otimes_{i\in S}(\sigma_i^3)^{\psi_i}\ket{G_S^\alpha}\label{eq:psi_binary_notation}
\end{eqnarray}
Note that the \textit{mixing probabilities} $\lambda_\psi(\alpha,s)$ are in general functions of the PMS, $\alpha$, and the type of the noise, $s$. These probabilities and subsequently the state $\tilde{\rho}_S$ can be calculated from the knowledge of the connectivity between the sets of qubits $S$ and $S_1^{\prime\prime}$, where $S_1^{\prime\prime}\subseteq S_1^\prime$ such that $s_j=1,2$ $\forall j\in S_1^{\prime\prime}$, i.e., the noise on $S_1^{\prime\prime}$ does not commute with the measurement, which is $\sigma^3$ for all qubits in $S_1^\prime$. A technical account of the prescription for calculating $\tilde{\rho}_S$ is given in the Appendix~\ref{app:GDstate} for the interested readers.

Note further that an application of Eq.~(\ref{eq:noisy_state_on_s}) on $\tilde{\rho}_S$ leads to another GD state on $S$.  To determine this, we (a) first obtain the GD state resulting from the application of Eq.~(\ref{eq:noisy_state_on_s}) on $\ket{\psi}\bra{\psi}\equiv\ket{0}\bra{0}=\ket{G^\alpha_S}\bra{G^\alpha_S}$ (using Eq.~(\ref{eq:pauli_map_identities})) as 
\begin{eqnarray}
\sum_{(S,s)} q_{S,s}  J_{S,s^{\prime\prime}} \ket{0}\bra{0} J_{S,s^{\prime\prime}}^\dagger = \sum_{\psi^\prime} \lambda_{\psi^\prime}(\alpha,s)\ket{\psi^\prime}\bra{\psi^\prime}, \nonumber\\
\label{eq:noisy_state_on_00}
\end{eqnarray}
and (b) then use Eqs.~(\ref{eq:tilde_rho_S}) and (\ref{eq:noisy_state_on_00}) in Eq.~(\ref{eq:noisy_state_on_s}), with the aid of Eq.~(\ref{eq:psi_binary_notation}), to obtain
\begin{eqnarray}
\tilde{\varrho}^k_S &=& \sum_{(S,s)} q_{S,s}  J_{S,s^{\prime\prime}} \left[ \sum_{\psi} \lambda_{\psi}\ket{\psi}\bra{\psi} \right] J_{S,s^{\prime\prime}}^\dagger\label{eq:noisy_state_on_GD1} \\ 
&=& \sum_{\psi,\psi^\prime} \lambda_{\psi}\lambda_{\psi^\prime} \ket{\psi\oplus \psi^\prime}\bra{\psi\oplus \psi^\prime}, \label{eq:noisy_state_on_GD5}
\end{eqnarray}
%\end{widetext}
where $\psi\oplus \psi^\prime$ represents modular two addition of the binary strings $\psi$ and $\psi^\prime$. Also to arrive at Eq.~(\ref{eq:noisy_state_on_GD5}), we use $\sigma^3 \sigma^j \sigma^3=\sigma^j,j\in\{1,2,3\}$ upto irrelevant multiplicative factors.
 
From Eq.~(\ref{eq:unitary_equivalence_under_noise}), it is clear that   
\begin{eqnarray}
E(\varrho^k_S)=E(\tilde{\varrho}^k_S).  
\label{eq:unitary_equivalence_outcome_independence}
\end{eqnarray}
Moreover, note that for two distinct outcomes $k_1$ and $k_2$ having their respective components in $S_1^{\prime\prime}$ as $l^{\prime}_1$ and $l^{\prime}_2$\footnote{Note that $l_1^\prime$ and $l_2^\prime$ can be seen as the parts of corresponding $l^\prime$s that belong to $S^{\prime\prime}_1$}, $l_2^{\prime}=l^{\prime}_1\oplus L^\prime$ in binary representation, where $L^\prime$ is another binary string of the same length as $l^{\prime}_1,l_2^\prime$. Therefore, a sum over $l_1^\prime$ (see  Eq.~(\ref{eq:measurement_outcomes_changed})\footnote{Note that the sum over $s$ in Eq.~(\ref{eq:measurement_outcomes_changed}) can be seen as an effective sum over all possible $l_1^\prime$} is equivalent to a sum over $l^{\prime}_2$ for a fixed $L^\prime$. This results in local unitary connected states for the two outcomes $k_1$ and $k_2$ as 
\begin{eqnarray}
\tilde{\varrho}^{k_1}_S=\mathbf{W}^{L^\prime}_S \tilde{\varrho}^{k_2}_S\mathbf{W}^{L^\prime}_S, 
\end{eqnarray}
where $\mathbf{W}^{L^\prime}_S$ is a local unitary operator acting on $S$, having the form,
\begin{equation}
\mathbf{W}^{L^\prime}_S=\prod_{j\in S_1^{\prime\prime}} \otimes_{i\in \mathcal{N}_j\cap S}(\sigma^3_{i})^{L_j^\prime},  
\end{equation}
where $L^\prime_j$ is the outcome on the $j$th qubit, $j\in S_1^{\prime\prime}$. 
Therefore, $E(\tilde{\varrho}^{k_1}_S)=E(\tilde{\varrho}^{k_2}_S)$ for any two distinct outcomes  $k_1,k_2$, which, along with Eq.~(\ref{eq:unitary_equivalence_outcome_independence}), leads to 
\begin{eqnarray}
    E_S(\varrho^{k_1}_S)=E_S(\varrho^{k_2}_S).
\end{eqnarray}
Hence the proof. 
\end{proof}

\section{Calculating localizable entanglement for the graph in Fig.~\ref{fig:forbid} under noise}
\label{app:demo}

To demonstrate the calculation of localizable entanglement in the presence of noise,  we consider the graph in Fig.~\ref{fig:forbid}(a), where each qubit is subjected to BPF noise.

\subsection{When no outcomes are forbidden}
\label{subsec:no_forbidden_set}

Let us consider $\sigma^3$ measurements performed on qubits $1$ and $2$ constituting $S^\prime$. Since  $U_j=V_j=I$ corresponding to qubits $j=1,2$ in the case depicted in Fig.~\ref{fig:forbid}(a), the reduced graph is equivalent to the original graph, and $s^{\prime\prime}=s^\prime=s$ in Eqs.~(\ref{eq:noisy_system_1})-(\ref{eq:measurement_outcomes_changed}). The change $l_1\rightarrow l_1^\prime$ and $m_2\rightarrow m_2^\prime$ in Eq.~(\ref{eq:measurement_outcomes_changed}) due to the values of $s$, according to Eq.~(\ref{eq:transformed_projector}), are tabulated in Table~\ref{tab:measurement_outcomes_changed}\textbf{A}. Note that in this specific example, 
all four measurement outcomes $k\equiv l_1m_2$ are equally probable, and remain so even after the change  $l_1\rightarrow l_1^\prime$ and $m_2\rightarrow m_2^\prime$. Since both qubits in $S^{\prime\prime}_1=\{1,2\}$ are connected to the two qubits in $S$, the normalized state $\tilde{\rho}_S$ (see Eq.~(\ref{eq:equiv_eq.70})) corresponding to outcome $k=(+1)(+1)\equiv 00$, occurring with a probability $1/4$, can be written in a row vector notation as
\begin{eqnarray}
\sum_{\psi} \lambda_{\psi}\ket{\psi}\bra{\psi}\equiv\bigg[1-q+\frac{q^2}{2},0,0,q-\frac{q^2}{2}\bigg]
\end{eqnarray}
where $\lambda_\psi$ is the value in the position $\psi$ of the row. Similarly, equivalent to Eq.~(\ref{eq:noisy_state_on_00}), one may write 
\begin{eqnarray}
\sum_{\psi^\prime} \lambda_{\psi^\prime}\ket{\psi^\prime}\bra{\psi^\prime}\equiv\bigg[1-q+\frac{q^2}{2},0,0,q-\frac{q^2}{2}\bigg]. \label{eq:effect_BPF_on_2qubit_S}
\end{eqnarray}
Combining the above equations in the case where $S_2^\prime=\emptyset$ and $k=l=(+1)(+1)\equiv 00$, one obtains, equivalent to  Eq.~(\ref{eq:noisy_state_on_GD5}),
\begin{eqnarray}
\tilde{\varrho}^{(+1)(+1)}_S &=& \bigg[ \left( 1-q+\frac{q^2}{2} \right)^2+\left(q-\frac{q^2}{2}\right)^2,0,0,\nonumber \\ 
&& 2\left(1-q+\frac{q^2}{2}\right)\left(q-\frac{q^2}{2}\right) \bigg].
\end{eqnarray}
The states $\tilde{\varrho}^{k}_S$ for $k=(+1)(-1), (-1)(+1),(-1)(-1)$ are also equally probable with probability $1/4$, and can be obtained from $\tilde{\varrho}^{(+1)(+1)}_S$ via local unitary transformations, where the local unitary operators can also be seen from table~\ref{tab:measurement_outcomes_changed}\textbf{A}. For example outcomes $k=(+1)(+1)\equiv 00$ and $k=(-1)(+1)\equiv 10$ are connected by $L=10$, and the corresponding states by the local unitary operator  $\mathbf{W}^L_S=\sigma_1^3\otimes \sigma^3_2$.

\begin{table*}
\centering
\begin{tabular}{c}
\textbf{A.} $\alpha\in\Gamma$ (all outcomes are allowed) \\
\\
\begin{tabular}{|c|c|c|c|c|}
\hline 
& $s_1=0,s_2=0$ & $s_1=0,s_2=2$ & $s_1=2,s_2=0$ & $s_1=2,s_2=2$ \\ 
\hline
$k\equiv l_1m_2$ & \multicolumn{4}{|c|}{$k^\prime\equiv l_1^\prime m_2^\prime$}\\
\hline 
$(+1)(+1)$ &  $(+1)(+1)$ & $(+1)(-1)$ & $(-1)(+1)$ & $(-1)(-1)$ \\ 
\hline
$(+1)(-1)$ & $(+1)(-1)$ & $(+1)(+1)$ & $(-1)(-1)$ & $(-1)(+1)$\\
\hline 
$(-1)(+1)$ & $(-1)(+1)$ & $(-1)(-1)$ & $(+1)(+1)$ & $(+1)(-1)$\\
\hline
$(-1)(-1)$ & $(-1)(-1)$ & $(-1)(+1)$ & $(+1)(-1)$ & $(+1)(+1)$\\
\hline
\end{tabular} \\
\\
\textbf{B.} $\alpha\in\overline{\Gamma}$ (some outcomes are allowed)\\
\\
\begin{tabular}{|c|c|c|c|c|}
\hline 
& $s_1=0,s_2=0$ & $s_1=0,s_2=2$ & $s_1=2,s_2=0$ & $s_1=2,s_2=2$ \\ 
\hline
$k\equiv l_1m_2$ & \multicolumn{4}{|c|}{$k^\prime\equiv l_1^\prime m_2^\prime$}\\
\hline 
$(+1)(+1)$ (a) & $(+1)(+1)$ (a) & $(+1)(-1)$ (f) & $(-1)(+1)$ (f) & $(-1)(-1)$ (a)\\
\hline
$(+1)(-1)$ (f) & $(+1)(-1)$ (f) & $(+1)(+1)$ (a) & $(-1)(-1)$ (a) & $(-1)(+1)$ (f) \\
\hline 
$(-1)(+1)$ (f) & $(-1)(+1)$ (f) & $(-1)(-1)$ (a) & $(+1)(+1)$ (a) & $(+1)(-1)$ (f)\\
\hline
$(-1)(-1)$ (a) & $(-1)(-1)$ (a) & $(-1)(+1)$ (f) & $(+1)(-1)$ (f) & $(+1)(+1)$ (a)\\
\hline
\end{tabular} \\
\end{tabular}
\caption{Change of outcomes due to Pauli noise on a qubit in the case of the example shown in Fig.~\ref{fig:forbid}, where BPF noise is applied to all qubits, and \textbf{A.} $\sigma^3$ and \textbf{B.} $\sigma^1$ measurements are performed on qubits $1$ and $2$. In the case of \textbf{B}, allowed and forbidden outcomes are labelled by (a) and (f).}
\label{tab:measurement_outcomes_changed}
\end{table*}

\subsection{When forbidden set of outcomes is present}
\label{subsec:forbidden_set}

For $\alpha\in\overline{\Gamma}$, there exists a set of outcomes that are forbidden (see Sec.~\ref{subsubsec:forbidden_sets}). In such cases, results discussed in the previous section hold except for the calculation of $\tilde{\rho}_S$, and the explicit calculation depends on the fact that the change $lm\rightarrow l^\prime m^\prime$ may result in a transition between allowed and  forbidden sets of outcomes. We demonstrate this using the graph shown in Fig.~\ref{fig:forbid}(a), where each qubit is subjected to BPF noise, and $\sigma^1$ measurements are performed on qubits $1$ and $2$ constituting $S^\prime$. Since  $U_j=H$ corresponding to qubits $j=1,2$, and $V_1=H,V_2=I$ in this case, one has to work with the reduced graph shown in Fig.~\ref{fig:forbid}(b). Also, $s^{\prime\prime}=s^\prime=s$ in Eqs.~(\ref{eq:noisy_system_1})-(\ref{eq:measurement_outcomes_changed}), since BPF does not change under Hadamard operation. The transformations $l_1\rightarrow l_1^\prime$ and $m_2\rightarrow m_2^\prime$ in Eq.~(\ref{eq:measurement_outcomes_changed}) due to the values of $s$, according to Eq.~(\ref{eq:transformed_projector}), are tabulated in Table~\ref{tab:measurement_outcomes_changed}\textbf{B}. Since $S_2^\prime=\{1\}$ and $S^{\prime}_1=S^{\prime\prime}_1=\{2\}$ are connected to the two qubits in $S$, $\tilde{\rho}_S$ corresponding to the outcome $k=lm=(+1)(+1)\equiv 00$ occurring with probability $(1-q+q^2/2)/2$, can be written as (equivalent to Eq.~(\ref{eq:equiv_eq.70}))
\begin{eqnarray}
\sum_{\psi} \lambda_{\psi}\ket{\psi}\bra{\psi}&\equiv&\bigg[\frac{1-q+\frac{q^2}{4}}{1-q+\frac{q^2}{2}},0,0,
\frac{\frac{q^2}{4}}{1-q+\frac{q^2}{2}}\bigg],\nonumber\\
\end{eqnarray}
It is worthwhile to point out that in contrast to the case discussed in Sec.~\ref{subsec:no_forbidden_set}, the probabilities are no longer independent of $q$ -- an artefact due to the transition between the outcomes belonging to the allowed and forbidden sets. Proceeding as before, one can write the normalized $\tilde{\varrho}^{(+1)(+1)}_S$ as
\begin{eqnarray}
\tilde{\varrho}^{(+1)(+1)}_S&=& \bigg[  1-q+\frac{q^2}{4} + \frac{(q-\frac{q^2}{2})\frac{q^2}{4}}{1-q+\frac{q^2}{2}},0,0,\nonumber\\ 
&& \frac{q^2}{4} + \frac{(q-\frac{q^2}{2})(1-q+\frac{q^2}{4})}{1-q+\frac{q^2}{2}} \bigg].
\end{eqnarray}

In a similar fashion, one may also obtain $\tilde{\varrho}^{(+1)(-1)}_S$ as
\begin{equation}
    \tilde{\varrho}^{(+1)(-1)}_S=\bigg[\frac{1}{2},0,0,\frac{1}{2}\bigg],
\end{equation}
which clearly indicates that $\tilde{\varrho}^{(+1)(-1)}_S$ and  $\tilde{\varrho}^{(+1)(+1)}_S$ are not connected by local unitary operators. Thus, in general, a local unitary connection among states corresponding to all the allowed outcomes does not exists,  thereby taking away the advantage in the computation, as opposed to the case described in Sec.~\ref{subsec:no_forbidden_set}, specifically in situations where the system-size is large. Note also that the outcomes which are initially forbidden at $q=0$ need not necessarily remain forbidden for $q>0$. In the above example, $k=(-1)(+1)$ is forbidden at $q=0$, but acquires a non zero probability $q(1-q/2)/2$ for $q>0$.

\section{Determination of \texorpdfstring{$\tilde{\rho}_S$}{GD state on subsystem S}}\label{app:GDstate}

\begin{figure}
    \centering
    \includegraphics[width=0.9\linewidth]{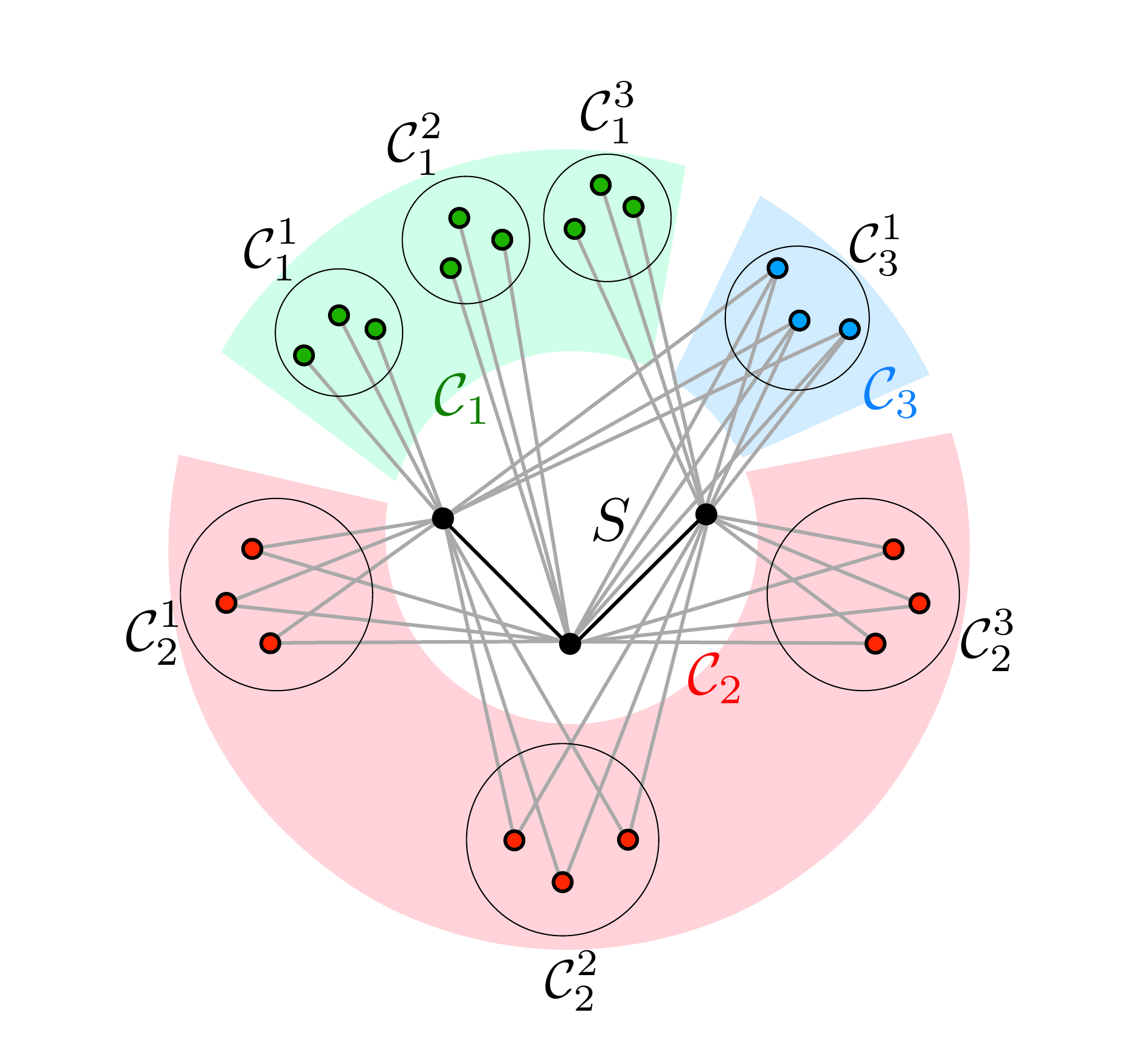}
    \caption{Schematic representation of the class-structure of the subsystem $S^{\prime\prime}$ of a graph, where entanglement is to be localized over a subsystem $S$ of size $3$. To keep the figure uncluttered, only the subsystems $S^{\prime\prime}_1$, $S$, the links between $S^{\prime\prime}_1$ and $S$, and the links between the nodes in $S$ are drawn.}
    \label{fig:class_division}
\end{figure}

Here we discuss the calculation of $\tilde{\rho}_S$ (see Sec.\ref{app:proof_proposition_2} and Sec.~\ref{subsec:no_forbidden_set}). Let us define $S_1^{\prime\prime}\subseteq S_1^{\prime}$ such that $S_1^{\prime\prime}$ consists of all nodes in $S_1^{\prime}$ that have $s^{\prime\prime}\in \{ 1,2 \}$. Let us further divide the nodes in $S^{\prime\prime}_1$ into $n$ \emph{classes}, $n$ being the size of $S$, such that the each of the nodes in the $m$th class, $\mathcal{C}_m$ ($m=1,2,\cdots,n$), is connected to $m$ nodes in $S$. Since $m$ nodes from $S$ can be chosen in $\genfrac(){0pt}{1}{n}{m}$ possible ways, the $m$th class can be further divides into $\genfrac(){0pt}{1}{n}{m}$ \emph{subclasses}, which we denote by $\mathcal{C}^r_m$, where each $\mathcal{C}_m^r$ has a size $|\mathcal{C}_m^r|$, $1\leq r\leq \genfrac(){0pt}{1}{n}{m}$.  See Fig.~\ref{fig:class_division} for an example of a possible class structure in the case of $n=3$. 

We once again use the binary representation to work using the graph state basis $\{\ket{\psi}\}$ (see Sec.~\ref{app:proof_proposition_2}), where $\ket{G_S}=\ket{0}$. The local unitary corrections (either $I$ or $\sigma^3$) applied to $\ket{G_S}$ can be represented in binary representation as 
\begin{eqnarray}
\ket{\psi}&=&\mathbf{W}^\psi\ket{0}, \text{ with }\mathbf{W}^\psi = \otimes_{j=1}^{n}(\sigma^3_j)^{\psi_j}.
\end{eqnarray}
In this representation, one can determine the \emph{effective correction} of each of the subclasses $\mathcal{C}^r_m$ in the $m$th class on $S$ by a \emph{parity} bit, denoted by $\gamma(r,m)$, such that for $\gamma(r,m)=0$ ($\gamma(r,m)=1$), there will be no correction (a $\sigma^3$ correction) applied to each node in $S$ that are connected to the nodes in $\mathcal{C}^r_m$ (See Eqs.(\ref{eq:unitary_z_0}),(\ref{eq:unitary_z_1})). The overall correction, denoted by $\mathbf{W}^\psi$, applied on $S$ will be the product of all corrections due to all subclasses from all of the classes:
\begin{equation}
    \mathbf{W}^{\psi}=\prod_{m=1}^{n}\prod_{r=1}^{\genfrac(){0pt}{1}{n}{m}} (\sigma^3_{\mathcal{N}_{(r,m)}})^{\gamma(r,m)},
\end{equation}
where $\mathcal{N}_{(r,m)}$ denotes the set of all nodes in $S$ connected to the subclass $C_m^r$.

Note that there may exist different combinations of $\gamma(r,m)$ giving rise to the same correction $\mathbf{W}^{\psi}$. For example, in the case of $n=3$ (Fig.(\ref{fig:class_division}),  both $\gamma(r,m)=0 \forall m,r$ and $\gamma(r,m)=1 \forall m,r$ give rise to the  trivial correction $I\otimes I\otimes I$ in S. Defining the total number of possible subclasses with $m\geq2$ as 
\begin{equation}
    D=\sum_{m=2}^{n}{\genfrac(){0pt}{1}{n}{m}},
    \label{eq:app_C__define_D}
\end{equation}
there will be a total of $n+D$ subclasses for $S$ since $\genfrac(){0pt}{1}{n}{1}=n$. The purpose of  this division into $D$ and $n$ will be evident soon. We now denote a binary string of $n+D$ bits as 
\begin{eqnarray}
\gamma=\gamma(1,1)\gamma(2,1)\dots\gamma(n,1)\gamma(1,2)\gamma(2,2)\dots\gamma(n,2)\nonumber\\
\dots\gamma(1,n)\gamma(2,n)\dots\gamma(1,n).\nonumber\\ 
\end{eqnarray}
The first $n$ of the parity bits in $\gamma$ are reserved for $m=1$ class, and the rest $D$ bits are assigned to the remaining $D$ subclasses of the classes $m=2,\cdots,n$. In this notation mixing probabilities, $\lambda_\psi$, are given by,
\begin{eqnarray}
    \lambda_\psi=\sum_{\gamma=0}^{2^D-1}\prod_{m=1}^{n}\prod_{r=1}^{\genfrac(){0pt}{1}{n}{m}} P_{\gamma(r,m)}(r,m)
    \label{eq:app_C_lambda_U}
\end{eqnarray}
with 
\begin{eqnarray}
    P_{\gamma(r,m)}(r,m)=\frac{1}{2}[1+(-1)^{\gamma(r,m)}(1-2q_{n})^{|\mathcal{C}^r_m|}],\nonumber \\
    \label{eq:app_C_class_div_prob}
\end{eqnarray}
and  
\begin{equation}
    \gamma(i,1)=\psi_i\oplus \bigoplus_{\substack{\{(r,m)\},m>1\\ i\in \mathcal{N}_{(r,m)}}}  \gamma_{(r,m)} \label{eq:app_C_gamma_rule}
\end{equation}
where $i$ denotes a qubit in $S$, $i\in \mathcal{N}_{(r,m)}$ condition picks up only those subclasses in which a connection exists between $i$ and $\mathcal{C}_m^r$,  $\oplus$ denotes modular $2$ addition, and $q_n$ is the total probability of occurrence of change of outcome on a single qubit, $l_j\rightarrow l_j^\prime$(see Eq.~\ref{eq:transformed_projector}), for various noises considered in this paper. For an example $q_n=q/2$ for Markovian BF and $q_n=0$ for Markovian PD noises. An explicit calculation with $n=2$ can be found in~\cite{amaro2020}. 

We point out here that the above calculation assumes identical noise strengths and non-Markovianity parameters on each qubit, irrespective of the type of noise present on the qubit in a specific subclass. However, one can also generalize to a scenario where $(q_i,\epsilon_i)$ on different qubits in a subclass are different, which originates from different $(q_i,\epsilon_i)$ on different qubits in the original graph. This leads to a form of $P_{\gamma(r,m)}(r,m)$ given by 
    \begin{eqnarray}
P_1(r,m)&=&\sum_{\substack{\beta=1,\\ \oplus \beta_i=1}}^{2^{|\mathcal{C}_m^r|}}\prod_{i\in \mathcal{C}_m^r} p_i^{\beta_i}(1-p_i)^{1\oplus \beta_i}, \nonumber \\ 
P_0(r,m)&=&1-P_1(r,m),
\label{eq:new_eqn}
    \end{eqnarray}
where $\beta=\beta_1\beta_2...\beta_{|\mathcal{C}_m^r|}$ is a multi-index with $\beta_i\in\{0,1\}$, and $p_i$ is the probability of occurrence of noise on qubit $i$, given by 
\begin{eqnarray} 
p_i=\frac{q_i}{2}\left[1+\epsilon_i\left(1-\frac{q_i}{2}\right)\right] 
\end{eqnarray}
for the BF and the BPF noise, and 
\begin{eqnarray} 
p_i=\frac{q_i}{4}\left[1+3\epsilon_i\left(1-\frac{3q_i}{4}\right)\right]
\end{eqnarray} 
for the DP noise. An example of this scenario can be found in Sec.~\ref{subsec:gme_examples_mixed_graphs}.

\section{Entanglement in graph states under noise}
\label{app:entanglement_in_gd_states}

In the presence of noise, and for states with more than two qubits, the problem of quantifying entanglement is significantly more complicated. For a given bipartition of a graph state under Pauli noise, the bipartite entanglement can be quantified by negativity~\cite{vidal2002}. On the other hand, for certain class of mixed states obtained when graph states are subjected to noise, such as the graph-diagonal states, useful criteria for detecting multipartite entanglement directly from the density matrix of the state exists~\cite{Guhne_2010,Guhne_2011}. One can also compute a \emph{genuine multiparty concurrence}~\cite{Hashemi2012} for graph states under noise with an $X$-state as the density matrix. We briefly discuss these in this section.

\subsection{Negativity as a bipartite measure for a graph-diagonal state}
\label{app:negativity}

For a graph-diagonal state $\rho$ representing a graph with two partitions $A$ and $B$, the negativity between $A$ and $B$ is defined as~\cite{vidal2002} 
\begin{eqnarray}
\mathcal{N}_{AB}=\frac{||\rho^{T_A}||-1}{2},
\end{eqnarray}
where $T_A$ denotes the transposition of $\rho$ w.r.t. the subsystem $A$, and $||\mathcal{A}||=\text{Tr}\left[\sqrt{\mathcal{A}^\dagger\mathcal{A}}\right]$ for the matrix $\mathcal{A}$. 

\subsection{Genuine multiparty entanglement in graph diagonal states}
\label{app:multipartite_entanglement_gd}

Determination of the multipartite entanglement measures is difficult in the case of mixed states. However, in the case of specific types of mixed states, such as the graph diagonal states of certain types or number of parties, criteria for the state to be genuine multiparty entangled, or biseparable, or fully separable can be determined using the density matrix of the state~\cite{Guhne_2010}. We first consider the example of a mixed state of $N$ qubits which is diagonal in the graph state basis (see Eqs.~(\ref{eq:graph_state_basis})-(\ref{eq:GDstate})) corresponding to an $N$-qubit star graph, where 
\begin{eqnarray}
    \ket{G}&=&\frac{1}{\sqrt{2}}\left[\ket{0_0}\otimes_{i=1}^{N-1}\ket{+_i}+\ket{1_0}\otimes_{i=1}^{N-1}\ket{-_i}\right],\nonumber\\
\end{eqnarray}
with the node ``$0$" being the hub to which all other nodes $i=1,2,\cdots,N-1$, are attached. We refer to this state as the \emph{star-graph-diagonal} (SGD) state. Since an $N$-qubit star graph can be converted to the $N$-qubit GHZ state via single-qubit Hadamard operations on all qubits except the qubit ``$0$" of the star-graph, the SGD state can be transformed to a \emph{GHZ-diagonal} (GHZD) state, given by Eqs.~(\ref{eq:graph_state_basis})-(\ref{eq:GDstate}), with $\ket{G}$ replaced by the $N$-qubit GHZ state. Therefore, the biseparability criteria of an $N$ qubit GHZD state~\cite{Guhne_2010} is equivalent to that of the $N$-qubit SGD state. An $N$-qubit GHZD state $\rho$ is biseparable if the matrix elements $\rho_{(i,j)}$, $1\leq i,j\leq 2^N$, of $\rho$ in the computational basis satisfy 
\begin{equation}\label{eq:NGHZ_biseparability}
    \left|\rho_{(1,2^N)}\right|\leq \sum_{i=1}^{2^{N-1}-1} \left[\rho_{(1+i,1+i)}\rho_{(2^N-i,2^N-i)}\right]^{\frac{1}{2}}.  
\end{equation}
Violation of (\ref{eq:NGHZ_biseparability}) indicates presence of GME in the state.

\begin{figure}
    \centering
    \includegraphics[width=0.9\linewidth]{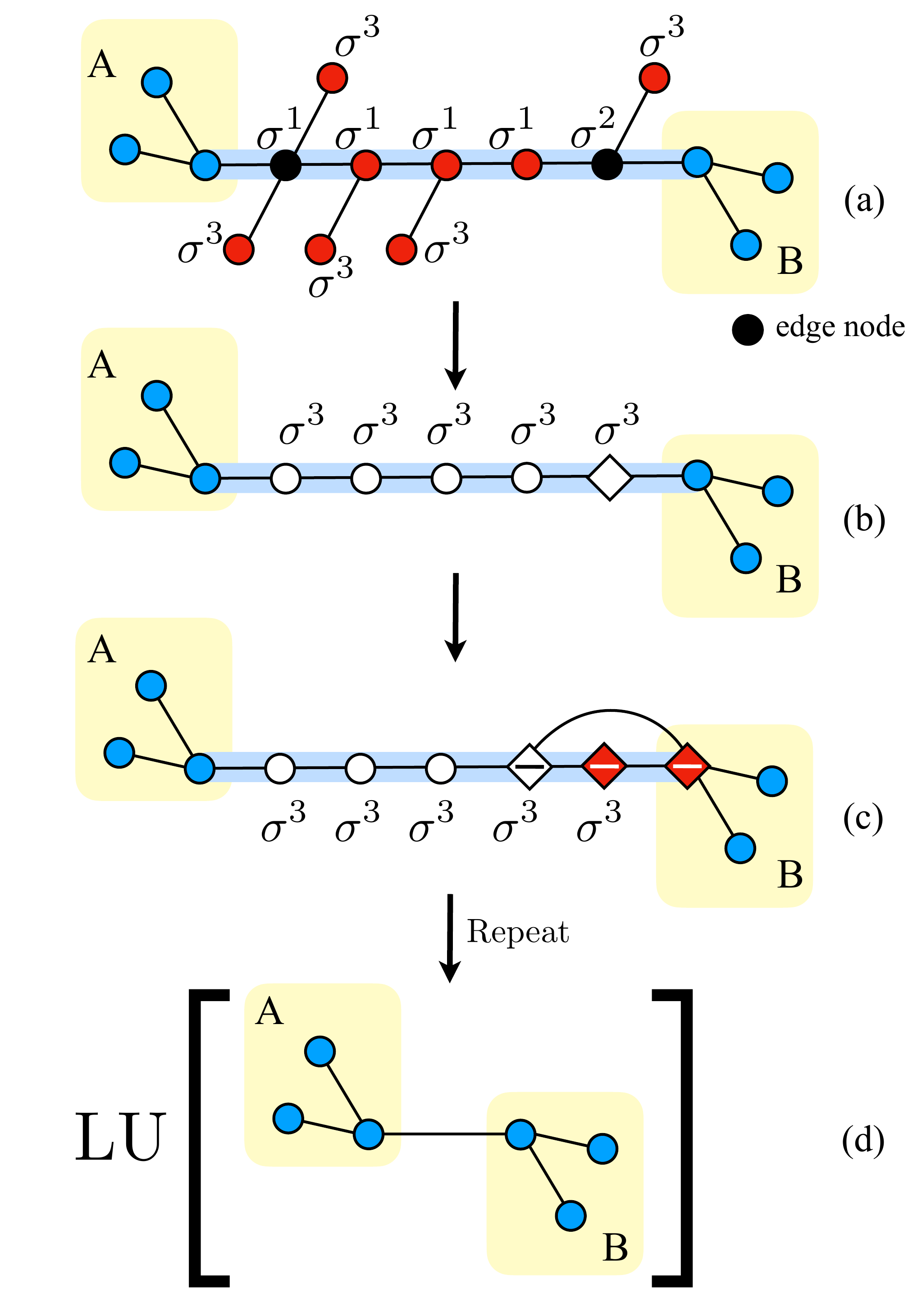}
    \caption{(a) Pauli measurement setup to create a direct link between two subsystems $A$, $B$ of a connected graph joined by a path (shaded), where the edge nodes are denoted by black dots. (b) Measurement transformations on each node on the path, including the edge nodes, where qubits outside the path are omitted for brevity. (c) The first of the five consecutive $\mathcal{B}_1$ operation on the nodes on the path that connects $A$ and $B$ during the graph reduction. (d) After the completion of graph reduction and the $\sigma^3$ measurements, $A$ and $B$ are connected by a direct link, and the corresponding graph state is equivalent to the state for this graph, up to local unitary operations (denoted by ``LU").}
    \label{fig:fig_app_H}
\end{figure}

Note that the above example includes a full description of separability of all GD states on a three-qubit graph, since all possible connected graphs on three qubits are local unitary equivalent to a star graph, representing a three-qubit GHZ state. Apart from the above example, the biseparability criteria for the case of four qubit GD states is also known~\cite{Guhne_2011}. Consider the four qubit graph state $\ket{G}_{1234}$, where the  qubits are labelled such that $1$ and $4$ are the edge qubits. For a diagonal state, $\rho$, in the graph state basis, where the fidelities in the said basis are defined as  $F_{ijkl}=\bra{ijkl}\rho \ket{ijkl}$ with $\ket{ijkl}=(\sigma^3)^i\otimes(\sigma^3)^j\otimes(\sigma^3)^k\otimes(\sigma^3)^l \ket{G}_{1234}$, $\forall i,j,k,l\in \{0,1\}$, the biseparability criteria is given by
\begin{eqnarray}
\label{eq:4cluster_biseparability_1}
F_{\alpha \beta \gamma \delta} \leq \frac{1}{2}\sum_{i,j} &&(F_{\alpha ij \delta} +F_{\overline{\alpha}ij \delta} + F_{\alpha ij \overline{\delta}}),\nonumber\\
\\
F_{\alpha \beta \gamma \delta}+F_{\overline{\alpha} \mu \nu \overline{\delta}} \leq \frac{1}{2}\sum_{i,j} && (F_{\alpha ij \delta} +F_{\overline{\alpha}ij \delta} + F_{\alpha ij \overline{\delta}}\nonumber\\
&&+F_{\overline{\alpha} ij\overline{\delta}}).
\label{eq:4cluster_biseparability_2}    
\end{eqnarray}
Here, $\overline{\alpha}=(1+\alpha)\mod2$. A violation of either of the inequalities (\ref{eq:4cluster_biseparability_1}) or (\ref{eq:4cluster_biseparability_2}) indicates existence of GME in the state $\rho$. 

\subsection{Genuine multiparty concurrence}

We also point out that the density matrix of a GHZD state has the form of an $X$ state.  For such states, one may compute the \emph{genuine multiparty concurrence}~\cite{Hashemi2012}(GMC) as a multipartite entanglement measure for $X$-states, which is defined as $C_{GM} = 2\max_j[0,\lambda]$ with 
\begin{eqnarray}
\lambda &=&\left|\rho_{(j,2^N-j-1)}\right|- \sum_{i=1,i\neq j}^{2^{N-1}-1} \left[\rho_{(1+i,1+i)}\rho_{(2^N-i,2^N-i)}\right]^{\frac{1}{2}}. \nonumber\\ 
\end{eqnarray}
We use this measure specifically in the case of the topological quantum codes under noise (see Sec.~\ref{sec:topological}). 

\section{Measurement protocol for localizing connected subsystem}
\label{app:xy_measurement_protocol}
In this section, we introduce a PMS through which any two subsystems $A,B$ of a connected graph $G$ can be directly connected by a link, given that there exists a path in the original graph $G$ connecting the two subsystems $A$ and $B$. Let us denote the nodes in the path that is connected to a qubit in either $A$ or $B$ as \emph{edge nodes} . The PMS, then, would be such that (see Fig.~\ref{fig:fig_app_H})
\begin{enumerate}
    \item[(a)] $\sigma^2$ measurement is performed on one of the edge nodes, while $\sigma^1$ measurement is performed on the rest of the nodes in the path, and
    \item[(b)] $\sigma^3$ measurement is performed on all other nodes in the graph.
\end{enumerate}
Note that the $\sigma^3$ measurements on qubits not in the path leave only $A$, $B$ and the path connected. Further, starting from the $\sigma^2$ measurement on the edge node, a series of $\mathcal{B}_1$ operations occur during the graph reduction (see Sec.~\ref{app:obtaining_GR}) along the path, so that $A$ and $B$, at the end, become connected by a direct link. Note also that if both the edge nodes share only one link each with $A$ and $B$ respectively, then the above PMS results in a direct link between $A$ and $B$ without changing the connectivity within $A$ and $B$. We use this PMS in the case of Fig.~(\ref{fig:toric_loop_qc_6}), with additional basis transformations according to Table~\ref{tab:pauli_transformation} due to the Hadamard operations on the control nodes connecting the stabilizer state from the toric code, and the local unitarily connected graph.

\bibliography{ref}

\end{document}